\documentclass[twocolumn,trackchanges]{aastex631}

\usepackage{graphicx}
\usepackage{subfigure}
\usepackage{multirow}
\usepackage{comment}
\usepackage{natbib}
\usepackage{hyperref}
\usepackage{mathtools}
\usepackage{longtable}
\usepackage{mathrsfs}
\usepackage{fontenc}
\usepackage{xcolor}
\usepackage{url}
\usepackage{gensymb}
\usepackage{pifont}

\newcommand{\sersic}{{S\'{e}rsic}}
\newcommand{\ledit}[1]{{{#1}}}

\newcommand{\ha}{\mbox{H$\alpha$}}
\newcommand{\hb}{\mbox{H$\beta$}}
\newcommand{\feii}{\mbox Fe\,{\sc ii}}

\makeatletter
\renewcommand{\doi}[1]{}
\makeatother

\shorttitle{High-redshift Quasar Host Galaxies}

\shortauthors{Li et al.}

\begin{document}

\title{The Dichotomy in the Nuclear and Host Galaxy Properties of High-redshift Quasars}

\author[0000-0001-8496-4162]{Ruancun Li}
\affil{Kavli Institute for Astronomy and Astrophysics, Peking University, Beijing 100871, China}
\affil{Department of Astronomy, School of Physics, Peking University, Beijing 100871, China}

\author[0000-0001-6947-5846]{Luis C. Ho}
\affil{Kavli Institute for Astronomy and Astrophysics, Peking University, Beijing 100871, China}
\affil{Department of Astronomy, School of Physics, Peking University, Beijing 100871, China}

\author[0009-0003-4721-177X]{Chang-Hao Chen}
\affil{Kavli Institute for Astronomy and Astrophysics, Peking University, Beijing 100871, China}
\affil{Department of Astronomy, School of Physics, Peking University, Beijing 100871, China}

\begin{abstract}

The early growth of high-redshift quasars and their host galaxies raises critical questions about their cosmic evolution. We exploit the angular resolution and sensitivity of NIRCam to investigate the host galaxies of 31 quasars at $4\lesssim z\lesssim7$ drawn from multiple JWST surveys. Using a new multiband forward-modeling code (\textsc{GalfitS}) that incorporates physically motivated priors, we securely detect and quantify the host emission in 30 objects, while simultaneously characterizing the nuclear spectral energy distribution. The host galaxies of high-redshift quasars are $\sim 0.3$~dex more compact than star-forming galaxies of comparable mass. A striking dichotomy emerges: luminous ``blue'' quasars ($L_{5100}\gtrsim10^{45}\,{\rm erg\,s^{-1}}$) reside in bulge-dominated galaxies (S\'ersic index $n \approx 5$) and exhibit a narrow range of ultraviolet nuclear slopes (median $\beta_{\rm UV} \approx -1.4$), while fainter ``red'' quasars inhabit disk-like hosts ($n\approx 1$) and display a broad range of slopes ($\beta_{\rm UV}\approx-2$ to 4). These two populations differ markedly in their black hole-to-stellar mass ratios, with high-luminosity quasars showing $M_{\mathrm{BH}}/M_\ast = 1.2\%$ compared to $4.7\%$ for lower luminosity sources, placing them collectively $\sim$0.6~dex above the local  $M_{\mathrm{BH}}-M_\ast$ relation. This offset likely reflects rapid black hole growth in early gas-rich environments, where feedback from the active galactic nucleus becomes effective only after substantial gas depletion. Our findings suggest that the observed dichotomy, whether due to intrinsic spectral differences or dust extinction, fundamentally shapes the coevolution of supermassive black holes and their host galaxies in the early Universe.

\end{abstract}


\section{Introduction}
\label{sec:sec1}

One of the most striking early discoveries from the James Webb Space Telescope (JWST) is the identification of numerous active galactic nuclei (AGNs) in the high-redshift ($z > 4$) Universe. Spectroscopic observations with JWST have revealed a population of broad-line AGNs with bolometric luminosities ranging from $\sim 10^{44}$ to $10^{46}\,{\rm erg\,s^{-1}}$, approximately 2 to 3 orders of magnitude lower than those of the bright quasars detected by ground-based surveys \citep{Harikane2023ApJ,Kocevski2023apjl,Maiolino2024AA,Matthee2024apj,Scholtz2025AA}. Powered by black holes (BHs) with masses of only $M_{\rm BH} \approx 10^6-10^8\,M_\odot$, these objects represent some of the least massive BHs known in the early Universe. \ledit{Photometric observations further show that a substantial subset of these faint broad-line AGNs, commonly referred to as ``little red dots'' (LRDs), are compact and exhibit unusually red rest-frame optical colors \citep[e.g.,][]{Greene2024ApJ,Kokorev2024apj,Akins2025ApJ,Kocevski2025ApJ,Labbe2025apj,Wang2025ApJ}. Although the overlap between photometrically selected LRDs and spectroscopically confirmed broad-line AGNs is not one-to-one and depends on the adopted selection criteria \citep[e.g.,][]{PerezGonzalez2024ApJ,Williams2024ApJ}, compact sources that morphologically resemble distant analogs of local quasars \citep{Wang2025ApJ} are very likely ($\sim 80\%$) to exhibit the ``V-shaped'' spectral energy distribution (SED) of LRDs \citep{Hviding2025AA}, characterized by a steep red rest-frame optical continuum and relatively blue ultraviolet (UV) colors \citep{Kokorev2024apj,Akins2025ApJ,Kocevski2025ApJ,Labbe2025apj}.} Their SEDs are often interpreted as the combined emission from a compact star-forming galaxy contributing to the UV and a dust-obscured active nucleus that dominates in the optical \citep{Greene2024ApJ,Wang2024ApJ}. Follow-up observations confirm broad emission lines in most photometrically selected LRDs \citep{Greene2024ApJ}, alongside detected variability \citep{Furtak2025AA}, reinforcing their classification as high-redshift AGNs with atypically red SEDs.

However, this evolutionary picture confronts two significant unresolved challenges. First, red AGNs appear ubiquitous in JWST deep fields, accounting for a few percent of galaxies at $z \approx 5$ \citep{Kokorev2024apj,Akins2025ApJ,Kocevski2025ApJ}. If all these objects are undergoing a prolonged, rapid growth phase (e.g., \citealp{Madau2025arXiv}), this could overshoot the BH mass function at the high-mass end \citep{Cavaliere1989ApJ}. Second, it remains unclear how the host galaxy evolves to bring the entire system into alignment with lower redshift observations. High-redshift quasars often deviate from the local $M_\mathrm{BH}–M_\ast$ scaling relation, hosting BHs that are overmassive relative to their galaxies, suggesting that central BHs grow rapidly before their hosts catch up. ALMA and JWST studies indicate that many quasars at $z \gtrsim 5$ harbor BHs that are $\sim 2–10$ times more massive relative to their galaxy’s stellar mass than predicted by local scaling relations \citep{Izumi2019PASJ,Pensabene2020AA,Maiolino2024AA,Yue2024ApJ}. In extreme cases, the BH mass rivals the stellar mass of the entire galaxy \citep{Goulding2023apjl,Stone2024ApJ,Chen2025}. Although the redshift evolution of the $M_\mathrm{BH}–M_\ast$ correlation remains uncertain \citep{Shen2015ApJ,Li2021ApJ,Zhuang2024ApJ,Silverman2025arXiv,Sun2025ApJ,Tanaka2025ApJ}, and some offsets may stem from selection effects \citep[e.g.,][]{Venemans2016ApJ,Li2025ApJ}, the discrepancy seems to persist even after accounting for such biases \citep{Ding2020ApJ,Sun2025ApJ}.

A detailed study of the properties of both the AGN and its host galaxy is essential to disentangle fully the two observational challenges outlined above, namely understanding the evolutionary pathways of high-redshift quasars and explaining their offset from the local $M_\mathrm{BH}–M_\ast$ scaling relation. However, the observed SED comprises contributions from both components, complicating their separation. At lower redshifts, broad-band SEDs, especially when X-ray data are included, can decompose effectively the signal contributions from the AGN and the host galaxy \citep{Yang2020MNRAS}. In contrast, the SED of high-redshift quasars is typically dominated by the AGN, rendering the host galaxy’s spectral features nearly undetectable \citep[e.g.,][]{Marshall2020ApJ}. Pure SED decomposition becomes challenging for fainter high-redshift AGNs because their broad-band spectrum often appears significantly redder, sometimes exhibiting a V-shaped distribution \citep{Lin2024ApJ,Labbe2025apj}.

Previous studies often relied on gas kinematics to measure the dynamical mass of the host galaxy to provide an indirect estimate of the total stellar mass of the system \citep{Izumi2019PASJ, Nguyen2020ApJ, Pensabene2020AA}. With JWST, we can now observe high-redshift quasar hosts, for the first time, in the rest-frame optical, which better traces stellar light. The high angular resolution and sensitivity of JWST deliver exceptionally deep images of quasar host galaxies previously accessible only for nearby AGNs using the Hubble Space Telescope \citep[e.g.,][]{Kim2008ApJ, Kim2017, Kim2021, Zhao2021}. Nevertheless, challenges persist: the angular sizes of high-redshift hosts are smaller on account of the larger angular diameter distances and their intrinsically compact nature \citep[e.g.,][]{vanderWel2014ApJ}, rendering them morphologically blended with the AGN point source. Nothwithstanding these difficulties, image decomposition techniques have detected successfully numerous quasar hosts at $z \gtrsim 6$ and measured their $M_\ast$ \citep{Ding2022ApJ, Yue2024ApJ, Marshall2025AA}, often complemented by point-spread function (PSF) subtraction methods \citep{Stone2024ApJ}. However, these approaches typically rely on a single filter---usually the reddest feasible band---leading to potential inconsistencies in the SED derived across different filters. Without comprehensive multiband imaging, a significant fraction of AGN hosts remains undetected \citep{Harikane2023ApJ, Yue2024ApJ}. Moreover, these studies poorly constrain the morphology or internal structure of the host galaxy, often assuming a specific functional form, such as a \citet{Sersic1968} function with index $n = 1$ \citep{Stone2024ApJ, Yue2024ApJ}. In summary: to accurately disentangle the SEDs of the AGN and its host, investigate their physical properties, and elucidate the early evolutionary pathways of high-redshift quasars and their location relative to the $M_\mathrm{BH}–M_\ast$ scaling relation, a self-consistent analysis of high-redshift quasars is imperative. This requires a multiband imaging decomposition technique that ensures consistency across wavelengths.

This paper is structured as follows. Section~\ref{sec:sec2} defines our sample and data reduction procedure for the JWST NIRCam images. We discuss our method to analyze the multiband data and conduct extensive tests to assess the reliability of the AGN host galaxy measurements (Section~\ref{sec:sec3}). Section~\ref{sec:sec4} presents our main results, including the mass and morphology of the host galaxy, as well as the properties of the AGN. Section~\ref{sec:sec5} discusses the potential implications for the growth path of high-redshift supermassive BHs. The main conclusions are summarized in Section~\ref{sec:sec6}. For the adopted $\Lambda$CDM cosmology, $\Omega_{m} = 0.308$, $\Omega_\Lambda = 0.692$, and $H_0 = 67.8\, \rm \; km \, s^{-1} \, Mpc^{-1}$ \citep{Planck2016AA}.

\section{Sample and Data Reduction}
\label{sec:sec2}

To investigate the early growth phases of high-redshift BHs and their host galaxies, we construct a comprehensive AGN sample by combining observations from several JWST surveys. Our final sample consists of 31 AGNs, comprising 15 powerful quasars and 16 lower luminosity counterparts at $z \gtrsim 4$.

\subsection{Data}

The high-luminosity quasars are drawn from three JWST programs.

\begin{itemize}
\item We use data from the Subaru High-redshift Exploration of Low-luminosity Quasars (SHELLQs) survey, which focuses on quasars at redshifts $6.0 < z < 6.4$ discovered by the Hyper Suprime-Cam Subaru Strategic Program \citep{Izumi2019PASJ}. From the 12 SHELLQs quasars observed with JWST/NIRCam in the F150W and F356W filters (PID 1967, PI: M.~Onoue), we select two (J2236+0032 and J2255+0251) that have publicly available NIRSpec spectroscopy analyzed by \citet{Ding2023Nature}. We adopt directly from \citet{Ding2023Nature} the luminosities at 5100\,\AA\ ($L_{5100}$; Table~\ref{tab:sample}) prior to any host-galaxy correction to ensure consistency across our sample.

\item We include data from the Emission-line Galaxies and Intergalactic Gas in the Epoch of Reionization (EIGER) survey (PID 1243; PI: S.~Lilly), a Guaranteed Time Observation program targetting quasars at redshifts $6 \lesssim z \lesssim 7$ that provides deep NIRCam imaging in the F115W, F200W, and F356W filters and spectra from Wide-Field Slitless Spectroscopy (WFSS). \citet{Yue2024ApJ} analyzed the WFSS data to derive spectral properties and physical parameters, including $L_{5100}$ and $M_{\mathrm{BH}}$. EIGER provides five quasars, after excluding J0100+2802, which is saturated in all NIRCam bands.

\item The JWST program SPectroscopic Survey of Biased Halos In the Reionization Era (ASPIRE; PID 2078, PI: F.~Wang) targets fields surrounding $z \approx 6-7$ quasars with NIRCam imaging (F115W, F200W, and F356W) and WFSS spectroscopy. We select eight quasars with redshifts $6.5 < z < 6.8$ whose WFSS F356W spectra have been analyzed by \citet{Yang2023ApJ}.
\end{itemize}

The low-luminosity quasars comprise two subsets.

\begin{itemize}
\item The first includes AGNs identified by \citet{Harikane2023ApJ} through JWST/NIRSpec spectroscopy that show \ledit{significant broad \ha\ emission ($\rm FWHM\gtrsim 1000 \rm \, km\, s^{-1}$) with signal-to-noise ratio $> 5$.} Among the 12 type~1 AGNs identified, four have publicly available JWST imaging from the Cosmic Evolution Early Release Science (CEERS; \citealt{Finkelstein2025}) survey covering all seven NIRCam bands (F115W, F150W, F200W, F277W, F356W, F410M, and F444W). For consistency with the rest of our sample, we recalculate $L_{5100}$ from the broad \ha\ luminosity (without intrinsic reddening correction) and convert the observed full width at half maximum (FWHM) of broad \ha\ to that of broad \hb\ following \citet{Greene2005ApJ}. These CEERS sources span $5 < z < 6$ with $L_{5100} \gtrsim 10^{43.5}\,\mathrm{erg\,s^{-1}}$.

\item The second subset consists of \ledit{broad \ha\ emitters (BHAEs; $\rm FWHM \gtrsim 1000 \rm \, km\, s^{-1}$) identified} in WFSS F356W spectra within the ASPIRE fields \citep{Lin2024ApJ}. From their initial sample of 16 sources, we exclude four that have incomplete imaging coverage or suffer from contamination by nearby stars, to yield 14 sources with redshifts ($4 < z < 5$) slightly lower compared to the CEERS subsample but that have comparable luminosities ($L_{5100} \gtrsim 10^{43.5}\,\mathrm{erg\,s^{-1}}$). We similarly convert their observed broad \ha\ luminosities to $L_{5100}$ and $\mathrm{FWHM}_{\ha}$ to $\mathrm{FWHM}_{\hb}$ based on \citet{Greene2005ApJ}.
\end{itemize}

We uniformly calculate the BH masses ($M_{\mathrm{BH}}$; Table~\ref{tab:sample}) following the virial method described by \citet{Ho2015ApJ}. In the absence of detailed bulge classification at high redshift, we adopt the virial factor suitable for both classical and pseudo bulges.

\subsection{Data Reduction}
\label{sec:data}

The NIRCam images were retrieved from the JWST Science Archive. For the CEERS fields, we downloaded complete imaging data for pointings 3 and 6, which cover all the CEERS objects under consideration in this study. For other targets, we selected all relevant JWST exposures within $6^\prime$ of each object to ensure coverage by both NIRCam modules A and B. The level~2 calibrated images (\texttt{*cal.fits*}) of each filter were combined to maximize signal-to-noise ratio.

Following JWST pipeline version 1.8.4, we first corrected the short-wavelength images (F115W, F150W, F200W) for ``wisp'' features and $1/f$ noise. Subsequently, large-scale background subtraction was performed using iterative masking and the \texttt{Background2D} routine in \textsc{photutils} according to the procedures outlined in \citet{Bagley2023ApJ}. Astrometric calibration involved aligning individual images using \textsc{Tweakwcs}, followed by absolute calibration against the Gaia DR3 catalog \citep{Gaia2023AA}. Individual images were inspected to remove those with severe pixel saturation or artifacts near target sources, which predominantly affect the EIGER and ASPIRE data. Cleaned images were then drizzled using the Image3Pipeline. To sample the PSF optimally, we adopt a pixel scale of $0\farcs015$ for shorter wavelengths (FWHM $=$ $0\farcs060 - 0\farcs075$) and $0\farcs03$ for longer wavelengths (FWHM $=$ $0\farcs119 - 0\farcs160$), based on the JWST PSF characterization by \citet{Zhuang2024ApJb}. This strategy ensures a Nyquist sampling rate approximately twice the instrumental PSF, which facilitates accurate morphological analysis.


\startlongtable
\begin{deluxetable*}{lcrrccllcc}
\tabletypesize{\tiny}
\setlength{\tabcolsep}{2.5pt}
\tablecaption{High-redshift Quasar Sample} \label{tab:sample}
\tabletypesize{\footnotesize}
\tablehead{
\colhead{Subsample}	    &
    \colhead{Name}          &
    \colhead{R.A.}  &
    \colhead{Decl.} &
    \colhead{$z$}           &
    \colhead{Dataset}       &
    \colhead{$\log{L_{5100}}$} &
    \colhead{$\rm FWHM_{H\beta ,\,  broad}$} &
    \colhead{$\log{M_\mathrm{BH}}$} &
    \colhead{Reference}    \\
&
&
(deg.) &
(deg.) &
&
&
\colhead{($\mathrm{erg\, s^{-1}}$)} &
\colhead{($\rm km \, s^{-1}$)} &
\colhead{($M_\odot$)} &
   \\
&
\colhead{(1)} &
\colhead{(2)} &
\colhead{(3)} &
\colhead{(4)} &
\colhead{(5)} &
\colhead{(6)} &
\colhead{(7)} &
\colhead{(8)} &
\colhead{(9)} 
}
\startdata
 High-luminosity   & J2236+0032        & 339.1858 & 0.54910  & 6.40 & SHELLQs      & $45.38\pm0.01$\tablenotemark{$\star$}     & $6290\pm560$  & $9.24\pm0.36$  & 1   \\
                   & J2255+0251        & 343.9085 &  2.8573  & 6.34 & SHELLQs      & $45.20\pm0.01$                            & $2560\pm110$  & $8.37\pm0.35$  & 1   \\
                   & J0148+0600        &  27.1568 &  6.0055  & 5.98 & EIGER        & $46.41\pm0.01$                            & $7828\pm485$  & $9.98\pm0.36$  & 2   \\
                   & J1030+0524        & 157.6130 &  5.4153  & 6.30 & EIGER        & $46.31\pm0.01$                            & $3669\pm15$   & $9.27\pm0.35$  & 2   \\
                   & J159+02           & 159.2258 & $-$2.5439  & 6.38 & EIGER        & $46.22\pm0.01$                            & $3493\pm29$   & $9.18\pm0.35$  & 2   \\
                   & J1120+0641        & 170.0061 &  6.6900  & 7.08 & EIGER        & $46.27\pm0.01$                            & $3337\pm111$  & $9.17\pm0.35$  & 2   \\
                   & J1148+5251        & 177.0693 & 52.8639  & 6.42 & EIGER        & $46.56\pm0.01$                            & $5370\pm81$   & $9.73\pm0.35$  & 2   \\
                   & J0109$-$3047        & 17.4714  & $-$30.7906 & 6.79 & ASPIRE       & $45.83\pm0.01$                            & $3033\pm117$  & $8.85\pm0.35$  & 3   \\
                   & J0218+0007        & 34.696   & 0.1209   & 6.77 & ASPIRE       & $45.85\pm0.01$                            & $3030\pm117$  & $8.86\pm0.35$  & 3   \\
                   & J0224$-$4711        & 36.1106  & $-$47.1915 & 6.52 & ASPIRE       & $46.49\pm0.01$                            & $3936\pm78$   & $9.43\pm0.35$  & 3   \\
                   & J0226+0302        & 36.5078  & 3.0498   & 6.54 & ASPIRE       & $46.26\pm0.01$                            & $3131\pm129$  & $9.11\pm0.35$  & 3   \\
                   & J0244$-$5008        & 41.0042  & $-$50.1482 & 6.73 & ASPIRE       & $46.20\pm0.01$                            & $2969\pm158$  & $9.03\pm0.35$  & 3   \\
                   & J0305$-$3150        & 46.3205  & $-$31.8489 & 6.61 & ASPIRE       & $45.86\pm0.01$                            & $3020\pm314$  & $8.86\pm0.37$  & 3   \\
                   & J2002$-$3013        & 300.6733 & $-$30.2227 & 6.69 & ASPIRE       & $46.16\pm0.01$                            & $2981\pm79$   & $9.01\pm0.35$  & 3   \\
                   & J2232+2930        & 338.2298 & 29.5089  & 6.67 & ASPIRE       & $45.65\pm0.01$                            & $6044\pm201$  & $9.35\pm0.35$  & 3   \\
\hline
    Low-luminosity & CEERS 007465      & 214.8091 & 52.8685  & 5.62 & CEERS        & $43.45\pm0.06$\tablenotemark{$\dagger$}   & $1803\pm162$  & $7.13\pm0.36$  & 4   \\
                   & CEERS 006725      & 214.8897 & 52.833   & 5.67 & CEERS        & $43.72\pm0.09$                            & $2419\pm278$  & $7.53\pm0.37$  & 4   \\
                   & CEERS 027825      & 214.8235 & 52.8303  & 5.24 & CEERS        & $44.13\pm0.28$                            & $2788\pm272$  & $7.87\pm0.39$  & 4   \\
                   & CEERS 003975      & 214.8362 & 52.8827  & 6.00 & CEERS        & $44.23\pm0.34$                            & $1892\pm352$  & $7.59\pm0.44$  & 4   \\
                   & J0109$-$3047-BHAE-1 &  17.4772 & $-$30.8353 & 4.36 & ASPIRE-BHAE  & $44.01\pm0.04$\tablenotemark{$\ddagger$}  & $3114\pm364$  & $7.90\pm0.37$  & 5   \\
                   & J0218+0007-BHAE-1 &   34.695 &   0.0874 & 4.23 & ASPIRE-BHAE  & $43.99\pm0.04$                            & $2207\pm187$  & $7.59\pm0.36$  & 5   \\
                   & J0224$-$4711-BHAE-1 &  36.0952 & $-$47.2007 & 4.06 & ASPIRE-BHAE  & $43.60\pm0.09$                            & $1946\pm494$  & $7.27\pm0.44$  & 5   \\
                   & J0229$-$0808-BHAE-1 &  37.4083 &  $-$8.1459 & 4.36 & ASPIRE-BHAE  & $43.82\pm0.04$                            & $1858\pm222$  & $7.35\pm0.37$  & 5   \\
                   & J0229$-$0808-BHAE-2 &  37.3964 &  $-$8.1954 & 5.04 & ASPIRE-BHAE  & $44.05\pm0.06$                            & $2576\pm404$  & $7.76\pm0.39$  & 5   \\
                   & J0430$-$1445-BHAE-1 &  67.6898 & $-$14.7873 & 4.09 & ASPIRE-BHAE  & $44.58\pm0.07$                            & $1847\pm15$   & $7.75\pm0.35$  & 5   \\
                   & J0923+0402-BHAE-1 & 140.9658 &   4.0545 & 4.87 & ASPIRE-BHAE  & $44.56\pm0.14$                            & $2242\pm133$  & $7.91\pm0.36$  & 5   \\
                   & J1526$-$2050-BHAE-2 & 231.6513 & $-$20.8566 & 4.87 & ASPIRE-BHAE  & $44.10\pm0.05$                            & $3763\pm466$  & $8.11\pm0.37$  & 5   \\
                   & J1526$-$2050-BHAE-3 &  231.671 & $-$20.8434 & 4.88 & ASPIRE-BHAE  & $43.60\pm0.15$                            & $2452\pm810$  & $7.47\pm0.49$  & 5   \\
                   & J2232+2930-BHAE-1 & 338.2275 &  29.4940 & 4.13 & ASPIRE-BHAE  & $44.29\pm0.05$                            & $2594\pm320$  & $7.89\pm0.37$  & 5   \\
                   & J2232+2930-BHAE-2 & 338.2381 &  29.5225 & 4.47 & ASPIRE-BHAE  & $43.64\pm0.05$                            & $1131\pm151$  & $6.82\pm0.38$  & 5   \\
                   & J2232+2930-BHAE-3 & 338.2325 &  29.5094 & 4.71 & ASPIRE-BHAE  & $43.72\pm0.08$                            & $1985\pm461$  & $7.35\pm0.42$  & 5   \\
\enddata
\tablecomments{Our high-redshift quasar sample is divided into two subsamples of high-luminosity and low-luminosity sources.
Col. (1): Name of the quasar.
Col. (2): Right Ascension (J2000).
Col. (3): Declination (J2000).
Col. (4): Spectroscopic redshift.
Col. (5): Dataset from which the quasar was selected.
Col. (6): Luminosity at 5100 \AA\ derived from rest-frame optical spectra.
Col. (7): FWHM of the broad \hb\ emission line. For the high-luminosity quasar subsample, $\rm FWHM_{H\beta ,\,  broad}$ is directly measured from the rest-frame optical spectrum, while for low-luminosity quasars we convert it based on the observed $\rm FWHM_{H\alpha ,\,  broad}$ following \citet{Greene2005ApJ}.
Col. (8): Black hole mass derived from $\rm FWHM_{H\beta ,\,  broad}$ and $L_{5100}$ based on \citet{Ho2015ApJ}.
Col. (9): Reference for the optical spectroscopy: 1 \citep{Ding2023Nature}; 2 \citep{Yue2024ApJ}; 3 \citep{Yang2023ApJ}; 4 \citep{Harikane2023ApJ}; 5 \citep{Lin2024ApJ}.
\tablenotetext{\star}{We adopt ${L_{5100}}$ without host galaxy subtraction from \citet{Ding2023Nature}, to be consistent with the EIGER and ASPIRE datasets.}
\tablenotetext{\dagger}{We adopt $L_{\rm H \alpha}$ from \citet{Harikane2023ApJ} before dust attenuation correction, to be consistent with the ASPIRE-BHAE dataset. We convert $L_{\rm H \alpha}$ to $L_{5100}$ and calculate $M_\mathrm{BH}$ based on the scaling relation of \citet{Greene2005ApJ}.}
\tablenotetext{\ddagger}{We obtain ${L_{5100}}$ using $L_{\rm H \alpha}$ from \citet{Lin2024ApJ}, following  \citet{Greene2005ApJ}. }
}
\end{deluxetable*}

\section{Method}
\label{sec:sec3}

To characterize robustly the physical properties of both AGNs and their host galaxies using multiband imaging, we perform a self-consistent image decomposition coupled with SED modeling. For this purpose, we utilize the newly developed code \textsc{GalfitS} \citep{GalfitS}, a forward-modeling decomposition tool designed explicitly for analyzing multiband imaging data\footnote{\ledit{Online documentation: \url{https://ruancunli.github.io/GalfitS/}.}}. Unlike traditional single-band imaging decomposition tools such as \textsc{Galfit} \citep{Peng2002AJ,Peng2010AJ} or \textsc{GaLight} \citep{Ding2020ApJ}, \textsc{GalfitS} simultaneously fits multi-filter images by physically linking morphological parameters across different bands, thereby maintaining model consistency with well-motivated astrophysical priors. Furthermore, \textsc{GalfitS} does not require uniform world coordinate system transformations or identical pixel sampling, as mandated by tools such as \textsc{GalfitM} \citep{Haussler2013MNRAS} or \textsc{ProFuse} \citep{Robotham2022MNRAS}. Crucially, \textsc{GalfitS} incorporates physically motivated SED models into the multiband fitting process, enabling simultaneous extraction of the physical properties of the AGN and its host galaxy across all available imaging data at their optimal pixel scales.

Below we describe our method to construct empirical PSFs tailored to each individual source, our procedure for AGN-host decomposition, and extensive input-output tests conducted to evaluate the reliability of the image decomposition and host galaxy detections. We then present the fitting results and discuss our approach to interpreting physically the derived AGN SEDs.

\begin{figure*}
\centering
\includegraphics[width=\textwidth]{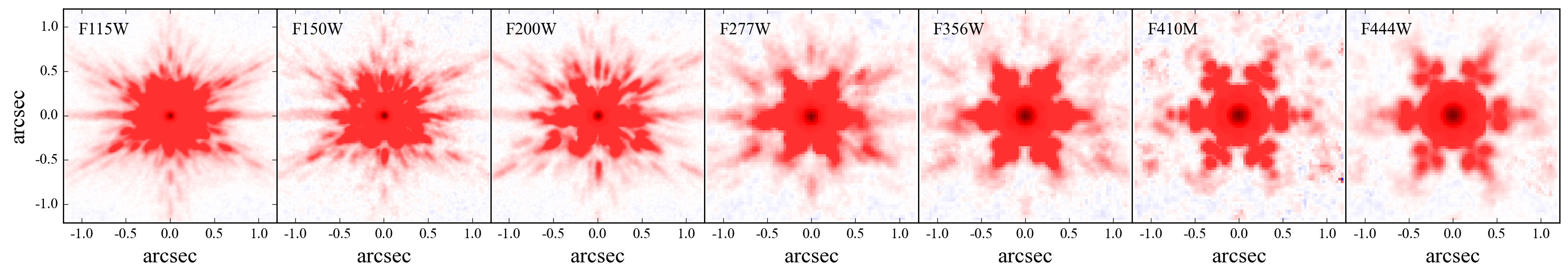}
\includegraphics[width=\textwidth]{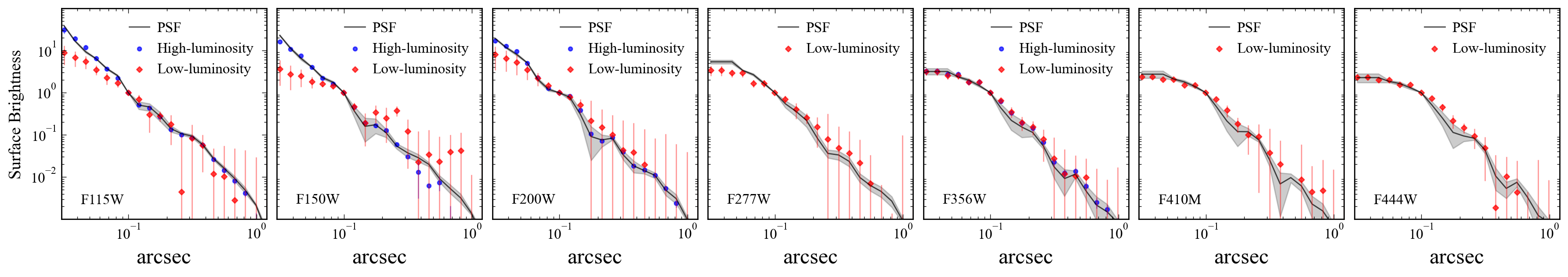}
\caption{Top row: examples of PSF models constructed for the seven NIRCam bands from CEERS pointing 1 data. Bottom row: median surface brightness profiles for fitted PSF models across all target fields (black lines), with the shaded grey region indicating one standard deviation. Median surface brightness profiles for high-luminosity and low-luminosity quasars are shown as blue and magenta points, respectively, with error bars representing one standard deviation. All profiles are normalized by their total flux.}
\label{fig:psf}
\end{figure*}

\subsection{Point-spread Function}
\label{sec:PSF}

Accurate PSF modeling is critical for robust AGN-host decomposition. For each target, we construct empirical PSFs from unsaturated, bright stars within the same NIRCam field. This approach is known to produce superior PSF models compared to purely theoretical methods such as \texttt{WebbPSF} (e.g., \citealp{Zhuang2024ApJb}). We select suitable stars in each band using \texttt{SExtractor} \citep{Bertin1996AAS}, following criteria recommended by \citet{Zhuang2024ApJb}. Candidate stars are identified as sources that have high signal-to-noise ratio (\texttt{SNR\_WIN} $>$ 100) and unblended (\texttt{FLAGS} $<$ 2), star-like (\texttt{CLASS\_STAR} $>$ 0.8) morphologies with elliptical elongations less than 1.5 and no problematic pixels (\texttt{IMAFLAGS\_ISO} $=$ 0). We then apply $3\,\sigma$-clipping around the median FWHM of the selected sources to exclude saturated stars. Importantly, we do not impose brightness limits relative to our quasars, for fainter stars naturally carry less weight in subsequent fitting procedures. Owing to variations in image size and field-of-view across different bands, the selection of PSF stars is performed independently in each filter. \ledit{Consequently, the number of PSF stars varies from band to band, typically 6--8 but ranging from as few as 4 to about 20.}

There are several commonly used methods to create empirical PSFs, such as those implemented in \textsc{PSFEx} and \textsc{photutils}, and hybrid approaches that combine the empirical core of an observed star with the theoretical wings from \textsc{WebbPSF} \citep{Yue2024ApJ,Zhuang2024ApJb,Chen2025}. Appendix~\ref{app:psfs} uses CEERS data to compare these approaches against a new direct-fitting method implemented within \textsc{GalfitS}. Our direct-fitting method consistently yields the lowest reduced chi-square values among the tested stars, and we adopt it for the subsequent analysis. Example PSF models fitted to seven NIRCam bands from CEERS pointing 1 are shown in the top panel of Figure~\ref{fig:psf}.

As a preliminary assessment of host galaxy detection, Figure~\ref{fig:psf} (bottom panel) compares the median one-dimensional surface brightness profile of the AGNs with that of their respective PSF, derived using concentric circular apertures. While the powerful quasars exhibit profiles nearly identical to those of the PSFs, consistent with their high luminosities ($L_{5100}\gtrsim10^{45}\,\mathrm{erg\,s^{-1}}$), the radial profile of lower luminosity AGNs ($L_{5100}\lesssim10^{45}\,\mathrm{erg\,s^{-1}}$) depart significantly from that of the PSF at \ledit{shorter wavelengths (F115W to F277W). This deviation diminishes progressively toward longer wavelengths (F356W and beyond),} indicating higher host-to-AGN flux ratios in the UV and relative to rest-frame optical wavelengths. It is interesting to note that this trend is the opposite of that traditionally observed \citep{Zhao2021} or assumed \citep{Ding2022ApJ} in most prior AGN host galaxy studies.

\subsection{AGN-host Decomposition}
\label{sec:galfits}

Given that our AGN sample comprises sources from five datasets, each different in the number of available photometric bands and range of intrinsic luminosities probed, the technical setup for AGN-host decomposition must be optimized carefully. Running \textsc{GalfitS} involves defining both a morphological and an SED model, potentially supplemented by astrophysical priors. Below we specify our \textsc{GalfitS} model configuration, summarize the priors applied during the fitting procedure, and describe how we sample the posterior distributions to determine the best-fit parameters.


\begin{deluxetable*}{lcccccccc}
		\tablecaption{\textsc{GalfitS} Setups for AGN-host Decomposition} \label{tab:gssetup}
		\tablehead{ \colhead{Dataset}
			& \colhead{$N_\mathrm{band}$}
      & \colhead{Galaxy SED model}
      & \colhead{AGN SED model}
      & \multicolumn{4}{c}{Priors}
			& \colhead{Note} \\
			&
			&
			&
			& \colhead{MSR}
			& \colhead{MER}
			& \colhead{MZR}
			& \colhead{SFH}
			&
		}
		\startdata
		SHELLQs        &  2     &  Kroupa IMF$+$Non-parametric SFH  & Non-parametric &  \ding{52} &  \ding{52} &  \ding{52} &  \ding{52} & \\
		EIGER          &  3     &  Kroupa IMF$+$Non-parametric SFH  & Non-parametric &  \ding{52} &  \ding{52} &  \ding{52} &  \ding{52} & \\
		ASPIRE         &  3     &  Kroupa IMF$+$Non-parametric SFH  & Non-parametric &  \ding{52} &  \ding{52} &  \ding{52} &  \ding{52} & \\
		CEERS          &  7     &  Kroupa IMF$+$Non-parametric SFH  & Non-parametric &  \ding{56} &  \ding{56} &  \ding{56} &  \ding{56} & \\
		A-BHAE         &  3     &  Kroupa IMF$+$Non-parametric SFH  & Non-parametric &  \ding{52} &  \ding{52} &  \ding{52} &  \ding{52} & \\
		\hline
		Galaxy         & 1/3/7    &  Kroupa IMF$+$Non-parametric SFH  & --- &  \ding{56} &  \ding{56} &  \ding{56} &  \ding{56} & \texttt{run1} \\
		Mock AGN       & 1/3/7    &  ---  & Non-parametric &  \ding{56} &  \ding{56} &  \ding{56} &  \ding{56} & \texttt{run2} \\
		       & 1/7    &  Kroupa IMF$+$Non-parametric SFH  & Non-parametric &  \ding{56} &  \ding{56} &  \ding{56} &  \ding{56} & \texttt{run3} \\
		       & 3      &  Kroupa IMF$+$Non-parametric SFH  & Non-parametric &  \ding{52} &  \ding{52} &  \ding{52} &  \ding{52} & \texttt{run3} \\
		       & 7      &  Kroupa IMF$+$Non-parametric SFH  & Typical quasar &  \ding{56} &  \ding{56} &  \ding{56} &  \ding{56} & \texttt{run4} \\
        Mock J1526$-$2050-BHAE-2    & 3    &  ---  & Non-parametric &  \ding{56} &  \ding{56} &  \ding{56} &  \ding{56} & \texttt{uplim} \\
         & 3      &  Kroupa IMF$+$Non-parametric SFH  & Non-parametric &  \ding{52} &  \ding{52} &  \ding{52} &  \ding{52} & \texttt{uplim} \\
	  \enddata
		\tablecomments{
		    We provide the optimal \textsc{GalfitS} setups for the AGN-host decomposition across the five datasets (Section~\ref{sec:galfits}). We also list the setups of our input-output test with mock AGN images (Section~\ref{sec:Merror}).
		}
		\label{tab:three-methods}
	\end{deluxetable*}

\begin{figure*}
\centering
\includegraphics[width=0.47\textwidth]{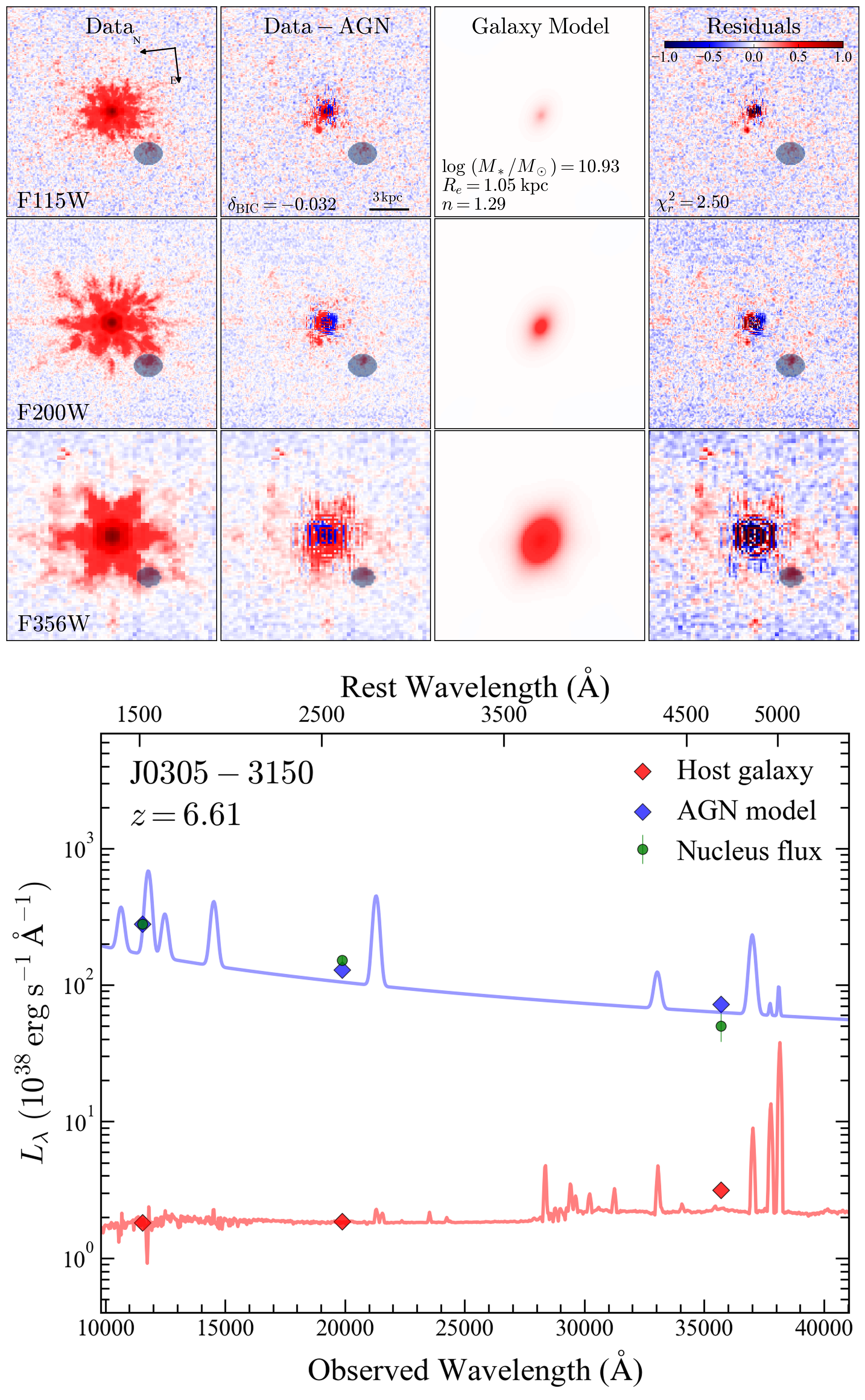}
\hspace{5mm}
\includegraphics[width=0.47\textwidth]{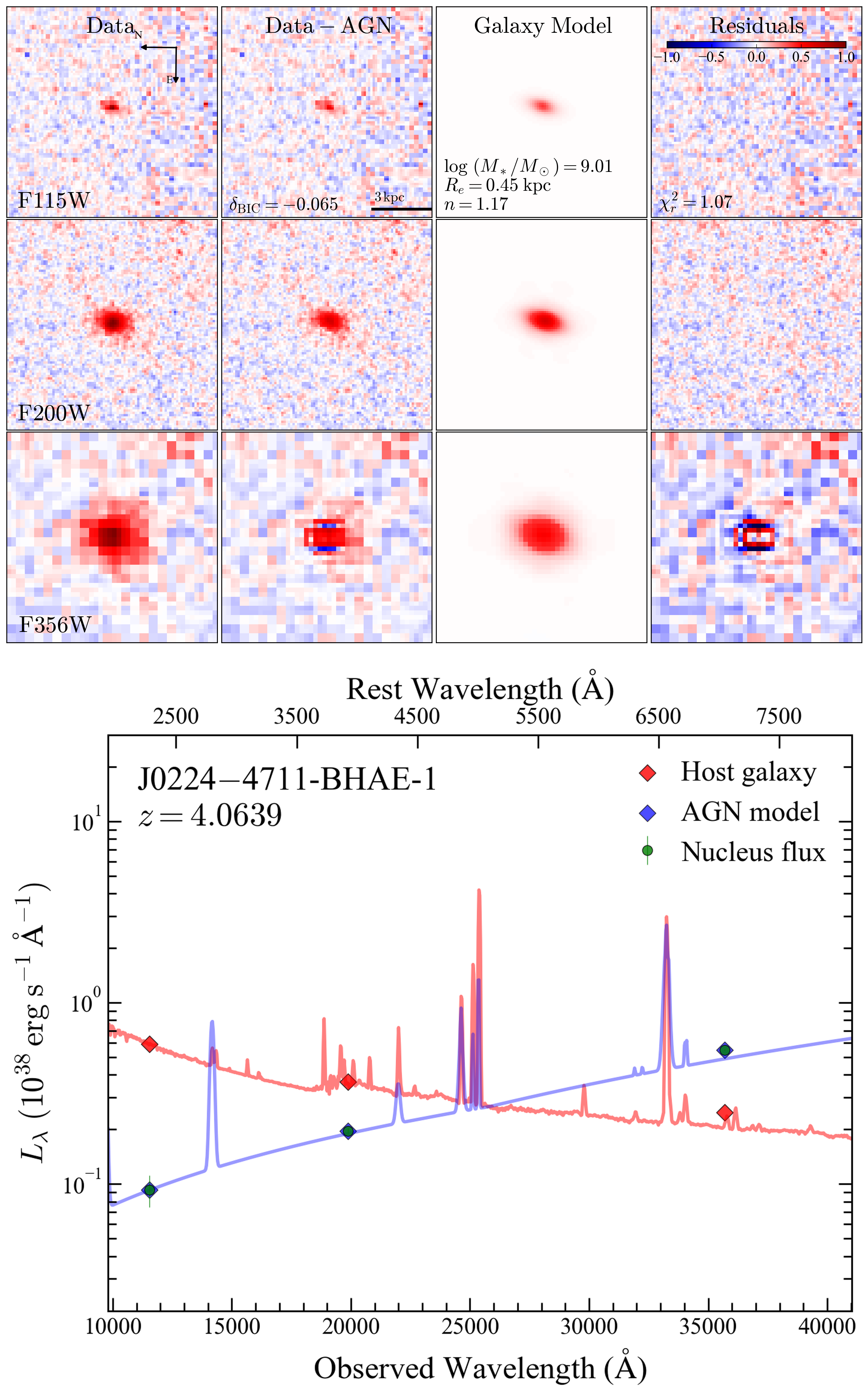}\\[1em]
\caption{Examples of \textsc{GalfitS} AGN–host decomposition for J0305$-$3150 (left) and J0224$-$4711-BHAE-1 (right), following the procedure detailed in Section~\ref{sec:galfits}. The top panels illustrate the imaging decomposition results for three filters (F115W, F200W, F356W), showing the original image, data with the best-fit AGN subtracted, best-fit galaxy model, and the residuals. Key structural parameters derived from the fits (stellar mass $M_\star$, effective radius $R_e$, and \sersic\ index $n$) are listed, along with the change in Bayesian information criterion ($\delta_{\rm BIC} < 0$ indicates host galaxy detection) and reduced chi-squared $\chi_r^2$. The bottom panel gives the observed-frame non-parametric flux for the nucleus (green circles), the host galaxy model (red curve), and the AGN SED fit (blue curve; Section~\ref{sec:agnsed}). The complete set of plots for all 31 AGNs is provided as an online figure set (Figure Set~1).}
\label{fig:galfits}
\end{figure*}

\begin{figure*}
\centering
\includegraphics[width=0.47\textwidth]{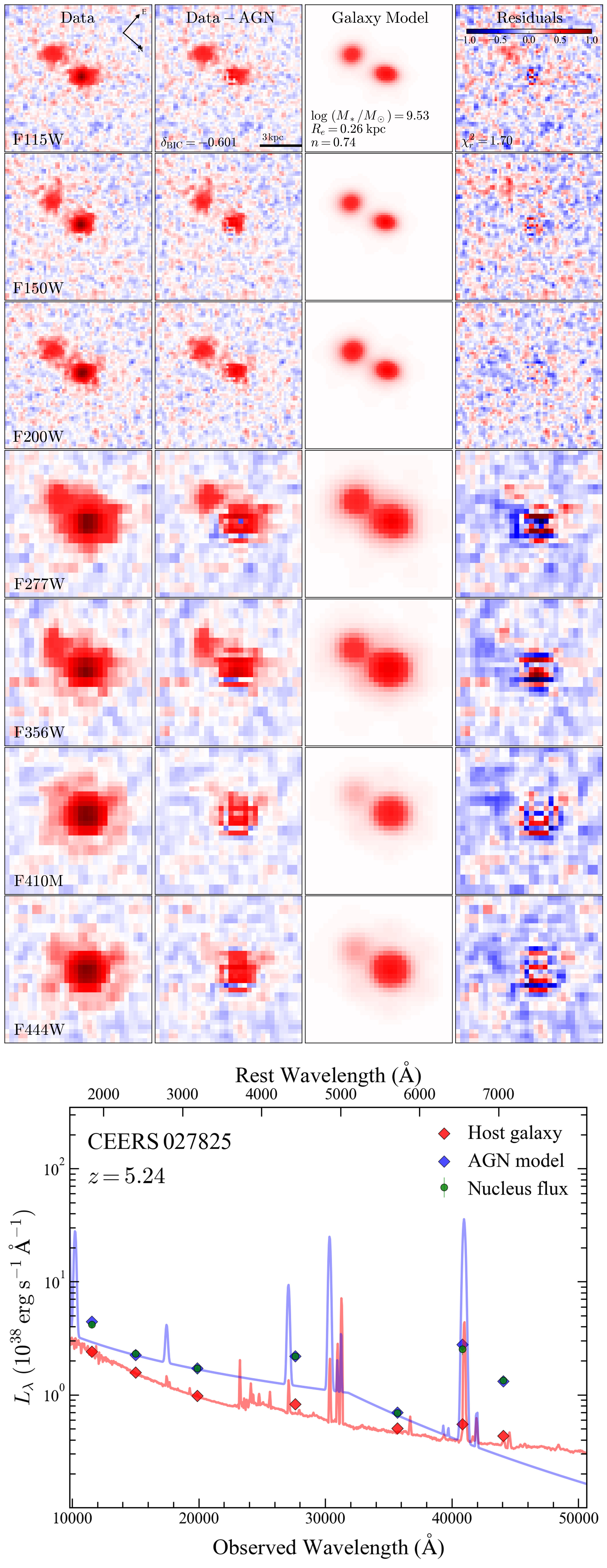}
\hspace{5mm}
\includegraphics[width=0.47\textwidth]{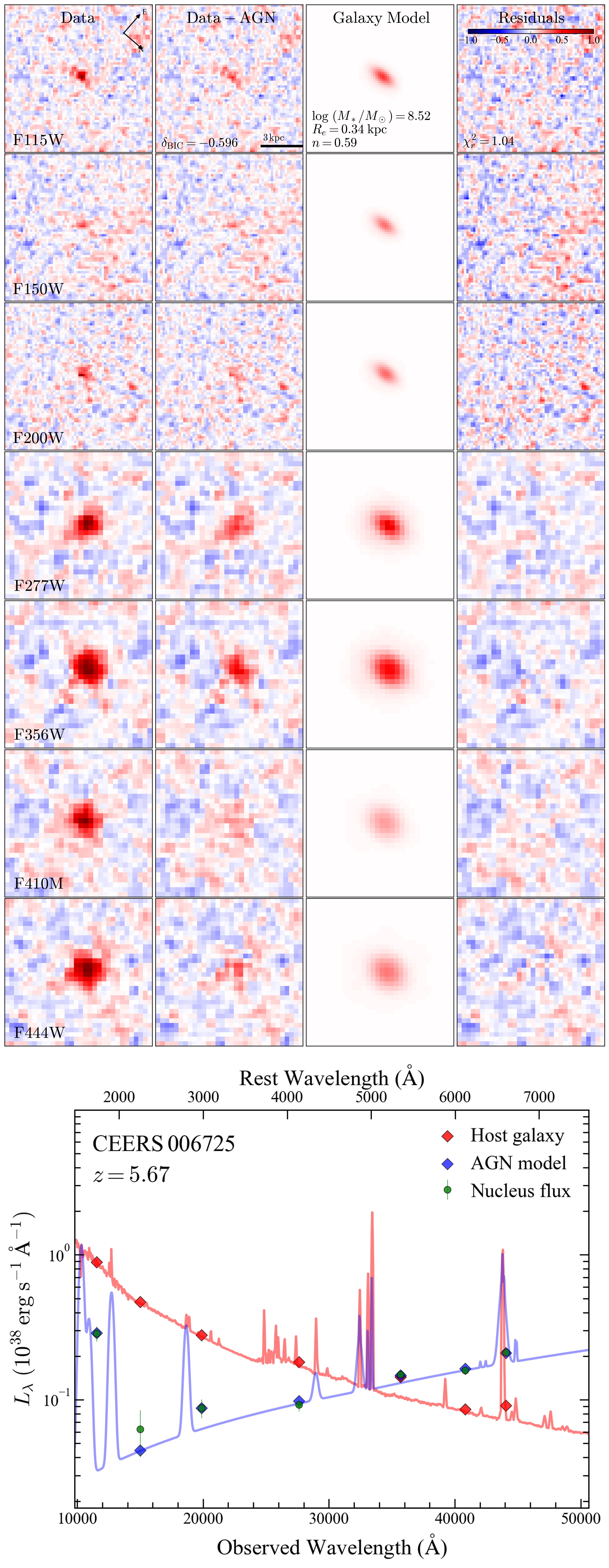}\\[1em]
\caption{Same as Figure~\ref{fig:galfits}, but for CEERS~027825 (left) and CEERS~006725 (right), using all seven NIRCam filters.}
\label{fig:galfits_exp7b}
\end{figure*}

\subsubsection{\textsc{GalfitS} Model Setup}
\label{sec:gssetup}

Using \textsc{GalfitS} to decompose multiband images of a galaxy into physically distinct subcomponents requires the specification of both morphological and SED models. \ledit{We use all available NIRCam images, fitting each target within a uniform $3\arcsec$ cutout that is slightly larger than our PSF model. All pixels are weighted equally. A three-band fit yields, for example, a total of $\sim 90{,}000$ degrees of freedom.} Considering the spatial resolution limit of NIRCam, which is $\sim 0\farcs06$ for the F115W filter \citep{Zhuang2024ApJb}, we represent the underlying stellar mass distribution of the host galaxy with a simplified model of a single \sersic\ component, whose parameters are the galaxy centroid position, \sersic\ index ($n$), effective radius ($R_e$), axis ratio ($q$), and position angle ($\Theta$). While \textsc{GalfitS} permits wavelength-dependent variations in galaxy morphology, which would be a manifestation of, for example, radial gradient in stellar population or dust extinction  (e.g., \citealt{Jin2025}), for simplicity we assume that the SED is constant with radius. We model the SED of the host galaxy using stellar population synthesis with a \cite{Kroupa2001MNRAS} stellar initial mass function and a non-parametric star formation history (SFH) motivated by the complex trends revealed by recent JWST observations of high-redshift galaxies (e.g., \citealt{Wang2024ApJS, Wright2024ApJ}. The stellar population parameters include stellar mass ($M_\ast$), star formation rate (SFR) within discrete time bins, metallicity ($Z$), and $V$-band dust extinction ($A_V$) assuming a \citet{Calzetti2000ApJ} extinction curve.

Figure~\ref{fig:galfits} gives two examples of fitted host galaxy models to illustrate the simultaneous derivation of morphological and SED fits. The AGN overwhelmingly dominates the host galaxy in J0305$-$3150, whereas in  J0224$-$4711-BHAE-1 the nucleus and galaxy are of comparable brightness. In both cases we only have three filters. In situations when nearby companions potentially contaminate the target (e.g., CEERS~027825; left panel of Figure~\ref{fig:galfits_exp7b}), we introduce an additional \sersic\ component to model the neighboring galaxy. In the absence of redshift information for the neighbor, we treat its redshift as a free parameter during the fit. Notes for individual objects are provided in Appendix~\ref{app:notes}.

We represent the AGN with a pure point source described by the empirical PSF. \textsc{GalfitS} offers both parametric and non-parametric options for modeling the nuclear SED. The parametric approach relies on predefined emission templates commonly employed in AGN spectroscopic analyses \citep[e.g.,][]{Ho2009,Li2022ApJ}. However, given the limited number of filters (typically three per source in our dataset), we opt for a non-parametric model, independently fitting the luminosities in each band. Even for the four CEERS AGNs with more extensive photometric coverage, our input-output tests (Appendix~\ref{app:iotest}) demonstrate that parametric models do not give significant improvements. Examples of the best-fit non-parametric nuclear SEDs are shown as green circles in Figures~\ref{fig:galfits} and \ref{fig:galfits_exp7b}. Table~\ref{tab:gssetup} summarizes the specific \textsc{GalfitS} modeling configurations for each dataset of our sample.

\subsubsection{\textsc{GalfitS} Prior Setup}
\label{sec:priors}

\textsc{GalfitS} allows the user the option to set up priors during the posterior sampling process to avoid unphysical solutions. Previous studies have encountered major major challenges when decomposing AGNs to derive the physical properties of their host galaxies. When the AGN light---described by the PSF---overshines the host galaxy, it can bias the solution toward an unphysically small effective radius ($R_e$) for the \sersic\ profile \citep{Stone2024ApJ} of the host. Moreover, the limited number of available filters (e.g., only two for the SHELLQs quasars and three for the EIGER and ASPIRE AGNs) hampers adequate sampling of the host galaxy SED. This, in turn, results in significant degeneracy among the SED parameters, particularly the SFH, metallicity, and dust extinction \citep{Conroy2013ARAA}. \textsc{GalfitS} addresses these challenges by incorporating astrophysical priors during the fitting process, as follows.

\ledit{First, to prevent the galaxy model from becoming artificially compact, we adopt a weak prior that AGN host galaxies occupy a similar region of parameter space in the stellar mass--size plane as mass-matched inactive galaxies at comparable redshifts. This assumption is empirically supported out to $z\lesssim 3.5$, where X-ray--selected AGN hosts are found to follow a mass--size relation broadly similar to that of star-forming and quiescent galaxies \citep{Rosario2015AA,Zhuang2024ApJ}.} During posterior sampling, we impose the stellar mass-size relation (MSR) as a prior to constrain the effective radius of the \sersic\ component according to

\begin{equation}
\log\, \left( \frac{R_e}{\mathrm{kpc}} \right) = \log \, A + \alpha \left[ \log \, \left( \frac{M_\ast}{M_\odot} \right) - 10.7 \right] + \sigma.
\label{equ:msr}
\end{equation}

\begin{figure*}
\centering
\includegraphics[width=0.98\textwidth]{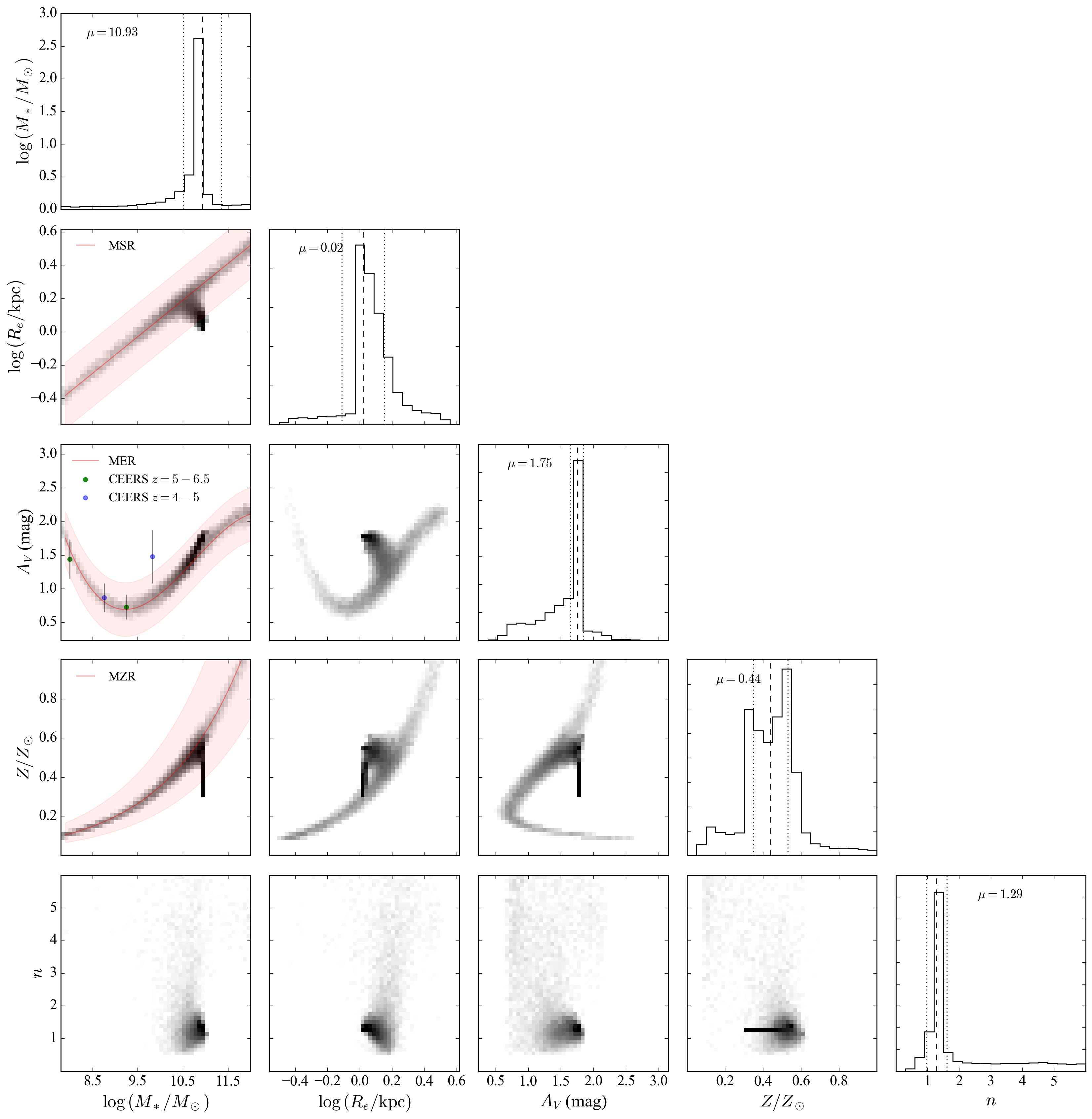}
\caption{Posterior distributions from our nested-sampling fit of the host galaxy of J0305$-$3150, showing the inferred stellar mass ($M_\ast$), effective radius ($R_e$), $V$-band extinction ($A_V$), metallicity ($Z$), and \sersic\ index ($n$). Top panels show the marginalized distributions with the best-fit values calculated based on the weighted mean: $\log \, (M_\star/M_\odot) = 10.93 \pm 0.42$, $\log \,  (R_\mathrm{e}/\mathrm{kpc}) = 0.02 \pm 0.13$, $A_V = 1.75 \pm 0.10$ mag, $Z = 0.44 \pm 0.09\, Z_\odot$, and $n=1.29 \pm 0.32$. The dashed lines give the weighted mean, and the dotted lines the $\pm1\,\sigma$ range of the full uncertainty listed in Table~\ref{tab:gsresultnp}. The first column shows the priors adopted during the posterior sampling: the mass-size relation (MSR, $0.2$~dex intrinsic scatter) from \citet{Allen2025AA}; the mass-dust extinction relation (MER, $0.4$~dex intrinsic scatter) from \citet{Maheson2024MNRAS}, with green and blue points marking typical CEERS galaxies at $z=5-6.5$ and $z=4-5$, respectively \citep{Shapley2023ApJ}, which are consistent with our inferred posterior constraints; and the mass-metalicity relation (MZR, $0.2$~dex intrinsic scatter) from \citet{Nakajima2023ApJS}.}
\label{fig:msr}
\end{figure*}

\noindent
Here, we adopt a slope of $\alpha = 0.22$ derived from JWST imaging analyses of high-redshift star-forming galaxies \citep{Allen2025AA}, which remains approximately constant for $3 \lesssim z \lesssim 7$. Consistent with previous results \citep{vanderWel2014ApJ}, the zero point varies with redshift as $\log \, A = -0.81\log\, (1+z)+0.95$ \citep{Allen2025AA}. For the luminous quasar sample at $6 \lesssim z \lesssim 7$, we adopt $\log \, A = 0.23$ based on their median redshift; for the less luminosity A-BHAE AGNs, we use $\log \, A = 0.35$ to reflect their lower redshifts. An intrinsic scatter of $\sigma = 0.2$~dex is assumed for both. In practice, as illustrated in Figure~\ref{fig:msr}, this implies that during sampling $R_e$ is drawn from a normal distribution with a median value of $\log \, A + \alpha\left[\log \, \left(\frac{M_\ast}{M_\odot}\right)-10.7\right]$ and a width given by $\sigma$.

Second, dust extinction in the host galaxy cannot be measured independently via the Balmer decrement (e.g., \citealp{Dominguez2013ApJ}) because our data do not cover both \ha\ and \hb\ in the observed wavelength range, not to mention of the difficulty of measuring the narrow lines in the presence of the much stronger broad lines from the AGN. As a result, $A_V$ becomes degenerate with other SED parameters that affect the colors, such as $Z$ and SFH. To mitigate this degeneracy, we adopt a scaling relation between dust extinction and stellar mass derived from the \ha/\hb\ decrement for nearly $10^6$ galaxies at $z\lesssim3$ \citep{Maheson2024MNRAS}. Assuming the extinction curve of \cite{Calzetti2000ApJ}, which is adopted in \textsc{GalfitS} (Section~\ref{sec:gssetup}), we convert \ha/\hb\ to $A_V$ and transform Equation~8 of \cite{Maheson2024MNRAS} to the stellar mass–dust extinction relation (MER)

\begin{equation}
\begin{aligned}
A_V &= a_0 + a_1 \log \, M_\ast + a_2 \bigl( \log \, M_\ast \bigr)^2 \\
&\quad + a_3 \bigl( \log \, M_\ast \bigr)^3 + a_4 \bigl( \log \, M_\ast \bigr)^4 + \sigma,
\label{equ:mar}
\end{aligned}
\end{equation}

\noindent
with $a_0 = 116.282$, $a_1 = -32.7843$, $a_2 = 2.98265$, $a_3 = -0.079877$, and $a_4 = -0.00059$.\footnote{The original \ha/\hb--$\log \, M_\ast$ relation was a third‐order polynomial, but converting \ha/\hb\ to $A_V$ introduces additional nonlinearity. We find that a fourth‐order function reproduces well the results of \citet{Maheson2024MNRAS}.} The MER should evolve with redshift on account of changes in metallicity with cosmic epoch. Moreover, Equation~\ref{equ:mar} pertains to galaxies at redshifts ($z\lesssim3$) lower than those of our quasar sample. However, overplotting recent JWST measurements of the Balmer decrement in galaxies at $4 \lesssim z \lesssim 6.5$ \citep{Shapley2023ApJ} reveals that they track our MER prior (Figure~\ref{fig:msr}). Instead of adopting the intrinsic scatter of $\sigma = 0.2$~dex from \citet{Maheson2024MNRAS}, we increase it to $\sigma = 0.4$~dex to better match the scatter observed in the JWST data.

Third, to mitigate degeneracies among metallicity, age, and SFH, we impose a stellar mass–metallicity relation (MZR) prior based on \citep{Nakajima2023ApJS}

\begin{equation}
\log\, (\mathrm{O/H}) + 12 = 5.74 + 0.25 \, \log \, M_\ast + \sigma,
\label{equ:mzr}
\end{equation}

\noindent
assuming an intrinsic scatter of $\sigma = 0.2$~dex. This relation is derived from JWST/NIRSpec data at redshifts $z = 4-10$, although it only covers the limited stellar mass range of $\log \, (M_\ast/M_\odot)=7.5-9.5$. The flattening of the MZR for $Z \gtrsim Z_\odot$ observed at $z=0$ \citep{Kewley2008ApJ} does not affect our fitting results because the metallicities of our sample are generallu sub-solar (Table~\ref{tab:gsresultnp}). Incorporating the MZR as a prior thus effectively constrains the metallicity based on stellar mass, reducing degeneracies with age and SFH \citep{Maiolino2008AA}.

\begin{figure}
\centering
\includegraphics[width=0.48\textwidth]{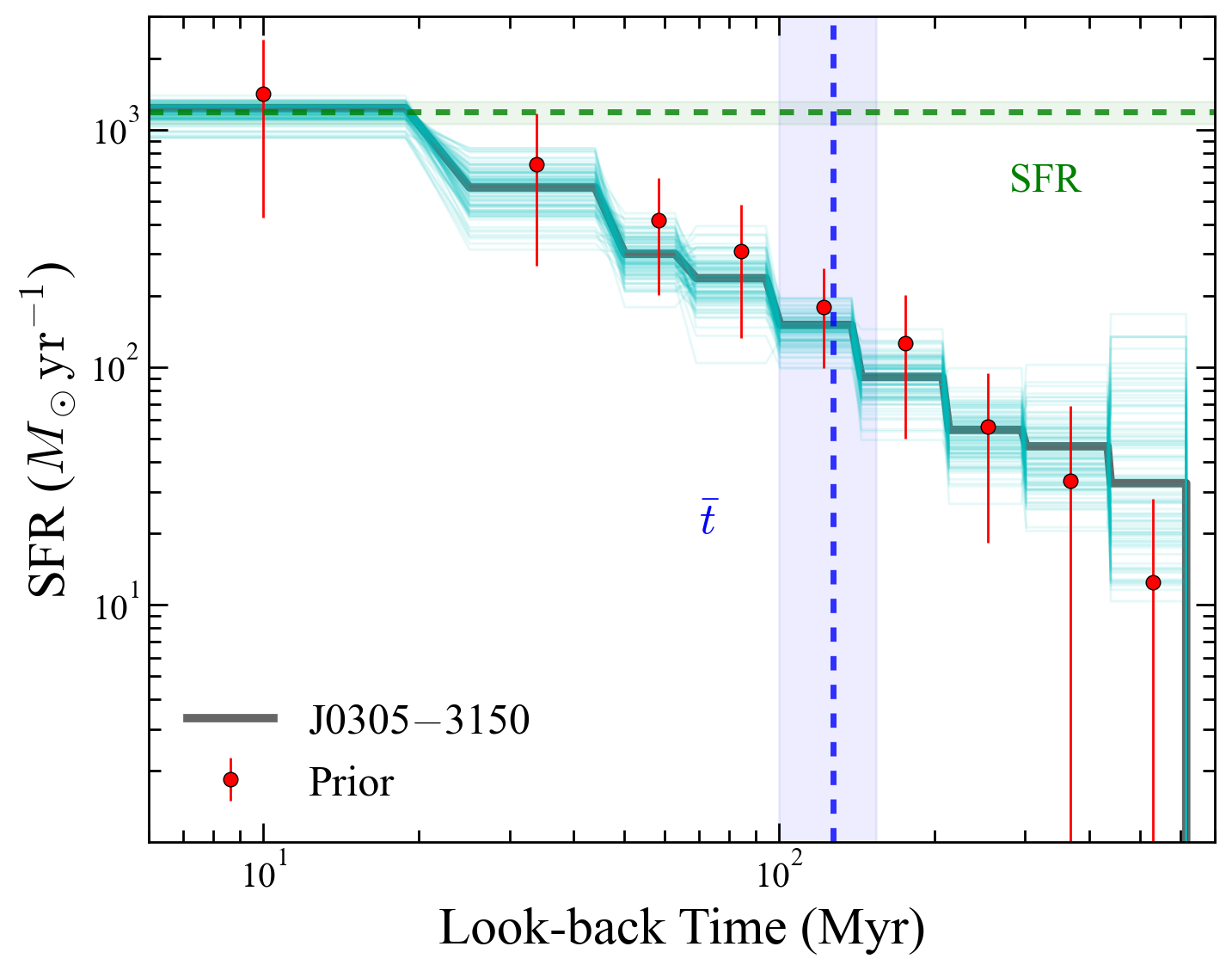}
\caption{Posterior star formation histories (SFHs) from our nested-sampling fit of the host galaxy of J0305$-$3150, plotted as a function of look-back time. The black curve shows the best-fit SFH, while the cyan curves represent random draws from the posterior. Red error bars indicate the prior star formation constraints used in the fit, derived by calculating the median observed SFR in each time bin from $\sim2500$ galaxies at similar redshifts \citep{Wang2024ApJS}. The error bars denote the 16\% and 84\% percentile values within each time bin. The green dashed line shows the SFR of the most recent time bin (20~Myr), and the blue dashed line shows the mass-weighted age of the galaxy (Equation~\ref{equ:bart}); the shaded region denotes the error. }
\label{fig:sfh}
\end{figure}

Lastly, as outlined in Section~\ref{sec:galfits}, we adopt a non‐parametric SFH \citep{Carnall2019ApJ} to avoid unphysical constraints that can arise from parametric models. In practice, however, most objects in our sample (except those in the CEERS subsample) have fewer than three photometric filters, which are insufficient to constrain robustly a non‐parametric SFH \citep{Leja2019ApJ}. Therefore, for all fits apart from those for the CEERS subsample, we impose a SFH prior derived from JWST galaxies observed at comparable redshifts. We define a uniform set of bins in look‐back time logarithmically spaced as [0.0, 0.02, 0.0477, 0.0689, 0.1, 0.144, 0.208, 0.301, 0.434, 0.628, 0.91]~Gyr, each truncated at a cosmic time of 0.2~Gyr to represent the approximate onset of galaxy formation \citep{Bouwens2015ApJ}. These SFH bins are applied to all objects, including those in the CEERS sample. For objects not in CEERS, we further refine the SFH by computing the median observed SFR in each bin, along with the $1\,\sigma$ scatter, using a sample of approximately 2500 JWST galaxies at $5.5 \lesssim z \lesssim 7.0$ \citep{Wang2024ApJS}\footnote{For the A-BHAE subsample, which occupies a lower redshift range, we use a slightly different set of time bins, namely [0, 0.02, 0.04, 0.062, 0.096, 0.149, 0.232, 0.359, 0.558, 0.865, 1.342]~Gyr, and we base the SFR priors on $\sim$4300 JWST galaxies at $4.0 \lesssim z \lesssim 5.0$ \citep{Wang2024ApJS}.}. During the fitting process, an overall normalization offset from these median SFRs is permitted, while the relative shape across the bins is constrained to match the observed distribution. An example of this procedure is illustrated in Figure~\ref{fig:sfh}.

We applied the aforementioned priors to the SHELLQs, EIGER, ASPIRE, and A-BHAE datasets (Table~\ref{tab:gssetup}), but not to the CEERS dataset because our input–output tests (Appendix~\ref{app:iotest}) indicate that the available filters are sufficient to constrain well the host galaxy parameters.

\subsubsection{Parameter Posterior Estimation}
\label{sec:posterior}

We use a nested‐sampling algorithm (e.g., \citealp{Skilling2004AIPC}) to estimate the posterior distributions because it explores high‐dimensional parameter spaces more efficiently than many traditional Markov chain Monte Carlo approaches (e.g., \citealt{ForemanMackey2013PASP}). For the galaxy stellar mass, we adopt a log‐uniform prior from $M_\ast = 10^{7.5}$ to $10^{12.5}\,M_\odot$. For $A_V$ and the SFR in different time bins, we use Gaussian priors (Section~\ref{sec:priors}) derived from JWST observations of real galaxies at similar redshifts. Regarding geometric parameters, \ledit{we allow the galaxy or PSF centroid to vary within $1 \arcsec\ $} of the image peak position, fitting the positions of the galaxy and PSF independently. The axis ratio has a uniform prior in the range $q = [0.2,1.0]$, motivated by the intrinsic thickness for typical galaxy morphologies \citep{Padilla2008MNRAS}, while the position angle is uniformly sampled over all possible orientations. The \sersic\ index has a uniform prior from $n = 0.5$ to $6$. For blended systems, we also allow the redshift of the nearby galaxy to vary from $0.5$ to $8.0$, as we generally lack information on whether the nearby source is projected or is physically associated. Detailed notes on the properties of the nearby sources appear in Appendix~\ref{app:notes}.

We use the Python package \texttt{dynesty} to carry out the nested‐sampling process, \ledit{employing $800$ live points (roughly 40 times the dimensionality of the parameter space) to ensure a well‐sampled posterior.} Figure~\ref{fig:msr} shows a corner plot of an example of the sampling for a subset of parameters. We take the weighted median for each parameter as our best‐fit value, with uncertainties derived from the diagonal elements of the weighted covariance matrix. Instead of giving the full parameters for the SFH modeling, we calculate the SFR during the recent $20\,$Myr. We also compute the mass-weighted look-back age,

\begin{equation}
\bar{t} = \frac{1}{M_\ast} \int_0^{t_{b}} \tau \, \mathrm{SFR}(\tau) \, d\tau ,
\label{equ:bart}
\end{equation}

\noindent
where $t_{b}$ represents the look-back time of the object when the cosmic age is 0.2~Gyr, and $\tau$ is the look-back time. \ledit{An example of the inferred SFR and the mass-weighted mean stellar age, $\bar{t}$, is shown in Figure~\ref{fig:sfh}. For J0305$-$3150, where the host is strongly affected by quasar overshining ($f_\mathrm{host}<0.1,f_\mathrm{AGN}$) and the number of available filters is limited, the best-fit SFH is driven largely by the adopted prior. This choice helps mitigate the age--metallicity--extinction degeneracy, as intended.\footnote{We also provide the model host galaxy magnitudes in Appendix~\ref{app:hostmag}, so that readers can directly adopt these values and apply their own SFH or SED assumptions to derive physical parameters.}} A summary of the properties of the host galaxies and the AGN is given in Tables~\ref{tab:gsresultnp} and \ref{tab:gsresultagn}, respectively.


\startlongtable
\begin{longrotatetable}
\begin{deluxetable*}{ccccccccccrcr}
\setlength{\tabcolsep}{2.5pt}
\tablecaption{Properties of the AGN Host Galaxies} \label{tab:gsresultnp}
\tabletypesize{\footnotesize}
\tablehead{
\colhead{Name}             &
\colhead{$n$} &
\colhead{$R_{e}$} &
\colhead{$q$} &
\colhead{$\Theta$} &
\colhead{$\log\,{M_\ast}$}   &
\colhead{SFR} &
\colhead{$\bar{t}$} &
\colhead{$A_V$} &
\colhead{$Z$} &
\colhead{$\log\,{f_\mathrm{F356W}}$} &
\colhead{$\chi^2_r$} &
\colhead{$\delta_\mathrm{BIC}$}\\
&
&
\colhead{(kpc) } &
&
\colhead{($^{\circ}$)} &
\colhead{($M_\odot$)} &
\colhead{($M_\odot\, \mathrm{yr^{-1}}$)} &
\colhead{(Gyr)} &
\colhead{(mag)}    &
\colhead{($Z_\odot$)} &
&
& \\
\colhead{(1)} &
\colhead{(2)} &
\colhead{(3)} &
\colhead{(4)} &
\colhead{(5)} &
\colhead{(6)} &
\colhead{(7)} &
\colhead{(8)} &
\colhead{(9)} &
\colhead{(10)} &
\colhead{(11)} &
\colhead{(12)} &
\colhead{(13)}
}
\startdata
J2236$+$0032         & $3.36\pm0.22$ & $0.34\pm0.31$ & $0.38\pm0.14$ & $137.65\pm17.63$ & $11.09\pm0.32$                    & $2024\pm241$  & $0.09\pm0.01$ & $1.40\pm0.14$ & $0.66\pm0.04$ & $0.32\pm0.10$  & $0.76$ & $-0.456$ \\
J2255$+$0251         & $1.12\pm0.33$ & $1.14\pm0.23$ & $0.68\pm0.10$ & $34.51\pm27.41$  & $10.28\pm0.32$                    & $226\pm39$    & $0.10\pm0.01$ & $1.62\pm0.27$ & $0.39\pm0.03$ & $0.34\pm0.05$  & $0.69$ & $-0.105$ \\
J0148$+$0600         & $5.97\pm0.43$ & $0.94\pm0.30$ & $0.31\pm0.21$ & $107.27\pm50.24$ & $11.08\pm0.50$                    & $1807\pm421$  & $0.12\pm0.10$ & $1.43\pm0.13$ & $0.66\pm0.07$ & $1.47\pm0.09$  & $4.09$ & $-0.017$ \\
J1030$+$0524         & $5.64\pm0.38$ & $0.52\pm0.28$ & $0.31\pm0.16$ & $59.48\pm52.71$  & $10.70\pm0.45$                    & $712\pm68$    & $0.09\pm0.01$ & $1.60\pm0.23$ & $0.53\pm0.18$ & $1.48\pm0.03$  & $2.88$ & $-0.005$ \\
J159$+$02            & $1.07\pm0.46$ & $0.31\pm0.19$ & $0.39\pm0.19$ & $130.59\pm41.51$ & $10.13\pm0.45$                    & $228\pm25$    & $0.10\pm0.01$ & $0.05\pm0.40$ & $0.34\pm0.09$ & $1.43\pm0.07$  & $2.02$ & $-0.113$ \\
J1120$+$0641         & $1.23\pm0.35$ & $0.35\pm0.10$ & $0.30\pm0.18$ & $155.04\pm20.72$ & $10.80\pm0.45$                    & $893\pm100$   & $0.10\pm0.01$ & $0.87\pm0.24$ & $0.52\pm0.03$ & $1.39\pm0.05$  & $1.39$ & $-0.091$ \\
J1148$+$5251         & $2.00\pm0.38$ & $1.39\pm0.27$ & $0.49\pm0.15$ & $124.91\pm30.30$ & $11.48\pm0.48$                    & $2651\pm460$  & $0.09\pm0.01$ & $1.63\pm0.27$ & $0.83\pm0.04$ & $1.66\pm0.03$  & $7.77$ & $-0.006$ \\
J0109$-$3047         & $4.96\pm0.33$ & $0.98\pm0.36$ & $0.55\pm0.13$ & $12.05\pm17.77$  & $11.27\pm0.43$                    & $2516\pm352$  & $0.08\pm0.01$ & $2.30\pm0.27$ & $0.55\pm0.09$ & $0.93\pm0.10$  & $0.89$ & $-0.028$ \\
J0218$+$0007         & $5.97\pm0.27$ & $0.67\pm0.37$ & $0.50\pm0.09$ & $167.16\pm9.14$  & $11.41\pm0.39$                    & $3811\pm433$  & $0.09\pm0.01$ & $1.78\pm0.11$ & $0.76\pm0.03$ & $0.90\pm0.04$  & $1.34$ & $-0.090$ \\
J0224$-$4711         & $5.85\pm0.33$ & $1.04\pm0.48$ & $0.39\pm0.09$ & $170.88\pm29.16$ & $11.64\pm0.43$                    & $3754\pm900$  & $0.08\pm0.01$ & $1.90\pm0.32$ & $0.26\pm0.12$ & $1.64\pm0.05$  & $3.61$ & $-0.107$ \\
J0226$+$0302         & $5.76\pm0.44$ & $1.22\pm0.11$ & $0.32\pm0.23$ & $39.96\pm45.29$  & $10.21\pm0.45$                    & $288\pm33$    & $0.09\pm0.01$ & $0.33\pm0.35$ & $0.40\pm0.02$ & $1.41\pm0.08$  & $3.48$ & $-0.036$ \\
J0244$-$5008         & $3.92\pm0.28$ & $3.12\pm0.61$ & $0.30\pm0.19$ & $136.59\pm29.21$ & $11.23\pm0.44$                    & $1909\pm303$  & $0.08\pm0.01$ & $2.28\pm0.28$ & $0.46\pm0.08$ & $1.33\pm0.03$  & $2.29$ & $-0.005$ \\
J0305$-$3150         & $1.29\pm0.32$ & $1.05\pm0.32$ & $0.66\pm0.07$ & $121.70\pm35.48$ & $10.93\pm0.42$                    & $1189\pm129$  & $0.13\pm0.03$ & $1.75\pm0.10$ & $0.44\pm0.09$ & $1.06\pm0.11$  & $2.50$ & $-0.032$ \\
J2002$-$3013         & $5.43\pm0.39$ & $1.41\pm0.34$ & $0.35\pm0.20$ & $175.10\pm25.56$ & $11.48\pm0.44$                    & $4468\pm396$  & $0.09\pm0.01$ & $2.31\pm0.19$ & $0.77\pm0.05$ & $1.31\pm0.05$  & $2.49$ & $-0.015$ \\
J2232$+$2930         & $5.31\pm0.29$ & $0.60\pm0.30$ & $0.46\pm0.09$ & $147.23\pm17.72$ & $10.96\pm0.41$                    & $1424\pm167$  & $0.09\pm0.01$ & $1.32\pm0.14$ & $0.56\pm0.03$ & $1.07\pm0.05$  & $1.64$ & $-0.054$ \\
CEERS 007465         & $0.51\pm0.36$ & $0.97\pm0.23$ & $0.38\pm0.14$ & $66.70\pm29.55$  & $8.43\pm0.12$                     & $7.3\pm2.7$   & $0.05\pm0.11$ & $0.31\pm0.82$ & $0.15\pm0.01$ & $-0.89\pm0.03$ & $1.33$ & $-0.604$ \\
CEERS 006725         & $0.59\pm0.41$ & $0.34\pm0.16$ & $0.32\pm0.22$ & $177.45\pm28.69$ & $8.52\pm0.15$                     & $2.9\pm4.2$   & $0.08\pm0.11$ & $0.34\pm1.26$ & $0.15\pm0.01$ & $-1.19\pm0.12$ & $1.04$ & $-0.596$ \\
CEERS 027825         & $0.74\pm0.29$ & $0.26\pm0.04$ & $0.57\pm0.15$ & $149.00\pm19.94$ & $9.53\pm0.18$                     & $19.7\pm14.1$ & $0.21\pm0.09$ & $0.59\pm0.14$ & $0.45\pm0.05$ & $-0.16\pm0.03$ & $1.70$ & $-0.601$ \\
CEERS 003975         & $0.60\pm0.19$ & $0.66\pm0.03$ & $0.56\pm0.06$ & $132.88\pm2.40$  & $9.92\pm0.13$                     & $92.9\pm49.1$ & $0.17\pm0.06$ & $0.30\pm0.04$ & $0.25\pm0.03$ & $0.07\pm0.04$  & $1.58$ & $-0.503$ \\
J0109$-$3047-BHAE-1  & $5.40\pm0.42$ & $0.29\pm0.18$ & $0.42\pm0.15$ & $42.01\pm33.51$  & $9.03\pm0.11$                     & $11.4\pm3.1$  & $0.14\pm0.02$ & $0.95\pm0.14$ & $0.21\pm0.01$ & $-0.97\pm0.04$ & $0.90$ & $-0.025$ \\
J0218$+$0007-BHAE-1  & $1.30\pm0.44$ & $0.27\pm0.14$ & $0.32\pm0.23$ & $146.00\pm34.63$ & $8.56\pm0.12$                     & $4.9\pm1.6$   & $0.14\pm0.02$ & $0.97\pm0.09$ & $0.16\pm0.01$ & $-0.83\pm0.05$ & $0.90$ & $-0.002$ \\
J0224$-$4711-BHAE-1  & $1.17\pm0.43$ & $0.45\pm0.17$ & $0.34\pm0.19$ & $15.19\pm17.96$  & $9.01\pm0.09$                     & $14.0\pm3.7$  & $0.15\pm0.02$ & $0.87\pm0.11$ & $0.20\pm0.01$ & $-1.11\pm0.05$ & $1.07$ & $-0.065$ \\
J0229$-$0808-BHAE-1  & $4.67\pm0.30$ & $1.16\pm0.18$ & $0.43\pm0.12$ & $62.88\pm25.28$  & $9.15\pm0.06$                     & $13.5\pm2.8$  & $0.18\pm0.02$ & $0.78\pm0.06$ & $0.22\pm0.01$ & $-1.13\pm0.03$ & $0.80$ & $-0.129$ \\
J0229$-$0808-BHAE-2  & $1.85\pm0.26$ & $0.35\pm0.25$ & $0.77\pm0.12$ & $132.29\pm52.79$ & $9.46\pm0.10$                     & $7.7\pm1.6$   & $0.27\pm0.02$ & $0.36\pm0.05$ & $0.30\pm0.01$ & $-1.24\pm0.03$ & $0.74$ & $-0.074$ \\
J0430$-$1445-BHAE-1  & $0.74\pm0.22$ & $1.03\pm0.11$ & $0.33\pm0.12$ & $140.50\pm5.54$  & $8.84\pm0.15$                     & $7.9\pm2.1$   & $0.17\pm0.02$ & $0.47\pm0.14$ & $0.18\pm0.01$ & $-0.59\pm0.05$ & $0.83$ & $-0.115$ \\
J0923$+$0402-BHAE-1  & $5.58\pm0.32$ & $0.08\pm0.08$ & $0.48\pm0.18$ & $111.85\pm45.39$ & $9.10\pm0.16$                     & $6.4\pm3.5$   & $0.17\pm0.02$ & $0.83\pm0.14$ & $0.19\pm0.01$ & $-0.43\pm0.04$ & $0.76$ & $-0.216$ \\
J1526$-$2050-BHAE-2  & $4.12\pm0.44$ & $0.63\pm0.12$ & $0.62\pm0.21$ & $179.87\pm59.96$ & $< 8.59$\tablenotemark{a}         & $2.4\pm0.8$   & $0.14\pm0.02$ & $1.17\pm0.09$ & $0.14\pm0.01$ & $-0.98\pm0.07$ & $1.41$ & $0.002$  \\
J1526$-$2050-BHAE-3  & $4.78\pm0.40$ & $0.73\pm0.11$ & $0.56\pm0.22$ & $1.91\pm33.23$   & $8.65\pm0.10$                     & $4.9\pm1.2$   & $0.14\pm0.02$ & $0.90\pm0.08$ & $0.17\pm0.01$ & $-1.12\pm0.06$ & $0.65$ & $-0.008$ \\
J2232$+$2930-BHAE-1  & $0.77\pm0.03$ & $1.28\pm0.21$ & $0.81\pm0.05$ & $41.46\pm12.46$  & $10.39\pm0.10$                    & $67.6\pm34.6$ & $0.21\pm0.02$ & $1.66\pm0.19$ & $0.47\pm0.02$ & $-1.87\pm0.04$ & $1.06$ & $-0.574$ \\
J2232$+$2930-BHAE-2  & $0.95\pm0.14$ & $0.32\pm0.28$ & $0.33\pm0.16$ & $163.33\pm23.58$ & $8.38\pm0.06$                     & $3.6\pm0.8$   & $0.13\pm0.01$ & $0.89\pm0.11$ & $0.22\pm0.02$ & $-1.43\pm0.03$ & $0.74$ & $-0.001$ \\
J2232$+$2930-BHAE-3  & $0.58\pm0.49$ & $0.31\pm0.14$ & $0.36\pm0.20$ & $47.31\pm37.90$  & $8.70\pm0.09$                     & $5.7\pm1.7  $ & $0.15\pm0.02$ & $0.75\pm0.10$ & $0.17\pm0.01$ & $-1.12\pm0.06$ & $1.02$ & $-0.008$ \\
\enddata
\tablecomments{Best-fit parameters from the \textsc{GalfitS} analysis of the high-redshift quasars.
Col. (1): Name of the quasar.
Col. (2): S\'ersic index of the galaxy profile.
Col. (3): Effective radius.
Col. (4): Axis ratio.
Col. (5): Position angle, measured counterclockwise from the North.
Col. (6): Stellar mass of the host galaxy.
Col. (7): Star formation rate within the recent 20~Myr time bin.
Col. (8): Mass-weighted mean age (Equation~\ref{equ:bart}).
Col. (9): Dust extinction in the $V$ band.
Col. (10): Stellar metallicity.
Col. (11): AGN-to-host flux ratio in the F356W filter.
Col. (12): Reduced chi-squared of the fit.
Col. (13): Relative difference of the Bayesian information criterion after adding the host galaxy component (Equation~\ref{equ:dbic}). 
\tablenotetext{a}{From the direct output of our nested-sampling fit, the stellar mass of J1526$-$2050-BHAE-2 is $\log\, (M_\ast/M_\odot) = 8.32 \pm 0.11$. However, with $\delta_\mathrm{BIC} > 0$, the host galaxy is considered undetected. We obtain a robust estimate of the upper limit of $\log\, (M_\ast/M_\odot) < 8.59$ (Section~\ref{sec:hostdec}). The rest of the parameters are the direct output of our nested sampling.}
}
\end{deluxetable*}
\end{longrotatetable}

\subsection{Host Galaxy Detectability and Uncertainty}
\label{sec:Merror}

We perform a series of input–output tests (Appendix~\ref{app:iotest}) using mock AGN images to assess the detectability of host galaxies with our method and to estimate upper limits on host galaxy properties (Section~\ref{sec:hostdec}). These empirical experiments also allow us to test the reliability of the key host galaxy parameters, as well as to quantify possible systematics in our fitting approach and the applied priors (Section~\ref{sec:reliability}). While the uncertainties derived from nested sampling reflect only the formal errors based on our chosen parametric model, empirical PSF, and astrophysical priors, additional systematic uncertainties arising from our choice of empirical PSF must also be considered (Section~\ref{sec:uncert}). We generate the mock images by artificially inserting a nucleus, represented by the PSF model, of varying strength onto real galaxy images at similar redshifts.

%
\begin{figure}
\centering
\includegraphics[width=0.45\textwidth]{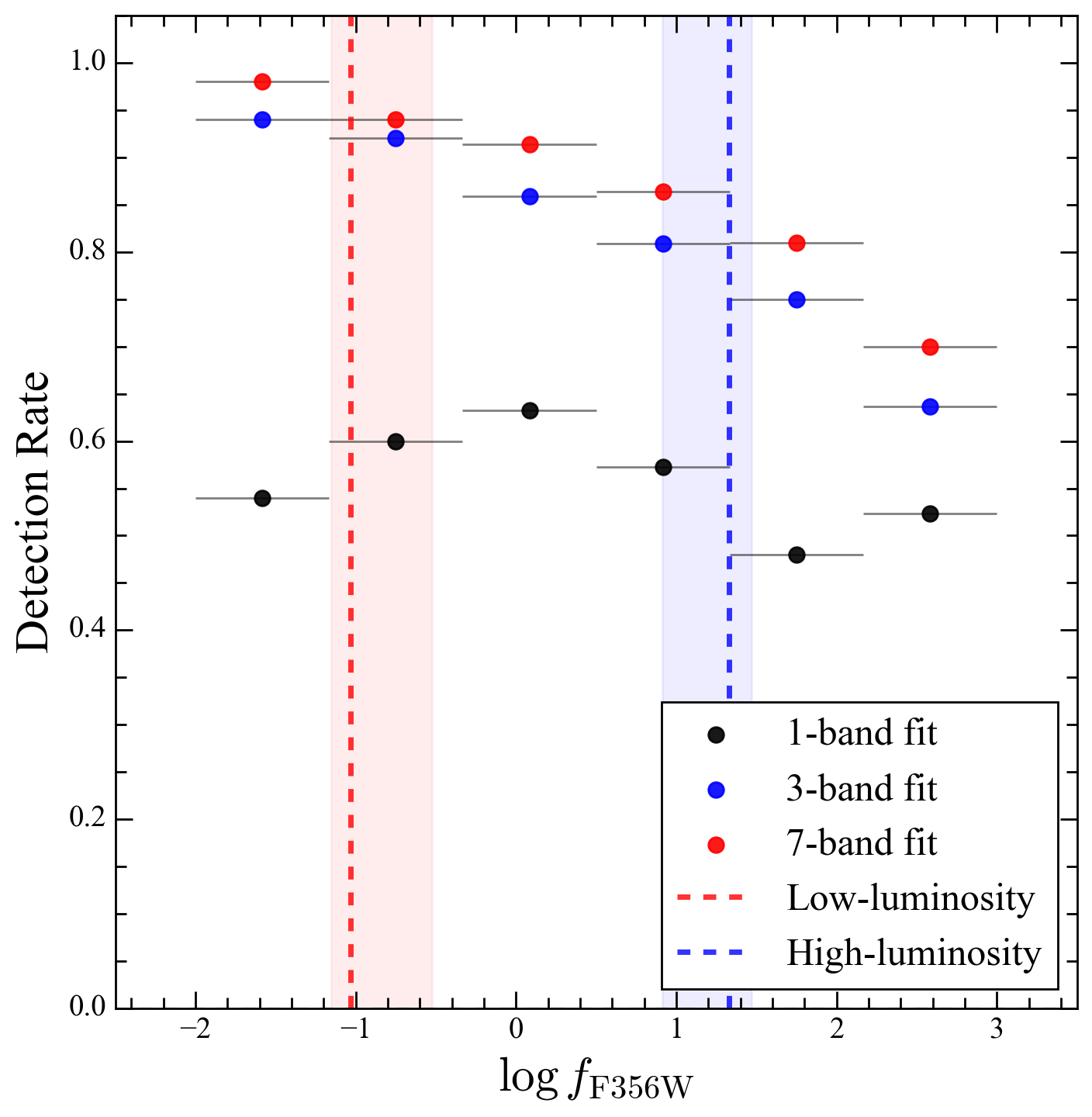}
\caption{Detection rate of simulated AGN host galaxies as a function of $f_{\mathrm{F356W}}$, the ratio of injected AGN flux to the host galaxy flux in the F356W filter. The black, blue, and red points represent the detection rates from 1‐band, 3‐band, and 7‐band fits, respectively, with horizontal error bars indicating the width of the $f_{\mathrm{F356W}}$ bins. The median $f_{\mathrm{F356W}}$ and the 16th and 84th percentiles are marked by the dashed line and shaded region for the low-luminosity (red) and high-luminosity (blue) quasars.}
\label{fig:detect}
\end{figure}

\begin{figure}
\centering
\includegraphics[width=0.48\textwidth]{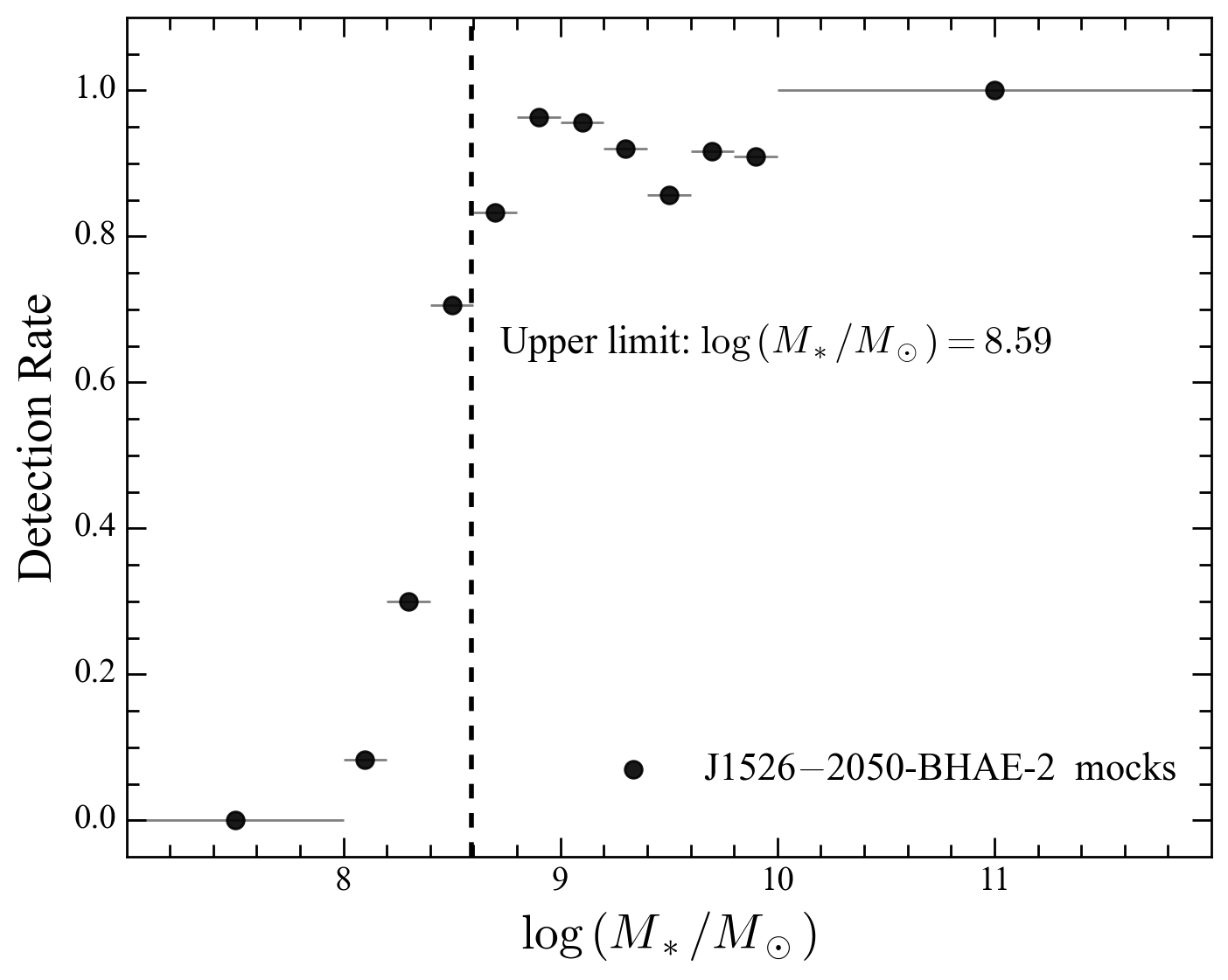}
\caption{Mock simulations to determine the upper limit of the stellar mass of the host galaxy of J1526$-$2050-BHAE-2.  Each point shows the host galaxy detection rate as a function of the stellar mass; error bars indicate the width for each mass bin. Above $\log\,  (M_\ast/M_\odot) \approx 8.4$ the detection rate exceeds $60\%$, and we adopt $\log\,  (M_\ast/M_\odot) = 8.59$ (dashed line) as the upper limit.}
\label{fig:uplim}
\end{figure}

\begin{figure*}
\centering
\includegraphics[height=0.33\textwidth]{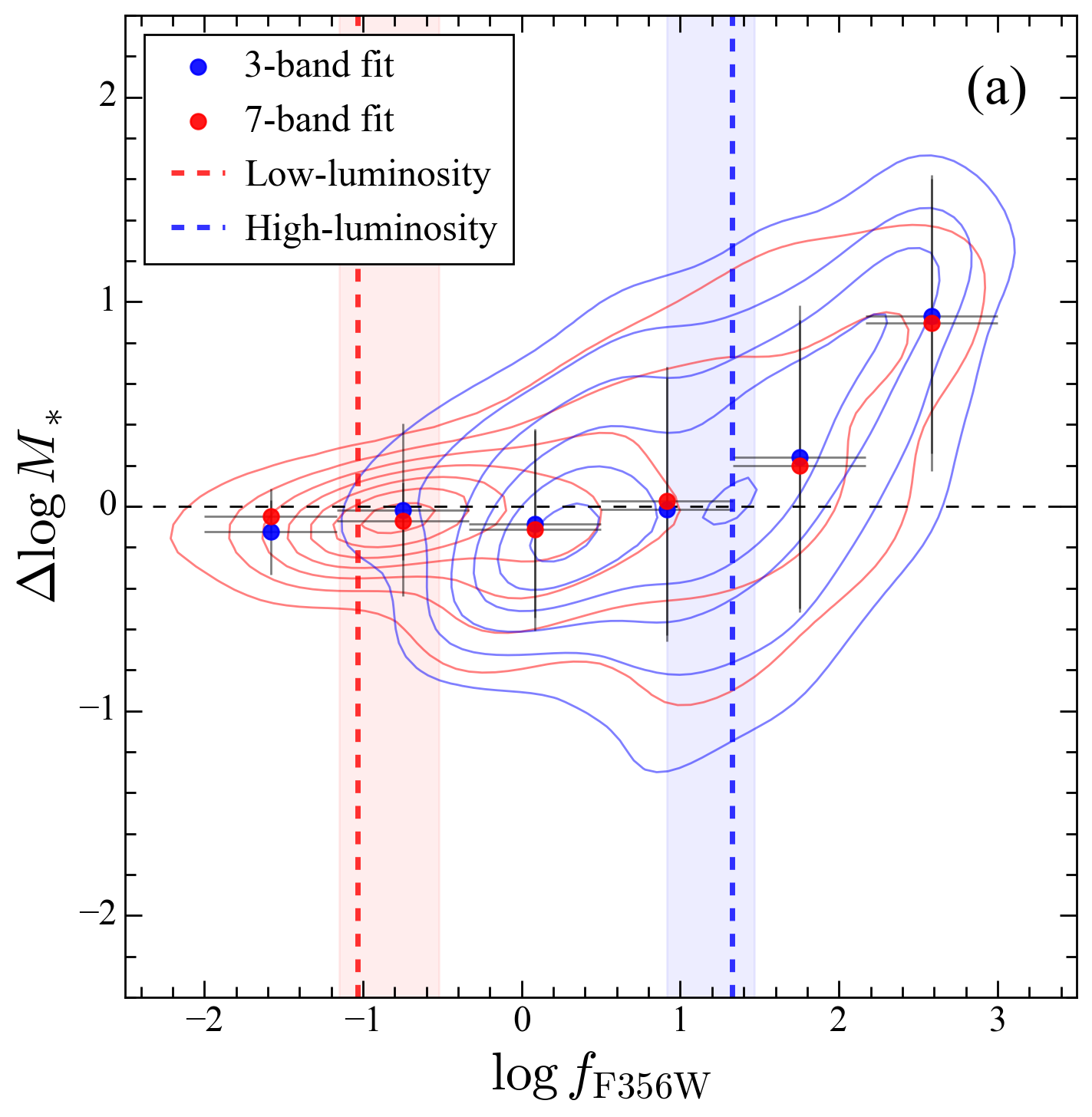}
\includegraphics[height=0.33\textwidth]{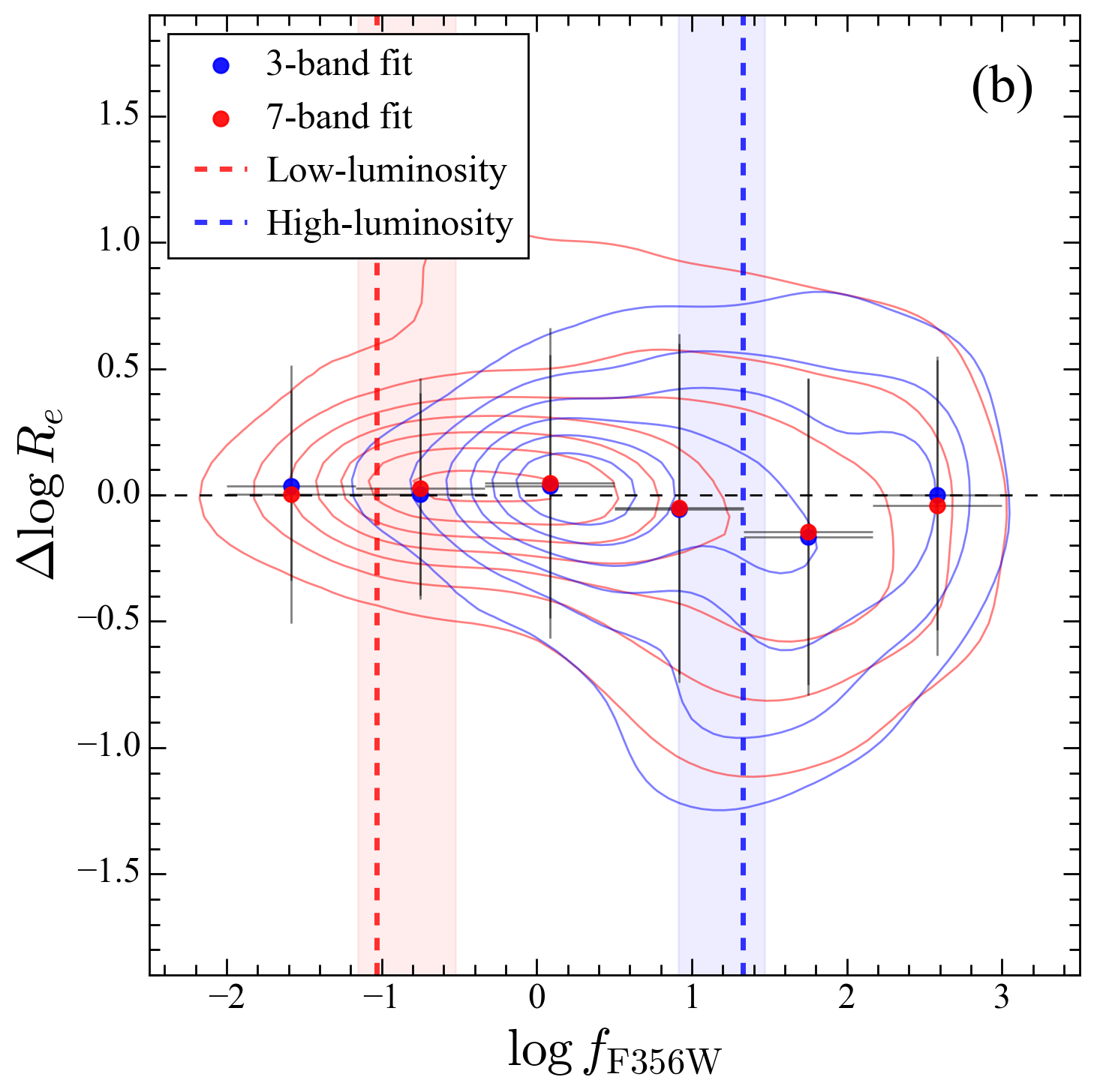}
\includegraphics[height=0.33\textwidth]{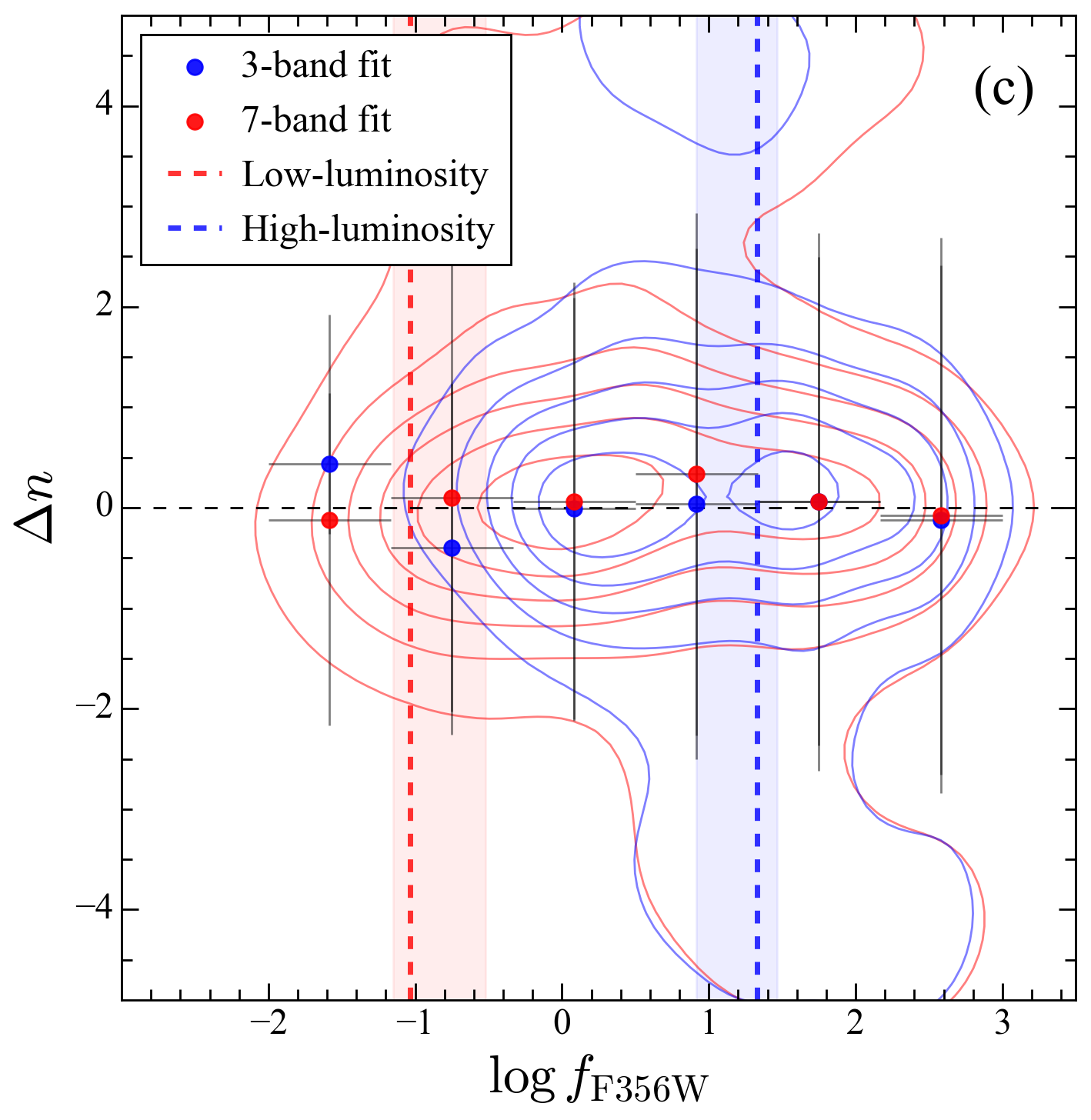}
\caption{
Comparison of the systematic bias in measurememts of host galaxy (a) stellar mass $M_*$, (b) effective radius $R_e$, and (c) \sersic\ index $n$ as a function of $f_{\rm F356W}$, the AGN-to-host flux ratio in the F356W band. The blue and red points/contours represent the 3‐band and 7‐band fits for the mocks, respectively, with error bars indicating the $1\,\sigma$ scatter. The median $f_{\mathrm{F356W}}$ and the 16th and 84th percentiles are marked by the dashed line and shaded region for the low-luminosity (red) and high-luminosity (blue) quasars. The horizontal dashed line in each panel indicates $\Delta = 0$. }
\label{fig:compare}
\end{figure*}

\subsubsection{Host Galaxy Detection}
\label{sec:hostdec}

We define a host galaxy as detected if the addition of a galaxy component results in an improved Bayesian information criterion (BIC) relative to a pure nucleus model, where

\begin{equation}
\mathrm{BIC} = \chi^2_r + k \ln N,
\end{equation}

\noindent
with $k$ the number of model parameters and $N$ the total number of unmasked pixels across all images. To compare the BIC of the nucleus plus galaxy model (i.e., the \textsc{GalfitS} model from Section~\ref{sec:galfits}) with that of a pure nucleus model, we remove the galaxy component and refit the data, allowing the PSF luminosity and position to vary independently in each band, and then compute

\begin{equation}
\delta_\mathrm{BIC} = \frac{\mathrm{BIC}_{\rm nuc+gal} - \mathrm{BIC}_{\rm nuc}}{\mathrm{BIC}_{\rm nuc}}.
\label{equ:dbic}
\end{equation}

\noindent
The values of $\delta_\mathrm{BIC}$ for our sample are listed in Table~\ref{tab:gsresultnp}. Adopting $\delta_\mathrm{BIC} < 0$ as the criterion for detection, we find evidence for a host galaxy in 30 out of the 31 AGNs in our sample. The only non-detection is J1526$-$2050-BHAE-2, which has $\delta_\mathrm{BIC} = 0.002$ and the lowest stellar mass ($M_\ast = 10^{8.35}\,M_\odot$) in the PSF+galaxy fit. \ledit{The smallest value we obtain is $\delta_\mathrm{BIC}=-0.001$, for J2232+2930-BHAE-2, which corresponds to a $\Delta\mathrm{BIC}\approx -90$. We therefore regard all of our host galaxy detections as robust, for even the weakest case yields a highly significant improvement in the absolute value of $\mathrm{BIC}$.}

We detect galaxy hosts in several objects that previously had failed to be detected \citep{Harikane2023ApJ, Yue2024ApJ}. This improvement arises from several advantages of our approach (Section~\ref{sec:galfits}) compared to earlier single-band image decomposition methods. First, \textsc{GalfitS} performs a simultaneous fit across multiple bands, which effectively boosts the host galaxy signal. Second, previous decomposition techniques \citep[e.g.,][]{Ding2023Nature, Yue2024ApJ} relied solely on the F356W band, which, despite offering a favorable AGN-to-host flux ratio (see Figure~\ref{fig:galfits}), has a spatial resolution that is about 2.4 times worse than F115W and 1.5 times worse than F200W. Our multiband fit yields a higher effective spatial resolution than fitting only the longest wavelength filter. Third, \textsc{GalfitS} incorporates observational priors (Section~\ref{sec:priors}) to exclude unphysical parameter spaces, thereby promoting convergence toward a more globally optimal solution.

To confirm that the higher detection rate indeed arises from our improved decomposition method, we carry out three types of \textsc{GalfitS} runs on the mock AGNs: (1) \texttt{run1}:~a pure‐galaxy, single‐\sersic\ fit on a carefully matched sample of inactive galaxies before manually adding the nucleus; (2) \texttt{run2}:~a point source‐only fit on the mock AGN images; and (3) \texttt{run3}:~a point source$+$galaxy fit on the mock AGN images, applying the astrophysical priors from Section~\ref{sec:priors}. We repeat these runs for three data configurations, using all NIRCam filters (7-band), only F115W+F200W+F356W (3‐band), and only F356W (1‐band). To mimic our treatment of the CEERS data, no astrophysical priors are applied for the 1‐band and 7-band fits in \texttt{run3}, while for the 3-band fits we adopt the priors in Section~\ref{sec:priors} to be consistent with our analysis of the EIGER, ASPIRE, and A-BHAE data. We then compare detection rates under these different setups as a function of $f_\mathrm{F356W}$, defined as the ratio between the injected AGN luminosity and the galaxy luminosity at F356W, from the single‐\sersic, pure‐galaxy fit (Figure~\ref{fig:detect}). The mock AGNs span $\log\, f_\mathrm{F356W} = -3$ to 3, covering both the high-luminosity ($\log\, f_\mathrm{F356W} = 1.33^{+0.14}_{-0.41}$) and low-luminosity ($\log\, f_\mathrm{F356W} = -1.03^{+0.51}_{-0.12}$) samples. For the 1‐band fits, the detection rate exceeds 60\% only for $0.1 \lesssim f_\mathrm{F356W} \lesssim 1$, whereas for the 3‐band or 7‐band fits the detection rate exceeds $80\%$ for $f_\mathrm{F356W}\lesssim100$. These results demonstrate the strong capability of our multi‐band approach for detecting AGN host galaxies across a wide dynamic range in AGN‐to‐host luminosity ratios.

We designed a dedicated set of mock AGNs for J1526$-$2050-BHAE-2 (hereafter “mock J1526$-$2050-BHAE-2”; Table~\ref{tab:gssetup}), whose host galaxy was undetected ($\delta_\mathrm{BIC} > 0$), in order estimate an upper limit on the stellar mass for this source, recognizing that its SED may differ from that of the general mock AGNs. Specifically, we select 190 additional CEERS galaxies from \citet{Sun2024ApJ} with $4 \lesssim z \lesssim 5$ to match the redshift range of the A-BHAE AGNs. We then inject a nucleus into these galaxies using the same empirical PSF from the CEERS field, scaling the point source fluxes to the observed F115W, F200W, and F356W fluxes of J1526$-$2050-BHAE-2 obtained from a point source–only fit. We use the “uplim” run (Table~\ref{tab:gssetup}) to estimate the detection rate as a function of stellar mass derived from the pure-galaxy fit. The detection rate exceeds 60\% in the bin $8.4 \le \log\, (M_\ast/M_\odot) \le 8.6$ (Figure~\ref{fig:uplim}), and thus we adopt the upper bound of this bin and set the stellar mass upper limit for J1526$-$2050-BHAE-2 at $\log\, (M_\ast/M_\odot) = 8.59$.

\subsubsection{Reliability of the Fitting Results}
\label{sec:reliability}

To evaluate the ability of \textsc{GalfitS} to recover host galaxy properties, we compare the results of \texttt{run1} and \texttt{run3} to study the systematic offsets in stellar mass ($\Delta \log\, M_\ast$), effective radius ($\Delta \log\, R_e$), and S\'ersic index ($\Delta n$) as a function of the AGN‐to‐host flux ratio (Figure~\ref{fig:compare}). Across a wide range of $f_{\mathrm{F356W}}$, the median offsets remain small, which indicates that even in the presence of a luminous AGN component the multiband decomposition reliably recovers the underlying host parameters. It is worth noting that for the 7‐band run, we adopt only the MZR prior, leaving $R_e$ entirely free, following our treatment for the CEERS AGNs. In contrast, the 3‐band run includes the MSR, MER, and SFH priors (Section~\ref{sec:priors}) to match our treatment of the other quasars. Nevertheless, Figure~\ref{fig:compare} shows that the recovered values $M_\ast$, $R_e$, and $n$ are consistent regardless of whether the MSR, MER, and SFH priors are imposed, reinforcing the general reliability of our decomposition method (Section~\ref{sec:galfits}).

\begin{figure}
\centering
\includegraphics[width=0.48\textwidth]{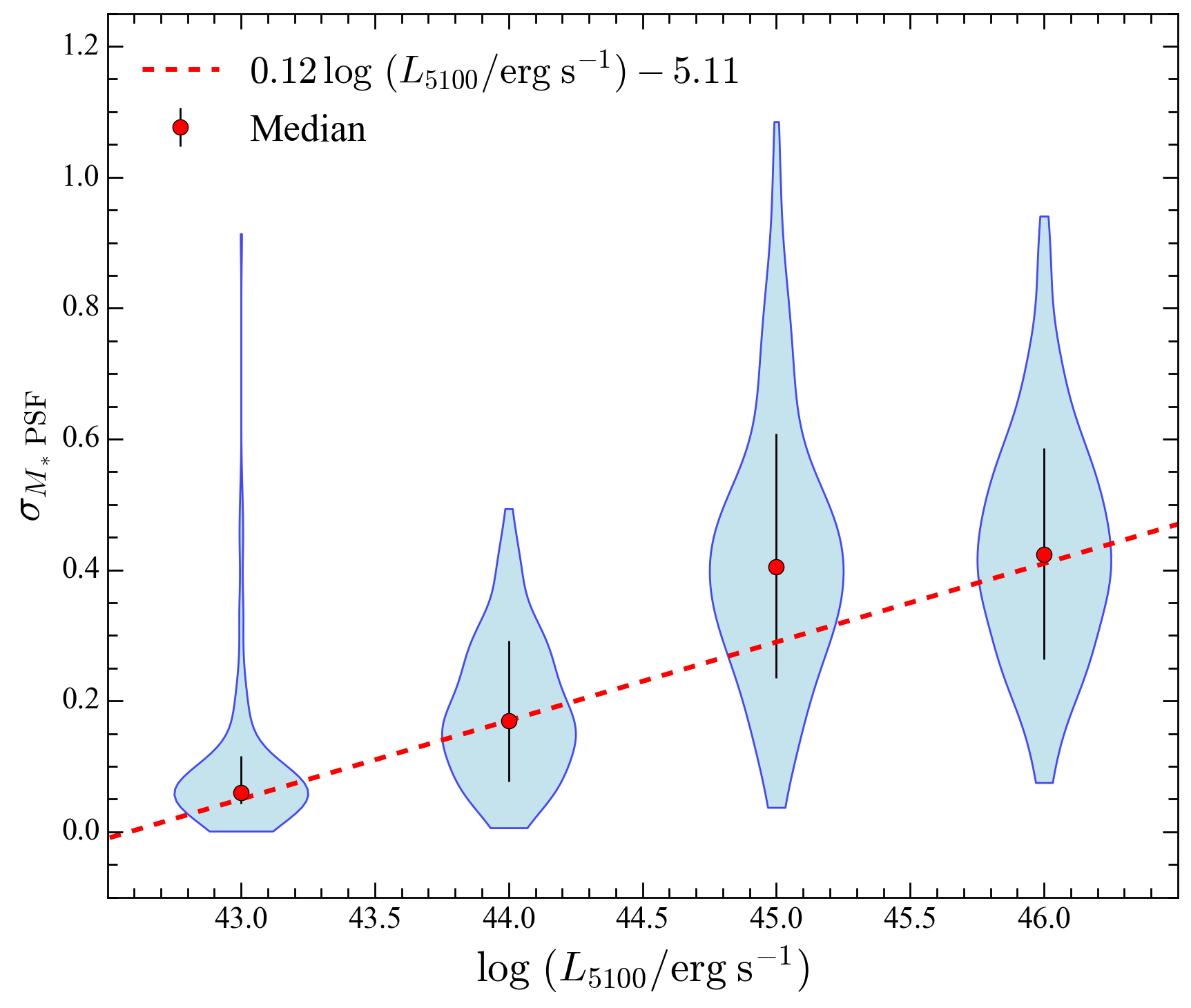}
\caption{The scatter ($\sigma_{M_\ast \, \rm PSF}$) in recovered stellar mass $M_{\ast}$ from five different PSF models, measured by analyzing mock AGNs, as a function of the input AGN luminosity $L_{5100}$. Each violin corresponds to one of the four luminosity bins used in the mock AGN simulations, illustrating the full spread of $\sigma_{M_\ast \, \rm PSF}$ values. The black vertical bars denote the central 16\% to 84\% range, while the red points mark the median. The red dashed line is the relation $\sigma_{M_\ast \, \rm PSF} = 0.12 \,\log\, (L_{5100}/\rm erg\, s^{-1}) - 5.11$. }
\label{fig:psferror}
\end{figure}

\subsubsection{Uncertainty from the PSF}
\label{sec:uncert}

As in many previous studies, our AGN–host decomposition method adopts a PSF model derived from field stars to represent the active nucleus. Care must be taken, however, to understand the potential impact of PSF variations that might arise from, for instance, differences in color and detector position \citep{Zhuang2024ApJb}. We employ \texttt{run3} of our mock AGN analysis to assess the sensitivity of $M_\ast$ to the choice of PSF. For each galaxy and for mock AGN images with nuclear luminosities $L_{5100} = 10^{43}-10^{46} \,\mathrm{erg\,s^{-1}}$, we generate independent realizations by adopting the PSF from five distinct field stars as the point source profile.  We then compute $\sigma_{M_\ast\, \rm PSF}$, the standard deviation of $\log\,  M_\ast$ obtained from fitting these images. Notably, $\sigma_{M_\ast\, \rm PSF}$ shows little dependence on the true galaxy mass but increases significantly with the input AGN luminosity (Figure~\ref{fig:psferror}), and over the range $\sim 0.05 - 0.4$~dex it can be approximated by $\sigma_{\log\, {M_\ast} \, \rm PSF} = 0.12\,\log\, (L_{5100}/\rm erg\, s^{-1}) - 5.11$. We estimate the final uncertainty of the stellar mass measurements ($\sigma_{\log M_\ast}$) as the quadrature sum of the PSF-induced uncertainty and the nested-sampling error derived from the posterior distribution (Section~\ref{sec:posterior}). For all other parameters, we estimate uncertainties by propagating the error in $\log M_\ast$.  If we let $\widehat{\log M_\ast}$ be the best‐fit value, we first select from the nested‐sampling “dead” points those satisfying $\widehat{\log M_\ast} - \sigma_{\log M_\ast} \;\le\;\log M_\ast\;\le\; \widehat{\log M_\ast} + \sigma_{\log M_\ast}$. This filtering effectively enlarges our posterior sample to encompass the $\pm\,1\,\sigma$ range in stellar mass \ledit{and propagates roughly 0.1\,dex uncertainty to other parameters}. We then define the uncertainty of each remaining parameter as the standard deviation of its values over this filtered posterior set. The resulting parameter uncertainties are presented in Figure~\ref{fig:msr} and summarized in Table~\ref{tab:gsresultnp} as the final parameter uncertainties.

\begin{figure}
\centering
\includegraphics[width=0.48\textwidth]{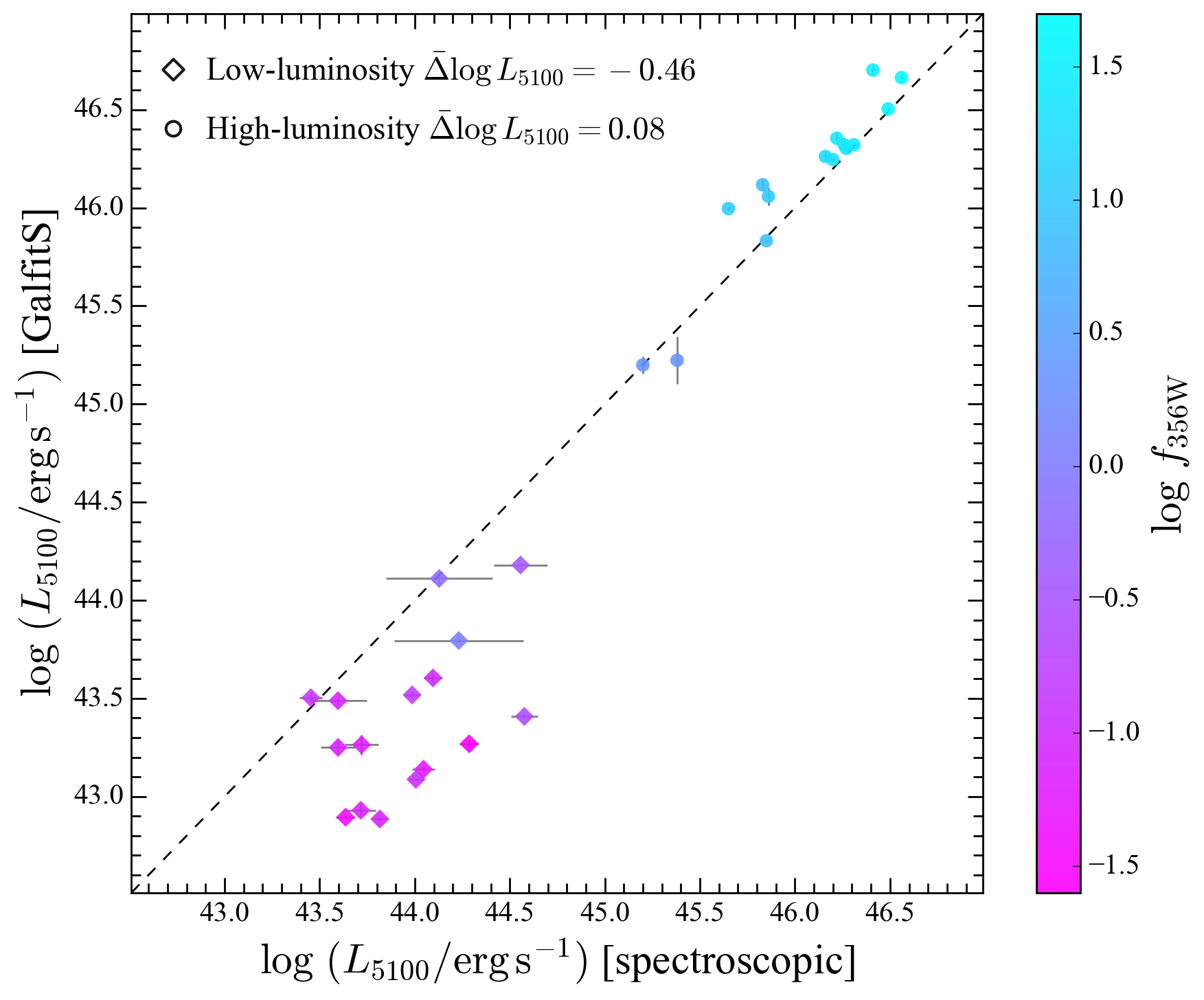}
\caption{Comparison of the rest‐frame $5100$\,\AA\ luminosities derived from our \textsc{GalfitS} SED modeling versus values obtained from spectroscopic analysis. The data points are color‐coded by $f_{\mathrm{F356W}}$, the AGN-to-host flux ratio in the F356W band. The dashed line indicates the one‐to‐one relation. For low-luminosity quasars, spectroscopic $L_{\mathrm{5100}}$ is systematically lower by $\bar{\Delta} \log \, L_{\mathrm{5100}} = -0.46$~dex, while for high-luminosity quasars our values are higher by $\bar{\Delta} \log \, L_{\mathrm{5100}} = 0.08$~dex. The discrepancy is discussed in Section~\ref{sec:agnsed}. }
\label{fig:L5100}
\end{figure}

\bigskip
\subsection{AGN SED Analysis}
\label{sec:agnsed}

The \textsc{GalfitS} AGN-host decomposition (Section~\ref{sec:galfits}) treats the observed AGN luminosity in each band as an independent parameter. Unlike the rest‐frame luminosity, the observed AGN luminosity is defined as $L_{\mathrm{AGN}} \equiv 4\pi f_{\mathrm{AGN}} D_{c}^2$, where $f_{\mathrm{AGN}}$ is the observed (fitted) flux in a given filter and $D_{c}$ is the comoving distance (Table~\ref{tab:gsresultagn}). \textsc{GalfitS} adopts a minimal‐parameter model to further characterize the broad‐band SED of the AGN.  Motivated by the composite UV-optical spectrum of quasars of \cite{VandenBerk2001AJ}, we employ a single power law ($f_\lambda \propto \lambda^{\beta}$) for the high-luminosity quasars and a broken power law that breaks at $5100$\,\AA for the less luminous subsample. This parameterization suffices because our longest observed filter only reaches $\sim4700$\,\AA\ in the rest-frame for the luminous sources at higher redshifts. The free parameters of this SED are the rest‐frame luminosity at $5100$\,\AA\ ($L_{\mathrm{5100}}$), the UV spectral index $\beta_{\mathrm{UV}}$ for $\lambda \leq 5100$\,\AA, and the optical spectral index $\beta_{\mathrm{opt}}$ for $\lambda \geq 5100$\,\AA. We include broad (O\,VI $\lambda 1038$, Ly$\alpha$, N\,V $\lambda 1242$, Si\,IV $\lambda 1400$, C\,IV $\lambda 1549$, He\,II $\lambda 1640$, C\,III] $\lambda 1909$, Mg\,II $\lambda 2800$, H$\gamma$, H$\beta$, H$\alpha$) as well as narrow (H$\beta$, [O\,III] $\lambda \lambda 4959,\, 5007$, [O\,I] $\lambda \lambda 6300,\, 6364$, [N\,II] $\lambda \lambda 6548,\, 6584$, H$\alpha$, [S\,II] $\lambda \lambda 6716,\, 6731$) emission lines, whose relative intensities are linked via empirically established scaling relations. In particular, the broad H$\alpha$ and H$\beta$ luminosities are tied to $L_{\mathrm{5100}}$ following \citet{Greene2005ApJ}\footnote{\ledit{The luminosities of the broad H$\alpha$ and H$\beta$ lines are allowed to deviate from the \citet{Greene2005ApJ} scaling relations. We fit for this offset with a Gaussian prior of $\sigma=0.2$\,dex, motivated by the intrinsic scatter reported in \citet{Greene2005ApJ}. For a given $M_\mathrm{BH}$, the line widths are tied to $L_{\mathrm{5100}}$ following \citet{Greene2005ApJ}. The line centroids are fixed to their laboratory rest-frame wavelengths.}}, and other lines are scaled according to the composite spectrum of \citet{VandenBerk2001AJ}. The [O\,III]\,$\lambda5007$ luminosity is further linked to broad H$\alpha$ based on \citet{Stern2012MNRAS}, while other narrow lines adhere to the ratios given by \citet{Stern2013MNRAS} for type~1 AGNs. The line widths are also constrained: for simplicity, the broad lines follow the virial relation from \citet{Ho2015ApJ}, whereas narrow lines have a fixed width of $300$\,km\,s$^{-1}$ \citep{Nelson1996ApJ}. We do not include a pseudo‐continuum (e.g., broad \feii\, emission and  Balmer continuum), given that our coverage is primarily in the rest‐frame UV and their strengths are not well constrained by the available data. In practice, the broken power law effectively accounts for both the accretion disk continuum and any unmodeled \feii\ emission or Balmer continuum. An example of our AGN SED model is illustrated in the bottom panels of Figure~\ref{fig:galfits}, and the best‐fit parameters appear in Table~\ref{tab:gsresultagn}.

Figure~\ref{fig:L5100} compares our $L_{\mathrm{5100}}$ measurements with independent spectroscopic estimates. For the low-luminosity quasars, $L_{\mathrm{5100}}$ is derived from the broad H$\alpha$ luminosities reported by \citet{Harikane2023ApJ} and \citet{Lin2024ApJ}, converted using the empirical scaling relation derived for nearby AGNs by \citet{Greene2005ApJ}. Our SED‐based measurements are systematically lower by $\bar{\Delta} \log\,L_{\mathrm{5100}}= -0.46$~dex, implying that the observed ratio of broad H$\alpha$ to continuum strength is larger than in local AGNs. This trend is consistent with the overall tendency for JWST‐selected AGNs at $z\geq4$ to exhibit broad H$\alpha$ equivalent widths roughly 3 times those in nearby AGNs \citep{Maiolino2025MNRAS}. For the luminous quasar subsample, we obtain slightly higher luminosities ($\bar{\Delta} \log\,L_{\mathrm{5100}}= 0.08$~dex) compared to spectral‐decomposition measurements \citep{Ding2023Nature, Yang2023ApJ, Yue2024ApJ}, an affect that can be plausibly attributed to our power‐law model, which includes as part of the continuum \feii\ emission that is known to be strong in these sources \citep{Yang2023ApJ}. We convert $L_{\mathrm{5100}}$ into a rough estimate of the mass accretion rate of the BH ($\dot{M}_{\rm BH}$; Table~\ref{tab:gsresultagn}), assuming that the AGN bolometric luminosity $L_{\mathrm{bol}} = 10.6\,L_{\mathrm{5100}}$ \citep{Richards2006ApJS} and a fiducial radiative efficiency of $\eta=0.1$ \citep{Soltan1982MNRAS,Yu2002MNRAS}.

We keep the AGN SED model intentionally simple, with only 2 to 3 free parameters designed to capture both the overall strength and spectral shape of the continuum. Despite its minimalism, it performs well for our sample of 31 AGNs. In principle, \textsc{GalfitS} can also be configured to fit the AGN SED simultaneously during the image decomposition. Instead of treating AGN luminosity in each band as an independent parameter, one can solve directly for $L_{\mathrm{5100}}$, $\beta_{\mathrm{UV}}$, and $\beta_{\mathrm{opt}}$. We tested this more constrained, parametric approach but found that, because of the fewer degrees of freedom and the assumption that all emission lines are fixed to a fraction of the continuum luminosity, the resulting $\chi^2_r$ of the fits often deteriorate by an average of $\sim 30\%$. To evaluate systematically which strategy recovers host galaxy properties more accurately, we performed an additional series of tests using on our mock AGNs (\texttt{run4}) \ledit{ by directly fitting the AGN+galaxy system with the above parametric SED model during the \textsc{GalfitS} decomposition}. Appendix~\ref{app:iotest} compares the derived 7‐band stellar masses from our baseline approach (\texttt{run3}) and \texttt{run4}. For systems with AGN‐to‐host flux ratios $f_{\mathrm{F356W}} \lesssim 2$, the free‐SED method (Section~\ref{sec:galfits}) exhibits less systematic bias, whereas the parametric-SED approach underestimates the stellar mass and shows larger scatter. Interestingly, at very high flux ratios ($f_{\mathrm{F356W}} \gtrsim 2$)---wherein the AGN strongly dominates---the parametric approach achieves near‐zero systematic bias. However, across the range of $f_{\mathrm{F356W}}$ relevant for our sample \texttt{run3} provides much better consistency. This outcome further supports the robustness of our chosen AGN–host decomposition strategy (Section~\ref{sec:galfits}).


\startlongtable
\begin{longrotatetable}
\begin{deluxetable*}{cccccccccccc}
\setlength{\tabcolsep}{2.5pt}
\tablecaption{Results of the \textsc{GalfitS} Analysis of AGN SED} \label{tab:gsresultagn}
\tabletypesize{\footnotesize}
\tablehead{
      \colhead{Name}             &
      \colhead{$\log{L_\mathrm{F\,115W}^{\mathrm{AGN}}}$} &
      \colhead{$\log{L_\mathrm{F\,150W}^{\mathrm{AGN}}}$} &
      \colhead{$\log{L_\mathrm{F\,200W}^{\mathrm{AGN}}}$} &
      \colhead{$\log{L_\mathrm{F\,277W}^{\mathrm{AGN}}}$} &
      \colhead{$\log{L_\mathrm{F\,356W}^{\mathrm{AGN}}}$} &
      \colhead{$\log{L_\mathrm{F\,410M}^{\mathrm{AGN}}}$} &
      \colhead{$\log{L_\mathrm{F\,444W}^{\mathrm{AGN}}}$} &
      \colhead{$\log{L_\mathrm{5100 ,\, GS}}$} & 
      \colhead{$\beta_\mathrm{UV}$} & 
      \colhead{$\beta_\mathrm{opt}$} & 
      \colhead{$\log{\dot{M}_\mathrm{BH}}$}\\
&
\colhead{$(\mathrm{erg\, s^{-1}\, \mathring{A}^{-1}})$} &
\colhead{$(\mathrm{erg\, s^{-1}\, \mathring{A}^{-1}})$} &
\colhead{$(\mathrm{erg\, s^{-1}\, \mathring{A}^{-1}})$} &
\colhead{$(\mathrm{erg\, s^{-1}\, \mathring{A}^{-1}})$} &
\colhead{$(\mathrm{erg\, s^{-1}\, \mathring{A}^{-1}})$} &
\colhead{$(\mathrm{erg\, s^{-1}\, \mathring{A}^{-1}})$} &
\colhead{$(\mathrm{erg\, s^{-1}\, \mathring{A}^{-1}})$} &
\colhead{$(\mathrm{erg\, s^{-1}})$} &
&
&
\colhead{($M_\odot \, \rm yr^{-1}$)}\\
\colhead{(1)} &
\colhead{(2)} &
\colhead{(3)} &
\colhead{(4)} &
\colhead{(5)} &
\colhead{(6)} &
\colhead{(7)} &
\colhead{(8)} &
\colhead{(9)} &
\colhead{(10)} &
\colhead{(11)} &
\colhead{(12)} 
}
\startdata
J2236+0032         & ---            & $39.52\pm0.15$ & ---            & ---            & $39.02\pm0.09$ & ---            & ---            & $45.22\pm0.12$ & $-1.40\pm0.53$ & ---            & $0.47\pm0.12$  \\
J2255+0251         & ---            & $39.48\pm0.08$ & ---            & ---            & $38.99\pm0.01$ & ---            & ---            & $45.20\pm0.04$ & $-1.38\pm0.24$ & ---            & $0.45\pm0.04$  \\
J0148+0600         & $41.07\pm0.01$ & ---            & $40.79\pm0.01$ & ---            & $40.15\pm0.07$ & ---            & ---            & $46.70\pm0.03$ & $-0.81\pm0.21$ & ---            & $1.95\pm0.03$  \\
J1030+0524         & $40.90\pm0.01$ & ---            & $40.49\pm0.01$ & ---            & $40.11\pm0.01$ & ---            & ---            & $46.32\pm0.01$ & $-1.47\pm0.02$ & ---            & $1.57\pm0.00$  \\
J159+02            & $40.70\pm0.01$ & ---            & $40.44\pm0.03$ & ---            & $40.08\pm0.06$ & ---            & ---            & $46.35\pm0.03$ & $-0.79\pm0.21$ & ---            & $1.60\pm0.03$  \\
J1120+0641         & $40.56\pm0.01$ & ---            & $40.42\pm0.01$ & ---            & $40.01\pm0.01$ & ---            & ---            & $46.30\pm0.01$ & $-0.96\pm0.02$ & ---            & $1.55\pm0.00$  \\
J1148+5251         & $41.19\pm0.01$ & ---            & $40.85\pm0.01$ & ---            & $40.29\pm0.01$ & ---            & ---            & $46.66\pm0.01$ & $-1.39\pm0.05$ & ---            & $1.91\pm0.01$  \\
J0109$-$3047         & $40.43\pm0.01$ & ---            & $40.13\pm0.01$ & ---            & $39.56\pm0.08$ & ---            & ---            & $46.12\pm0.03$ & $-0.85\pm0.25$ & ---            & $1.37\pm0.03$  \\
J0218+0007         & $40.41\pm0.01$ & ---            & $40.11\pm0.01$ & ---            & $39.53\pm0.01$ & ---            & ---            & $45.83\pm0.01$ & $-1.68\pm0.03$ & ---            & $1.08\pm0.01$  \\
J0224$-$4711         & $40.87\pm0.01$ & ---            & $40.68\pm0.02$ & ---            & $40.25\pm0.03$ & ---            & ---            & $46.50\pm0.02$ & $-0.77\pm0.11$ & ---            & $1.75\pm0.02$  \\
J0226+0302         & $41.00\pm0.01$ & ---            & $40.60\pm0.04$ & ---            & $40.05\pm0.07$ & ---            & ---            & $46.32\pm0.04$ & $-1.75\pm0.25$ & ---            & $1.57\pm0.04$  \\
J0244$-$5008         & $40.82\pm0.01$ & ---            & $40.46\pm0.01$ & ---            & $39.96\pm0.01$ & ---            & ---            & $46.25\pm0.01$ & $-1.54\pm0.05$ & ---            & $1.50\pm0.01$  \\
J0305$-$3150         & $40.45\pm0.01$ & ---            & $40.18\pm0.03$ & ---            & $39.70\pm0.10$ & ---            & ---            & $46.06\pm0.05$ & $-0.94\pm0.32$ & ---            & $1.31\pm0.05$  \\
J2002$-$3013         & $40.82\pm0.01$ & ---            & $40.51\pm0.02$ & ---            & $39.94\pm0.05$ & ---            & ---            & $46.26\pm0.03$ & $-1.56\pm0.17$ & ---            & $1.51\pm0.03$  \\
J2232+2930         & $40.63\pm0.01$ & ---            & $40.28\pm0.01$ & ---            & $39.70\pm0.02$ & ---            & ---            & $46.00\pm0.01$ & $-1.74\pm0.06$ & ---            & $1.25\pm0.01$  \\
CEERS 007465       & $37.39\pm0.02$ & $36.74\pm0.09$ & $37.14\pm0.02$ & $37.06\pm0.02$ & $37.47\pm0.01$ & $37.57\pm0.01$ & $37.71\pm0.01$ & $43.50\pm0.02$ & $2.25\pm0.28$  & $2.87\pm0.07$  & $-1.25\pm0.02$ \\
CEERS 006725       & $37.46\pm0.05$ & $36.80\pm0.15$ & $36.94\pm0.06$ & $36.97\pm0.04$ & $37.17\pm0.02$ & $37.20\pm0.02$ & $37.33\pm0.01$ & $43.26\pm0.05$ & $1.23\pm0.55$  & $1.51\pm0.15$  & $-1.49\pm0.05$ \\
CEERS 027825       & $38.62\pm0.03$ & $38.37\pm0.02$ & $38.23\pm0.04$ & $38.34\pm0.01$ & $37.84\pm0.01$ & $38.41\pm0.05$ & $38.13\pm0.02$ & $44.11\pm0.03$ & $-1.02\pm0.24$ & $-3.99\pm0.08$ & $-0.64\pm0.03$ \\
CEERS 003975       & $38.63\pm0.01$ & $38.31\pm0.01$ & $38.08\pm0.01$ & $37.70\pm0.01$ & $38.05\pm0.01$ & $37.54\pm0.01$ & $37.41\pm0.01$ & $43.79\pm0.02$ & $-0.82\pm0.03$ & $-2.00\pm0.01$ & $-0.96\pm0.00$ \\
J0109$-$3047-BHAE-1  & $36.34\pm0.01$ & ---            & $37.03\pm0.01$ & ---            & $37.84\pm0.01$ & ---            & ---            & $43.09\pm0.01$ & $3.05\pm0.03$  & $5.15\pm0.02$  & $-1.66\pm0.01$ \\
J0218$+$0007-BHAE-1  & $37.35\pm0.02$ & ---            & $37.54\pm0.01$ & ---            & $38.00\pm0.01$ & ---            & ---            & $43.52\pm0.01$ & $0.90\pm0.08$  & $2.29\pm0.04$  & $-1.23\pm0.01$ \\
J0224$-$4711-BHAE-1  & $36.97\pm0.09$ & ---            & $37.29\pm0.02$ & ---            & $37.74\pm0.01$ & ---            & ---            & $43.25\pm0.04$ & $1.31\pm0.26$  & $1.87\pm0.11$  & $-1.50\pm0.04$ \\
J0229$-$0808-BHAE-1  & $37.17\pm0.01$ & ---            & $37.04\pm0.01$ & ---            & $37.68\pm0.01$ & ---            & ---            & $42.89\pm0.01$ & $-0.37\pm0.02$ & $5.56\pm0.01$  & $-1.87\pm0.00$ \\
J0229$-$0808-BHAE-2  & $37.85\pm0.02$ & ---            & $37.42\pm0.02$ & ---            & $37.45\pm0.01$ & ---            & ---            & $43.14\pm0.02$ & $-1.75\pm0.09$ & $4.79\pm0.05$  & $-1.61\pm0.02$ \\
J0430$-$1445-BHAE-1  & $37.32\pm0.02$ & ---            & $37.49\pm0.01$ & ---            & $38.24\pm0.01$ & ---            & ---            & $43.41\pm0.01$ & $0.50\pm0.07$  & $4.61\pm0.03$  & $-1.34\pm0.01$ \\
J0923$+$0402-BHAE-1  & $37.78\pm0.01$ & ---            & $37.50\pm0.01$ & ---            & $38.31\pm0.01$ & ---            & ---            & $44.18\pm0.01$ & $3.80\pm0.02$  & $0.57\pm0.01$  & $-0.57\pm0.00$ \\
J1526$-$2050-BHAE-2  & $37.00\pm0.05$ & ---            & $37.04\pm0.01$ & ---            & $37.75\pm0.01$ & ---            & ---            & $43.60\pm0.02$ & $3.02\pm0.16$  & $0.94\pm0.07$  & $-1.15\pm0.02$ \\
J1526$-$2050-BHAE-3  & $37.38\pm0.03$ & ---            & $36.89\pm0.03$ & ---            & $37.62\pm0.01$ & ---            & ---            & $43.49\pm0.02$ & $3.03\pm0.13$  & $1.00\pm0.06$  & $-1.26\pm0.02$ \\
J2232$+$2930-BHAE-1  & $37.34\pm0.04$ & ---            & $37.40\pm0.01$ & ---            & $37.23\pm0.02$ & ---            & ---            & $43.27\pm0.02$ & $0.18\pm0.14$  & $-2.50\pm0.07$ & $-1.48\pm0.02$ \\
J2232$+$2930-BHAE-2  & $37.63\pm0.02$ & ---            & $37.22\pm0.01$ & ---            & $37.45\pm0.01$ & ---            & ---            & $42.89\pm0.01$ & $-1.75\pm0.06$ & $4.00\pm0.03$  & $-1.86\pm0.01$ \\
J2232$+$2930-BHAE-3  & $37.35\pm0.05$ & ---            & $37.04\pm0.04$ & ---            & $37.64\pm0.01$ & ---            & ---            & $42.93\pm0.03$ & $-0.84\pm0.18$ & $6.99\pm0.09$  & $-1.82\pm0.03$ 
\enddata
\tablecomments{Best-fit parameters from the \textsc{GalfitS} analysis of the high-redshift quasars.
Col. (1): Name of the AGN.
Cols. (2)--(8): Observed AGN luminosity in different NIRCam filters (Section~\ref{sec:agnsed}).
Col. (9): Luminosity at 5100\,\AA\ of the power-law component of the AGN SED model.
Col. (10): UV spectral index for $\lambda \leq 5100$\,\AA.
Col. (11): Optical spectral index for $\lambda \ge 5100$\,\AA.
Col. (12): BH mass accretion rate estimated based on ${L_\mathrm{5100 ,\, GS}}$.
}
\end{deluxetable*}
\end{longrotatetable}

\section{Results}
\label{sec:sec4}

\begin{figure}
\centering
\includegraphics[width=0.45\textwidth]{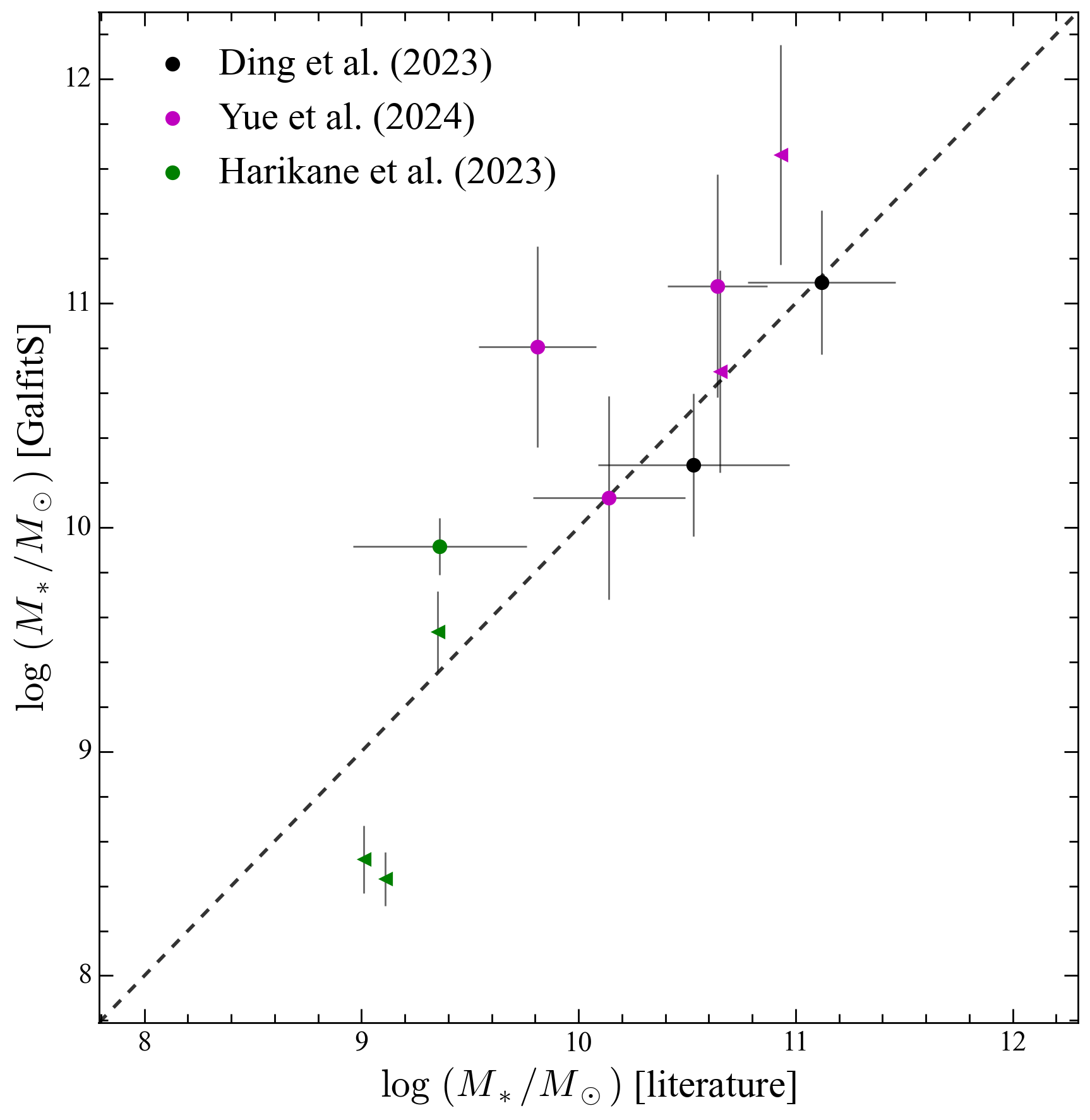}
\caption{
Comparison of stellar mass measurements between this work using \textsc{GalfitS} and previous studies: CEERS AGNs (green; \citealt{Harikane2023ApJ}), EIGER quasars (magenta; \citealt{Yue2024ApJ}), and SHELLQs quasars (black; \citealt{Ding2023Nature}). Upper limits are denoted with triangles. The dashed line indicates a 1:1 relation. }
\label{fig:masscomp}
\end{figure}

\begin{figure}
\centering
\includegraphics[width=0.45\textwidth]{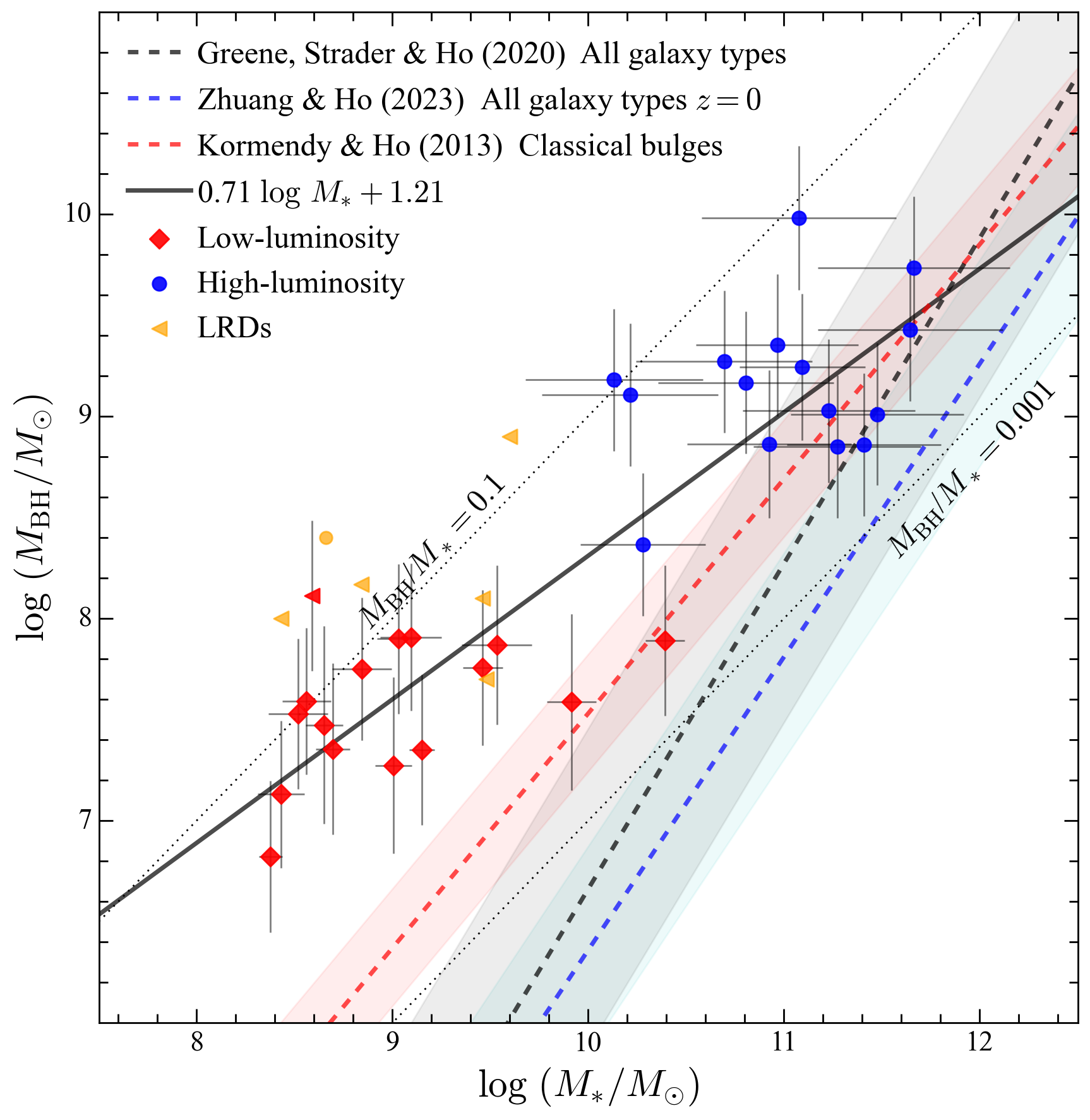}
\caption{Black hole mass ($M_{\mathrm{BH}}$) versus stellar mass ($M_\ast$) relation for our sample of 15 high-luminosity (blue) and 16 low-luminosity (red) quasars. The red triangle represents J1526$-$2050-BHAE-2, which only has an upper limit for stellar mass. The orange triangles indicate the LRDs from \citet{Chen2025}; the object plotted with an orange circle is the only LRD with a detected host galaxy. The blue dashed line marks the $z=0$ scaling relation for nearby type~1 AGNs \citep{Zhuang2023NatAs}, whose $0.51$~dex intrinsic scatter is shown as the blue shaded region, while the equivalent relation for inactive galaxies from \cite{Greene2020ARAA} is given as a black dashed line. The red dashed line corresponds to the $M_\mathrm{BH}-M_\mathrm{\ast ,\, bulge}$ relation of classical bulges, with the red shaded region representing its $0.28$~dex intrinsic scatter \citep{Kormendy2013ARAA}. The black solid line shows a linear fit to the quasars in this study: $\log \, M_{\mathrm{BH}} = 0.71\, \log \, M_\ast + 1.21$. Dotted lines denote $M_{\mathrm{BH}}/M_\ast = 0.1$ and 0.001.}
\label{fig:mmrelation}
\end{figure}

\subsection{Host Galaxy Stellar Mass}
\label{sec:mstar}

Stellar masses (or upper limits thereof) have already been published for 11 of the 31 AGNs in our sample, including two from SHELLQs \citep{Ding2023Nature}, five from EIGER \citep{Yue2024ApJ}, and four from CEERS \citep{Harikane2023ApJ}. These studies employed a two-step method to derive $M_\ast$: image fitting to detect possible host galaxy emission, followed by SED fitting of the decomposed host photometry. Comparison of our stellar masses with those from the literature reveals that the two sets of meaqsurements are generally consistent, with a slight tendency for our values to be higher ($\bar{\Delta} \log\,M_\ast = 0.21$~dex; Figure~\ref{fig:masscomp}). However, we emphasize that our stellar mass measurements are likely to be more robust. We use a multiband decomposition approach that ensures that the imaging model of the host galaxy is consistent across multiple bands and aligns with the SED model of the AGN. Extensive mock simulations (Section~\ref{sec:reliability}) validate our methodology, confirming that we can recover accurate stellar masses even in the presence of a luminous quasar.

Despite these improvements, stellar mass derivation remains sensitive to assumptions about the galaxy's SED. \citet{Yue2024ApJ} show that SED parameters are poorly constrained using \texttt{Prospector} \citep{Johnson2021ApJS} when only three photometric points are available. For instance, although our measurement of the host galaxy flux of J0148+0600, for which $\log \, f_{\rm 356W} = 1.47 \pm 0.07$, agrees well with the value of 1.4 reported by \citet{Yue2024ApJ} (also shown in Appendix~\ref{app:hostmag}), our derived stellar mass is 0.44~dex higher than their value of $\log \, (M_\ast/M_\odot) = 10.64$, likely as a consequence of differences in assumed dust extinction and SFH. In the case of J1148+5251, our stellar mass exceeds the upper limit from \citet{Yue2024ApJ}, even though, by our estimation, their detection limit is comparable to ours. \citet{Yue2024ApJ} calculated their $M_\ast$ upper limit by scaling the SED of J0148+0600, which, as mentioned above, yields a lower stellar mass than our measurement.

Figure~\ref{fig:mmrelation} examines the relationship between BH mass and total stellar mass for our sample of high-redshift quasars. For comparison, we overlay the $M_{\mathrm{BH}} - M_\ast$ relation for over 11,000 low-redshift type~1 AGNs, after applying a mild evolutionary correction to $z=0$ (blue dashed line; \citealt{Zhuang2023NatAs}), the equivalent relation for inactive local galaxies (black dashed line; \citealt{Greene2020ARAA}), as well as the relation that pertains only to nearby inactive classical bulges and ellipticals (red dashed line; \citealt{Kormendy2013ARAA}). A linear fit to the 30 high-redshift quasars with detected host galaxies yields $\log \, M_{\mathrm{BH}} = (0.71\pm0.07)\, \log \, M_\ast \, + \, (1.21\pm0.67)$. Notably, the slope of this relation is roughly only half of that of $z = 0$ galaxies, whether active (slope 1.46; \citealt{Zhuang2023NatAs}) or not (slope 1.26; \citealt{Greene2020ARAA}).

Among inactive classical bulges and elliptical galaxies in the local Universe, the central BH occupies $\sim 0.5\%$ (intrinsic scatter 0.28~dex) of the stellar mass budget for $M_{\mathrm{\ast, bulge}} \approx 10^{10}\, M_\odot$ \citep{Kormendy2013ARAA}. The 15 high-z quasars analyzed in this study (blue points) exhibit a systematically higher relative BH mass fraction (median $M_{\mathrm{BH}}/M_\ast = 2.3\%$; scatter 0.51~dex), but substantially lower than previous estimates that approach 10\% based on stellar masses inferred from ALMA-derived dynamical masses \citep[e.g.,][]{Venemans2017ApJ,Pensabene2020AA,Wangfeige2024ApJ}. For instance, two quasars in our sample, J0224$-$4711 and J2002$-$3013, have published dynamical masses of $M_{\mathrm{dyn}} = 10^{11.32}\,M_\odot$ and $10^{10.76}\,M_\odot$ \citep{Wangfeige2024ApJ}, which are lower than our measurements of $M_\ast = 10^{11.64}\,M_\odot$ and $10^{11.48}\,M_\odot$, respectively. What underlies this discrepancy? A likely culprit for the underestimated dynamical masses is that the [C\,II] $158\,\mu$m emission detected by ALMA does not reach yet the flat part of the rotation curve, which typically lies at $2-3$ times $R_e$ \citep{Sofue2001ARAA,Walter2008AJ}. According to our \textsc{GalfitS} analysis, J0224$-$4711 has $R_e = 1.04 \pm 0.48$~kpc and J2002$-$3013 $R_e = 1.41 \pm 0.34$~kpc (Table~\ref{tab:gsresultnp}), which are comparable to the [C\,II] disk sizes of $R_{\mathrm{[C\,II]}} = 1.42 \pm 0.08$ and $1.21 \pm 0.19$~kpc, respectively \citep{Wangfeige2024ApJ}.

More extreme BH-to-host mass ratios characterize the less luminous quasars (red points; Figure~\ref{fig:mmrelation}), for which $M_{\mathrm{BH}}/M_\ast = 4.7\%$ with a scatter of 0.50~dex, $\sim 4$ times higher than the value for the most luminous quasars. While these mass ratios are less extreme than the values reported by \citet{Chen2025} for LRDs (orange points)\footnote{\ledit{Table~1 of \citet{Chen2025} contains a typographical error. For source 20466, the BH mass, taken from \citet{Greene2024ApJ}, should be $\log (M_{\rm BH}/M_\odot) = 8.17 \pm 0.42$. The stellar mass upper limit should be  $\log (M_\ast/M_\odot) <8.84$, as seen in Figure~6 of \citet{Chen2025}.}}, our results contribute to the growing body of evidence that $z \gtrsim 4$ AGNs often host overmassive BHs \citep{Harikane2023ApJ,Kocevski2023ApJ, Maiolino2024AA}. Collectively, these findings support the notion that BH growth outpaced stellar mass assembly in high-redshift systems.


\begin{figure}
\centering
\includegraphics[width=0.45\textwidth]{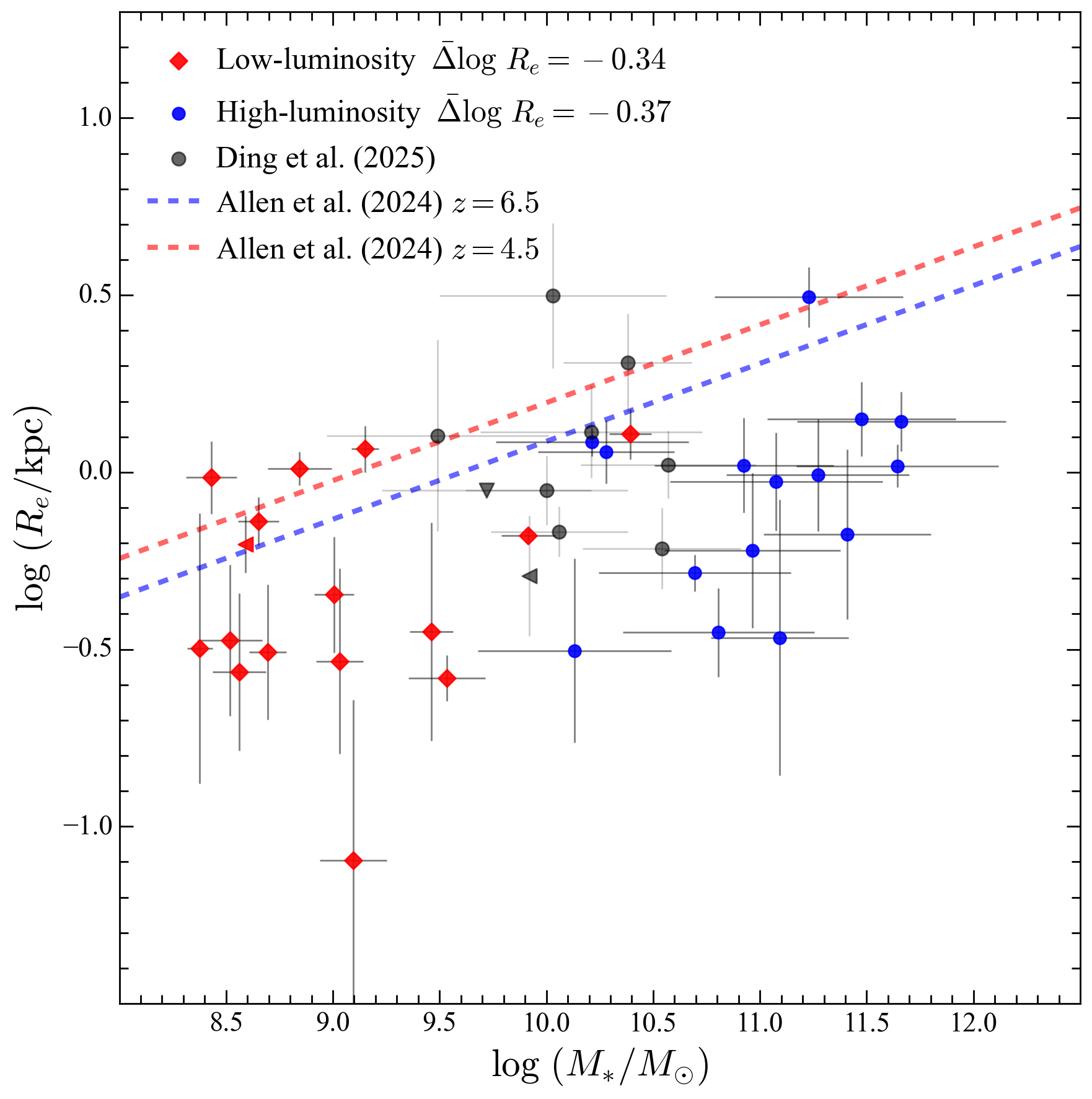}
\caption{Stellar mass-size relation for AGN host galaxies, consisting of 15 high-luminosity (blue) and 16 low-luminosity (red) quasars. The blue and red dashed lines correspond to the MSR from \citet{Allen2025AA} at $z = 6.5$ and 4.5, respectively. The high-luminosity quasars and low-luminosity quasars generally lie below the MSR at $z = 6.5$ and $z = 4.5$, with median deviations of $\bar{\Delta} \log{R_e} = -0.37$~dex and $-0.34$~dex, respectively. We also include 10 SHELLQs quasars (black circles; black triangles for upper limits) from \citet{Ding2025ApJ}.}
\label{fig:sizerelation}
\end{figure}

\begin{figure}
\centering
\includegraphics[width=0.45\textwidth]{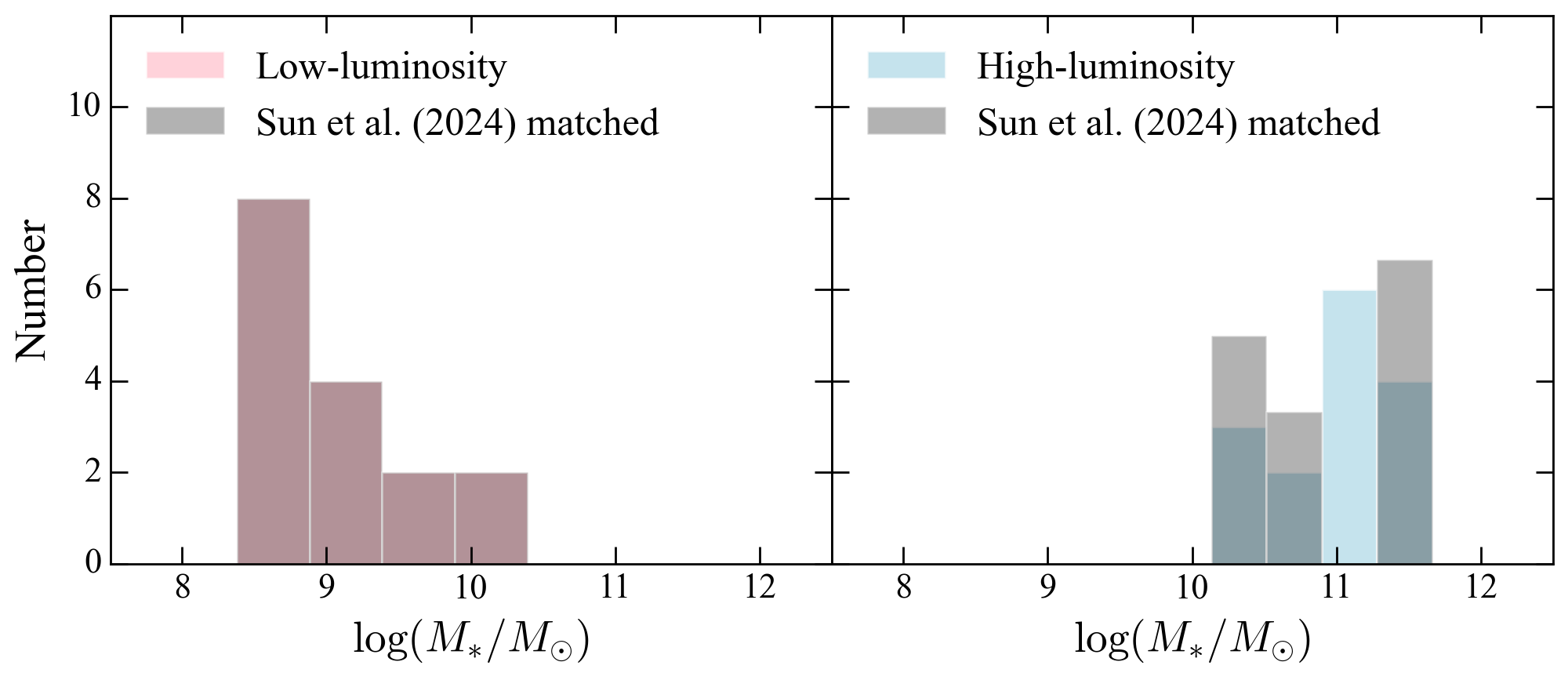}
\includegraphics[width=0.46\textwidth]{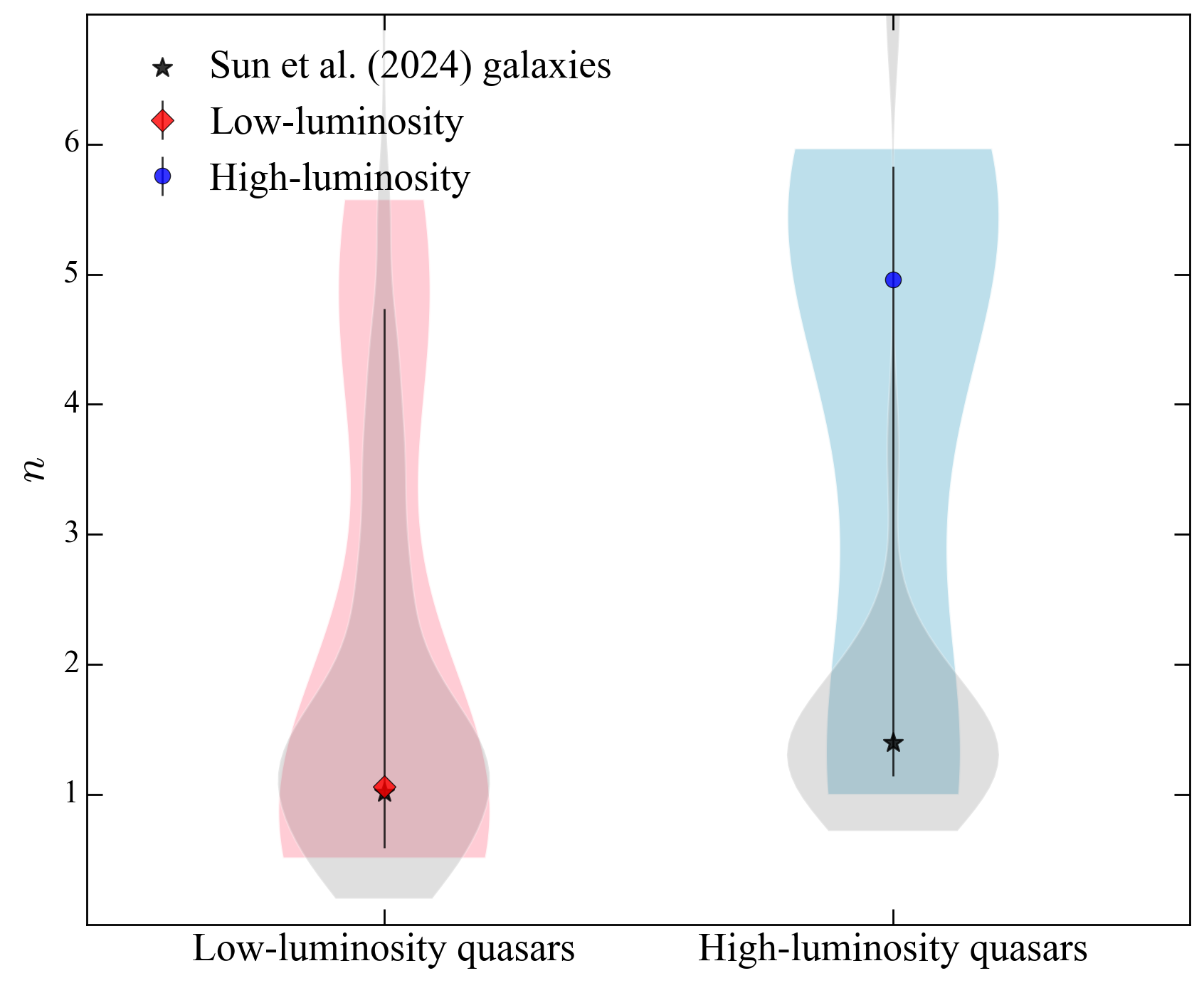}
\caption{Comparison of the distribution of \sersic\ index ($n$) of our quasar samples and with those of matched galaxies from \citet{Sun2024ApJ}. The upper panels show the stellar mass distributions of the low-luminosity (left) and high-luminosity (right) quasars, along with their respective comparison samples of galaxies matched in redshift and stellar mass (gray). The lower panel displays the corresponding violin plots for the distributions of $n$, highlighting the median values (filled circles) and their 16\% to 84\% ranges. The black stars denote the median $n$ of the matched galaxies.}
\label{fig:sindex}
\end{figure}

\subsection{Host Galaxy Morphology}
\label{sec:morpho}

Among the many empirical correlations that characterize the galaxy population, one of the most widespread is the relation between stellar mass and size (e.g., \citealt{Shen2003, vanderWel2014ApJ}). Our sample of high-redshift AGN host galaxies lies systematically below the MSR for galaxies at similar redshifts (Figure~\ref{fig:sizerelation}). \ledit{ Using the fiducial MSR at $z=4.5$ and $z=6.5$ as reference \citep{Allen2025AA}, the median size deviations of low-luminosity and high-luminosity quasars is $\bar{\Delta} \log\, {R_e} = -0.34$ and $-0.37$~dex, respectively.}

Apart from size, we also examine the distribution of \sersic\ indices of our quasar hosts with matched galaxy populations from \citet{Sun2024ApJ}, who used two-dimensional parametric fits to analyze a large sample ($N \simeq 350$) of $4 < z < 9.5$ galaxies with $M_\ast \approx 10^7 - 10^{11}\,M_\odot$ using seven-band NIRCam observations from the CEERS program. To enable a direct comparison, we divide the galaxy sample into two redshift bins, $4 \leq z \leq 6$ and $6 \leq z \leq 7$, matching their stellar masses to those of the high-luminosity and low-luminosity quasars, respectively (upper panel in Figure~\ref{fig:sindex}). Relative to the control sample of inactive galaxies, the host galaxies of high-luminosity quasars possess significantly larger median \sersic\ indices ($n \approx 5$), consistent with early-type morphologies. By contrast, the host galaxies of the less powerful quasars exhibit \sersic\ indices $n \approx 1$, similar to their matched counterparts and indicative of exponential, disk-like structures. These results strongly indicate that luminous quasars reside in more centrally concentrated, evidently bulge-dominated host galaxies distinctly different from both typical star-forming galaxies and lower luminosity AGNs at similar redshifts and comparable stellar mass. Like the most powerful quasars, their less luminous counterparts are also more compact at a given stellar mass than typical galaxies, but their stellar mass distribution or internal structure (concentration, \sersic\ $n$, bulge-to-disk ratio) is manifestedly different. Our results echo Hubble Space Telescope studies that find a higher incidence of early-type hosts among low-redshift quasars of higher luminosity and larger BH mass (e.g., \citealt{Dunlop2003MNRAS, Kim2017, Zhao2021}). Meanwhile, our observations are broadly consistent with theoretical expectations that luminous quasars at high redshift form in dense environments conducive to rapid gas inflows and gas-rich mergers. Such dissipative processes efficiently funnel gas toward galaxy centers, rapidly building dense stellar cores that host accreting supermassive BHs (e.g., \citealp{Dekel2014MNRAS,Zolotov2015MNRAS}).

While the comparison between the AGNs and the control sample of inactive galaxies points to plausible intrinsic differences between the two populations, some caution needs to be exercised because there are methodological differences between our decomposition technique and the approach employed to study the control sample. With the aid of \textsc{GalfitM} \citep{Haussler2013MNRAS}, \citet{Allen2025AA} performed single-component \sersic\ profile fits and showed that galaxy size increases toward longer rest-frame wavelengths (but see \citealt{Jin2025} for discussion about departures from this trend). In contrast, our \textsc{GalfitS} decomposition assumes a uniform spectral shape for the host galaxy without accounting for wavelength-dependent structural variations. Given the younger ages typical at high redshift, as reflected in our stellar population modeling results (see Table~\ref{tab:gsresultnp}), the sizes we measure are likely biased toward UV wavelengths. Although this effect potentially can explain part of the observed compactness of quasar host galaxies, we believe that it is unlikely to be the whole story. The effective radii of high-redshift galaxies is typically only $0.15$~dex smaller in the UV than in the optical \citep{Allen2025AA}, much less than the observed size discrepancy between quasar hosts and normal galaxies. Instead, our findings strongly imply that high-redshift quasar host galaxies are intrinsically more compact (have smaller $R_e$) than inactive galaxies of similar stellar mass at comparable redshift. This conclusion aligns with previous observational studies at intermediate redshifts ($1.6 \lesssim z \lesssim 3.5$; \citealt{Li2021ApJ,Ding2022ApJ}).  A similar caveat applies to the \sersic\ indices drawn from \citet{Sun2024ApJ}, which were also derived using \textsc{GalfitM}. Radial color gradients produce wavelength-dependent variations in \sersic\ $n$ \citep{Jin2025}, but not at a level that can account for the gross differences in concentration found between the high-luminosity quasars and control galaxies (lower panel in Figure~\ref{fig:sindex}).

\begin{figure}
\centering
\includegraphics[width=0.45\textwidth]{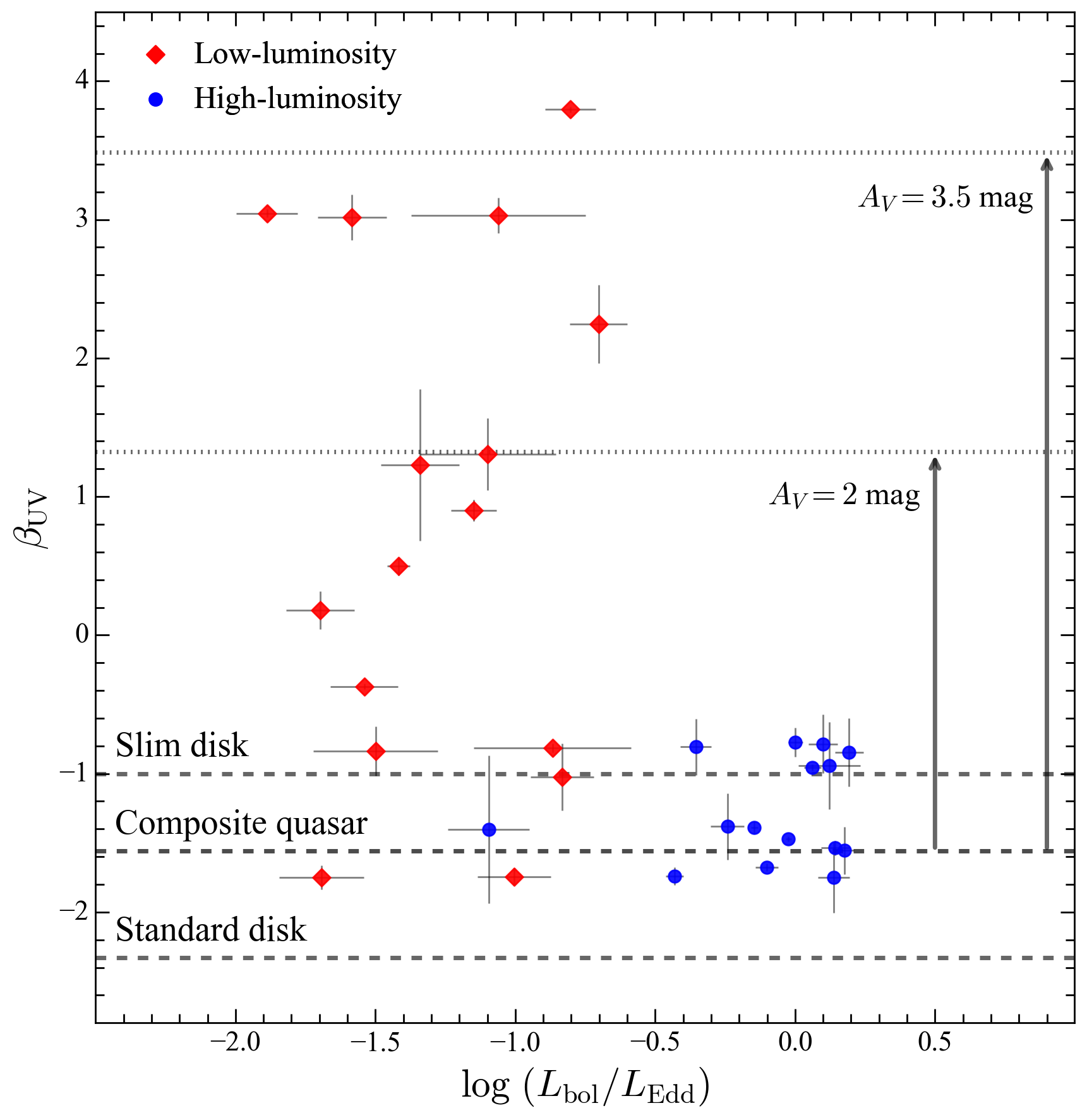}
\caption{The nuclear UV spectral index ($L_\lambda \propto \lambda^{\beta_\mathrm{UV}}$, for $\lambda \leq 5100$~\AA) as a function of the Eddington ratio ($L_{\rm bol}/L_{\rm Edd}$). The black dashed lines mark $\beta_\mathrm{UV} = -1$ for a slim accretion disk \citep{Abramowicz1988ApJ,Kato2008book}, $\beta_\mathrm{UV} = -1.56$ for the composite spectrum of SDSS quasars \citep{VandenBerk2001AJ}, and $\beta_\mathrm{UV} = -2.33$ for a standard accretion disk \citep{Shakura1973}. The two dotted lines denote the spectral index of the composite quasar SED dust-extincted by $A_V = 2$ and 3.5 mag. }
\label{fig:color}
\end{figure}

\subsection{A Bimodality in AGN Colors}
\label{sec:ncolor}

We fit the nuclear point source with a simple power‐law model (Section~\ref{sec:agnsed}) to derive two essential quantities, $L_\mathrm{5100}$ and the UV spectral index ($\beta_\mathrm{UV}$). The ``color-magnitude diagram" of $\beta_\mathrm{UV}$ versus Eddington ratio, $L_\mathrm{bol}/L_\mathrm{Edd} \equiv 10.6 \, L_\mathrm{5100}/L_\mathrm{Edd}$, reveals distinctly different distributions for our two subsamples (Figure~\ref{fig:color}). Whereas the most luminous quasars, characterized by $L_\mathrm{bol}/L_\mathrm{Edd} \gtrsim 0.3$, have a median spectral slope of ${\beta_\mathrm{UV}} = -1.39 \pm 0.36$ and cluster closely between the spectral slope of the composite spectrum of quasars ($\beta_\mathrm{UV} = -1.56$; \citealp{VandenBerk2001AJ}) and the theoretical expectation of a slim accretion disk ($\beta_\mathrm{UV} = -1$; \citealp{Abramowicz1988ApJ,Kato2008book}), the lower luminosity population, with $L_\mathrm{bol}/L_\mathrm{Edd} \approx 0.01-0.1$, occupies a surprisingly wide range of spectral slopes from $\beta_\mathrm{UV} \approx -1.7$ to $+3.8$. A typical luminous quasar such as J0305$-$3150 (Figure~\ref{fig:galfits}) has a blue spectrum that differs markedly from the red continuum of the less powerful nuclei in J0224$-$4711-BHAE-1 (Figure~\ref{fig:galfits}) and CEERS~006725 (Figure~\ref{fig:galfits_exp7b}).

Our decomposed AGN+host SED bears an intriguing resemblance to the popular interpretation of LRDs that attributes the blue portion of their SED to host galaxy emission and the red part to a heavily dust-reddened AGN (e.g., \citealp{Greene2024ApJ,Wang2024ApJ}). If such a scenario applies to the low-luminosity quasars under consideration, we can estimate roughly the required level of extinction $A_V$ by assuming that the intrinsic SED follows the composite quasar spectrum of \citet{VandenBerk2001AJ}, for which $\beta_\mathrm{UV} = -1.56$, and adopting the dust extinction curve from \citet{Calzetti2000ApJ}. Explaining the red slopes of the lower luminosity quasars would require dust extinctions of $A_V \approx 2–3.5$ mag (Figure~\ref{fig:color}).

We hasten to add that there is no shortage of alternative explanations for the V-shaped SED of LRDs, ones that place less reliance, if any, on dust reddening. Physical mechanisms that can produce intrinsically red optical spectra include a starburst accretion disk with an inner truncation radius \citep{Chen2024ApJ}, the gravitationally unstable outer region of an otherwise standard accretion disk \citep{Zhang2025}, blackbody radiation from a massive, optically thick gaseous envelope \citep{Kido2025}, and the circumbinary disk around a binary massive BH \citep{Inayoshi2025}. Additionally, the moderately low Eddington ratios ($L_\mathrm{bol}/L_\mathrm{Edd} \approx 0.03$) of the low-luminosity population may implicate a radiatively inefficient accretion flow \citep{Narayan1994ApJ}, which, by analogy with nearby low-luminosity AGNs \citep{Ho2008ARAA}, would lack of a strong ``big blue bump'' and naturally exhbit a redder continuum \citep{Ho1999}.

\section{Discussion}
\label{sec:sec5}

\begin{figure*}
\centering
\includegraphics[height=0.46\textwidth]{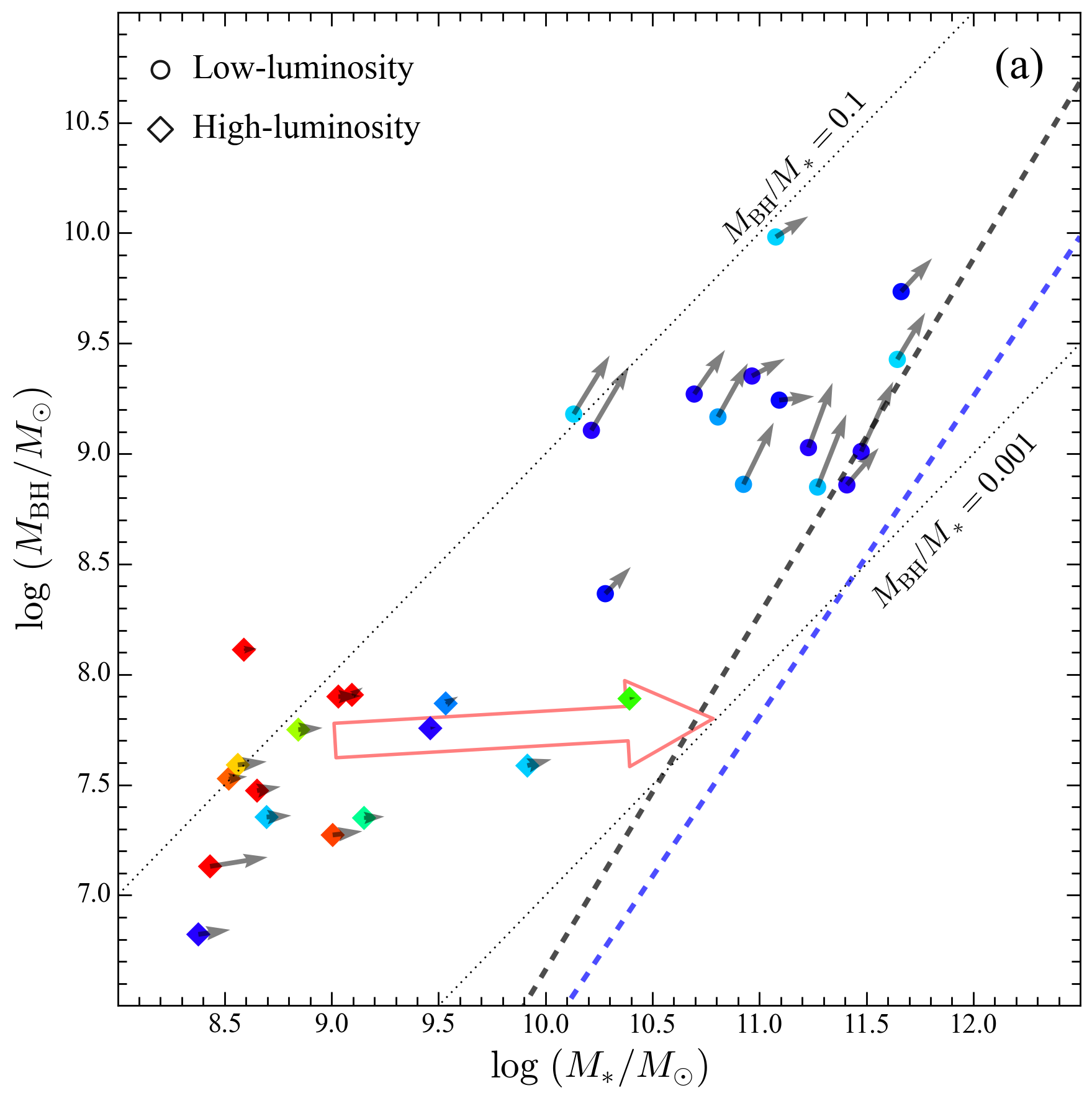}
\includegraphics[height=0.465\textwidth]{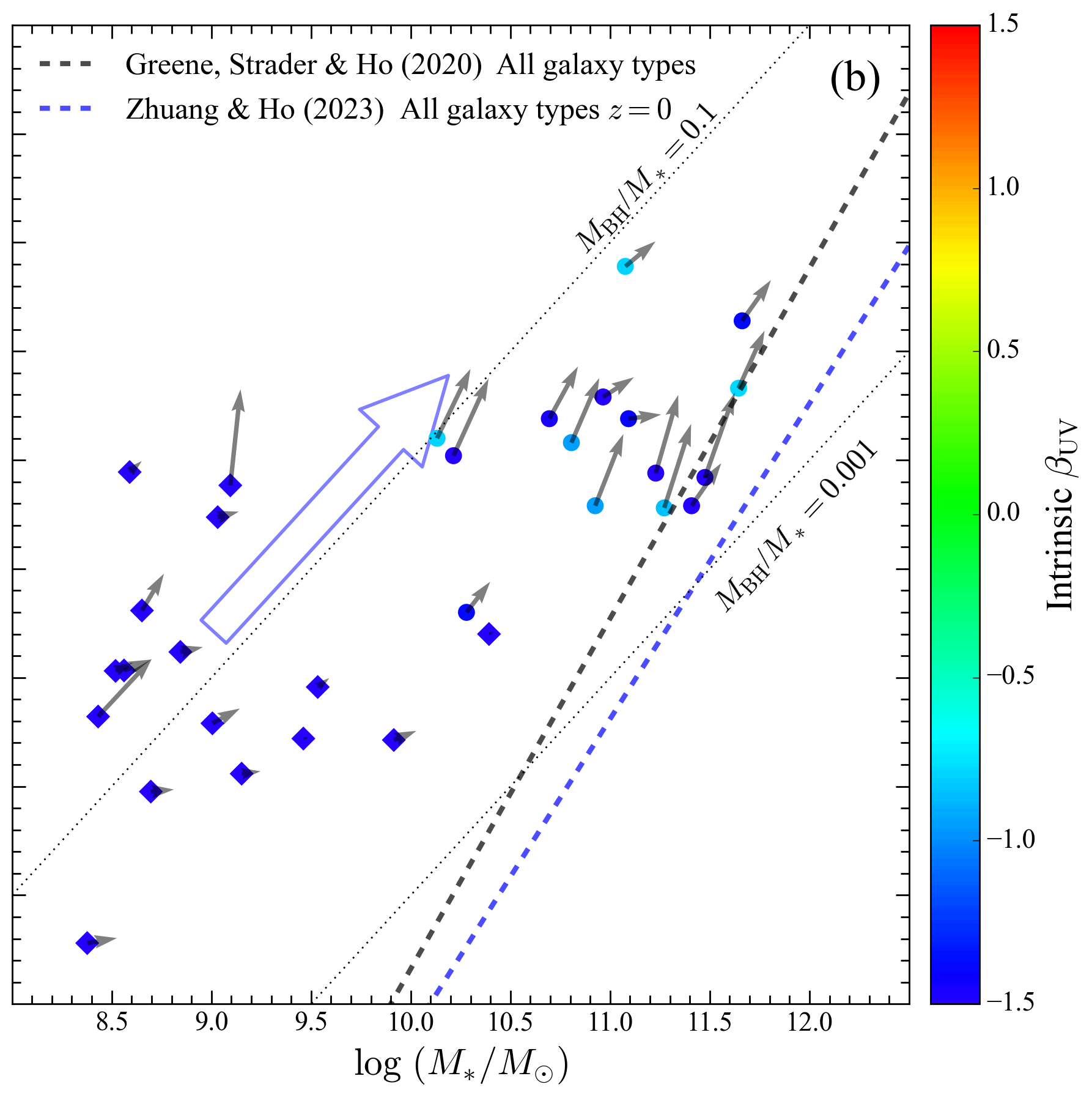}
\caption{Potential evolutionary tracks of quasars on the BH mass ($M_{\rm BH}$) versus stellar mass ($M_\ast$) plane. Panel (a): the sample is color-coded based on the observed UV spectral index $\beta_\mathrm{UV}$. The objects (see also Figure~\ref{fig:color}) outside the limits of the color bar are shown with red ($\beta_\mathrm{UV} \geq 1.5 $) and blue ($\beta_\mathrm{UV} \leq -1.5 $). Gray arrows indicate the shift in the $M_{\rm BH}–M_\ast$ plane over a timescale of 10~Myr, assuming the current values of SFR and $\dot{M}_{\rm BH}$ derived from our \textsc{GalfitS} measurements (Table~\ref{tab:gsresultnp}). The red arrow highlights a potential evolutionary pathway for the low-luminosity quasars. The dashed lines represent the local $M_{\rm BH}–M_\ast$ relation for AGN host galaxies (blue; \citealt{Zhuang2023NatAs}) and inactive galaxies (black; \citealt{Greene2020ARAA}). Dotted lines correspond to constant $M_{\rm BH}/M_\ast = 0.1$ and 0.001. Panel (b): same as panel (a), but here we assume that the low-luminosity quasars intrinsically follow a normal quasar SED ($\beta_\mathrm{UV} = -1.56$; \citealt{VandenBerk2001AJ}) and that their observed high values of $\beta_\mathrm{UV}$ result from dust reddening, using the \citet{Calzetti2000ApJ} extinction law. Both $M_{\rm BH}$ and $\dot{M}_{\rm BH}$ are corrected for extinction, which primarily affects the location and direction of movement of the low-luminosity quasars compared to panel (a). The blue arrow highlights the potential evolutionary pathway of the low-luminosity quasars under the assumption that they are not intrinsically red.}
\label{fig:mmvector}
\end{figure*}

\begin{figure*}
\centering
\includegraphics[width=0.98\textwidth]{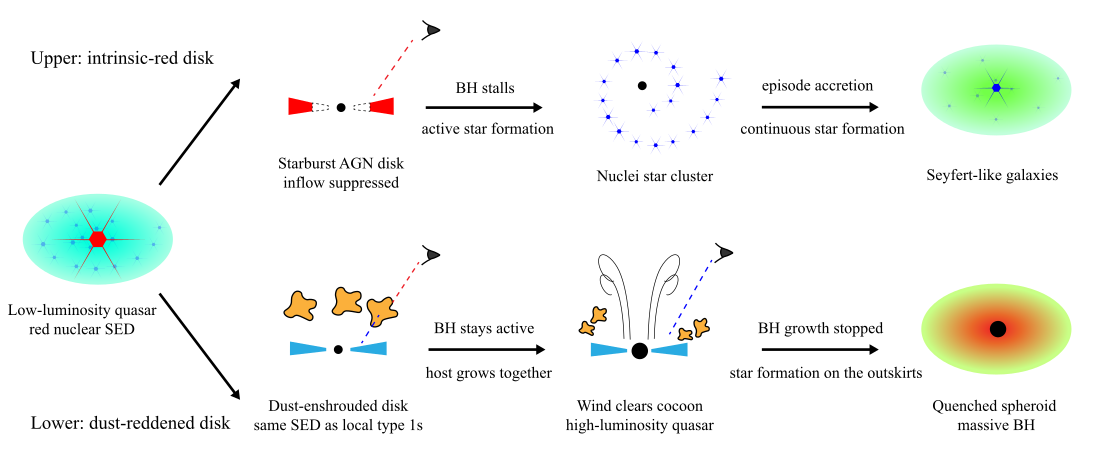}
\caption{Schematic illustration of the possible evolutionary pathways of low-luminosity quasars. The diagram outlines two scenarios. In the upper pathway, the nucleus is intrinsically red, with a starburst-driven accretion disk leading to suppressed BH growth, and it finally evolves into the population of Seyfert-like AGNs that follow the local $M_\mathrm{BH}–M_\ast$ relation. In the lower pathway, the nucleus has a standard disk that is initially obscured but later transitions into a luminous quasar phase because of energy-driven winds, ultimately evolving into an elliptical galaxy hosting an overmassive BH.}
\label{fig:cartoon}
\end{figure*}

\subsection{The Growth Path of High-redshift Supermassive BHs}
\label{sec:BHpath}

We have shown that high-redshift low-luminosity quasars lie significantly above the local $M_\mathrm{BH}–M_\ast$ relation (Figure~\ref{fig:mmrelation}). They are characterized by BH-to-stellar mass ratios of $M_{\mathrm{BH}}/M_\ast \approx 1\%-10\%$, values rarely found in the local Universe \citep{Zhuang2023NatAs} or even at cosmic noon (e.g., \citealt{Bongiorno2014MNRAS}), highlighting the importance of clarifying their growth pathways. Theoretical considerations predict that the first supermassive BHs form via heavy seeds ($M_{\rm BH} \approx 10^4-10^6\,M_\odot$) from ``direct collapse'' of metal-free gas \citep{Latif2016PASA,Regan2020MNRAS} or light seeds ($M_{\rm BH} \approx 10^2\,M_\odot$) of stellar origin that undergo super-Eddington accretion \citep[e.g.,][]{Natarajan2014GReGr,Pacucci2022MNRAS}. With $M_{\rm BH} \approx 10^7-10^8\,M_\odot$, however, the BHs in low-luminosity quasars are already quite massive and cannot be regarded as ``seeds'' in the conventional sense. Analogous to the LRDs (orange points in Figure~\ref{fig:mmrelation}), which are even more extreme in their near complete absence of any detectable host galaxy \citep{Chen2025}, the under-developed hosts of high-redshift low-luminosity quasars underscore that this population is caught during an early epoch of rapid evolution during which the growth of the BH and its host are not tightly synchronised.

Leveraging our \textsc{GalfitS} analysis, which yields robust measurements of the properly separated host galaxy and AGN components, we estimate the SFR and the BH accretion rate ($\dot{M}_{\rm BH}$) of each source. This enables a first-order forecast of the future evolution of $M_\ast$ and $M_{\rm BH}$. Figure~\ref{fig:mmvector}a illustrates the likely trajectory of each low-luminosity quasar over the next 10~Myr, roughly one \citet{Salpeter1964ApJ} e-folding time, assuming constant SFR and $\dot{M}_{\rm BH}$. Under steady conditions the sources will migrate toward the local $M_\mathrm{BH}–M_\ast$ relation (red arrow), consistent with the findings in \citet{Zhuang2023NatAs} that objects offset from the local relation tend to move back onto it. A major complication arises from the uncertain interpretation of the systematically redder SEDs of the low-luminosity quasars (Figure~\ref{fig:color}). Are the unusual SEDs intrinsically red or reddened? If the colors implicate dust reddening, then extinction correction would dramatically alter the evolutionary vector of the sources. Under this alternative scenario (Figure~\ref{fig:mmvector}b), wherein we assume that the intrinsic SED has a slope of $\beta_\mathrm{UV} = -1.56$ \citep{VandenBerk2001AJ} and deredden the observed spectral indices and luminosities using the \citet{Calzetti2000ApJ} extinction law, the low-luminosity quasars would grow along a path roughly consistent with $M_\mathrm{BH} \approx  0.1\,M_\ast$ (blue arrow), ultimately overlapping with the locus of luminous quasars at later epochs.

Figure~\ref{fig:cartoon} gives a schematic overview of how low-luminosity quasars might evolve, depending on whether the accretion disk is truly red or simply reddened by dust. We envisage two possible scenarios:

\begin{itemize}

\item \textit{Upper pathway:} The accretion disk is intrinsically red, with no significant dust extinction, because it
has entered a starburst phase that reduces its gas inflow. After an early period of rapid BH growth, via a massive ``heavy seed'' (e.g., \citealp{Volonteri2010AARv,Natarajan2014GReGr}), a light seed undergoing super-Eddington accretion (e.g., \citealp{Inayoshi2022ApJ}), or even ab initio from the core collapse of a self-interacting dark matter halo (e.g., \citealt{Jiang2025}), the BH attains a mass substantially above the local $M_\mathrm{BH}–M_\ast$ relation. At this point, the disk may become self-gravitating \citep{Chen2024ApJ}, with newly formed stars and their remnants depleting the inward gas supply. This process effectively truncates the disk at a radius of $\sim 100$ gravitational radii, producing a red SED similar to that seen in the low-luminosity quasars. With gas accretion onto the BH severely diminished, the BH barely grows further, while the host galaxy continues to form stars without strong AGN feedback. These systems evolve along the red arrow in Figure~\ref{fig:mmvector}a and gradually converge with the local $M_\mathrm{BH}–M_\ast$ relation. The central star-forming phase can culminate in the formation of a compact nuclear star cluster \citep{Neumayer2020}, and over time the galaxy may settle into a Seyfert-like system experiencing episodic low-level accretion \citep{Ho2008ARAA}. This scenario is reminiscent of nearby AGNs with evidence of nuclear star formation (e.g., \citealt{Davies2007ApJ}), and may be related to the relic stellar populations detected in the Milky Way’s center (e.g., \citealp{Pfuhl2011ApJ}).

\item \textit{Lower pathway:} If the nucleus is reddened by a dust-enshrouded environment, then after correcting for extinction the BHs in low-luminosity quasars would lie even farther above the local $M_\mathrm{BH}–M_\ast$ relation (blue arrow in Figure~\ref{fig:mmvector}b), in some cases reaching $M_\mathrm{BH} \approx M_\ast$ \citep[see also][]{Harikane2023ApJ,Maiolino2024AA}. Once the BH becomes sufficiently massive, energy-driven winds or radiative feedback (e.g., \citealp{Hopkins2008ApJS,King2015ARAA}) expel the surrounding gas, revealing a luminous quasar. Although the BH and the galaxy may continue to grow for some time, prolonged or repeated episodes of strong feedback can eventually quench star formation, leaving the BH either on or above the local scaling relation of massive, early-type galaxies.

\end{itemize}

Although both evolutionary pathways remain plausible, the obscured-nucleus scenario may overproduce massive BHs relative to the observed number densities at low redshift. The local number density of BHs with masses $M_{\rm BH} \geq 10^9\,M_\odot$ is $1.5 \times 10^{-5}\,\rm Mpc^{-3}$ \citep{Shankar2009ApJ,Kormendy2013ARAA}. \ledit{In contrast, at high redshift ($3.5 <z < 6$) the low-luminosity quasars are significantly more common, with a number density of $3.5 \pm 1.6 \times 10^{-4}\,\rm cMpc^{-3}$ for $M_{\rm BH} = 10^6-10^8\,M_\odot$ \citep{Taylor2025ApJ}.} If all the high-redshift systems were to evolve into massive systems, it would result in an overabundance of local $10^9\,M_\odot$ BHs. This apparent discrepancy suggests that the obscured-nucleus pathway, if applied uniformly to all high-redshift low-luminosity quasars, may not be viable without invoking additional mechanisms to reduce the final BH masses or alter the host galaxy properties during subsequent evolution.

\begin{figure*}
\centering
\includegraphics[height=0.43\textwidth]{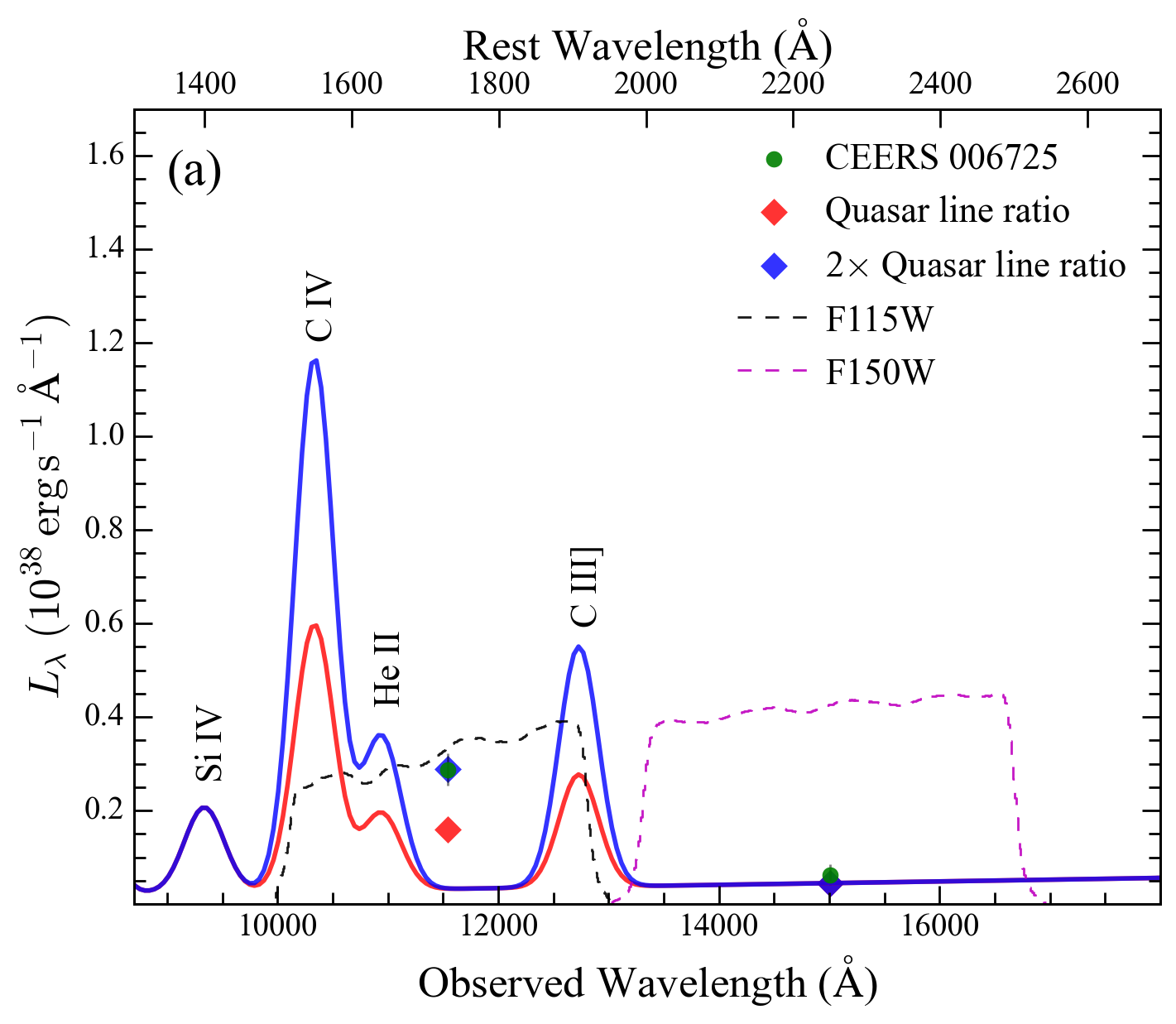}\includegraphics[height=0.43\textwidth]{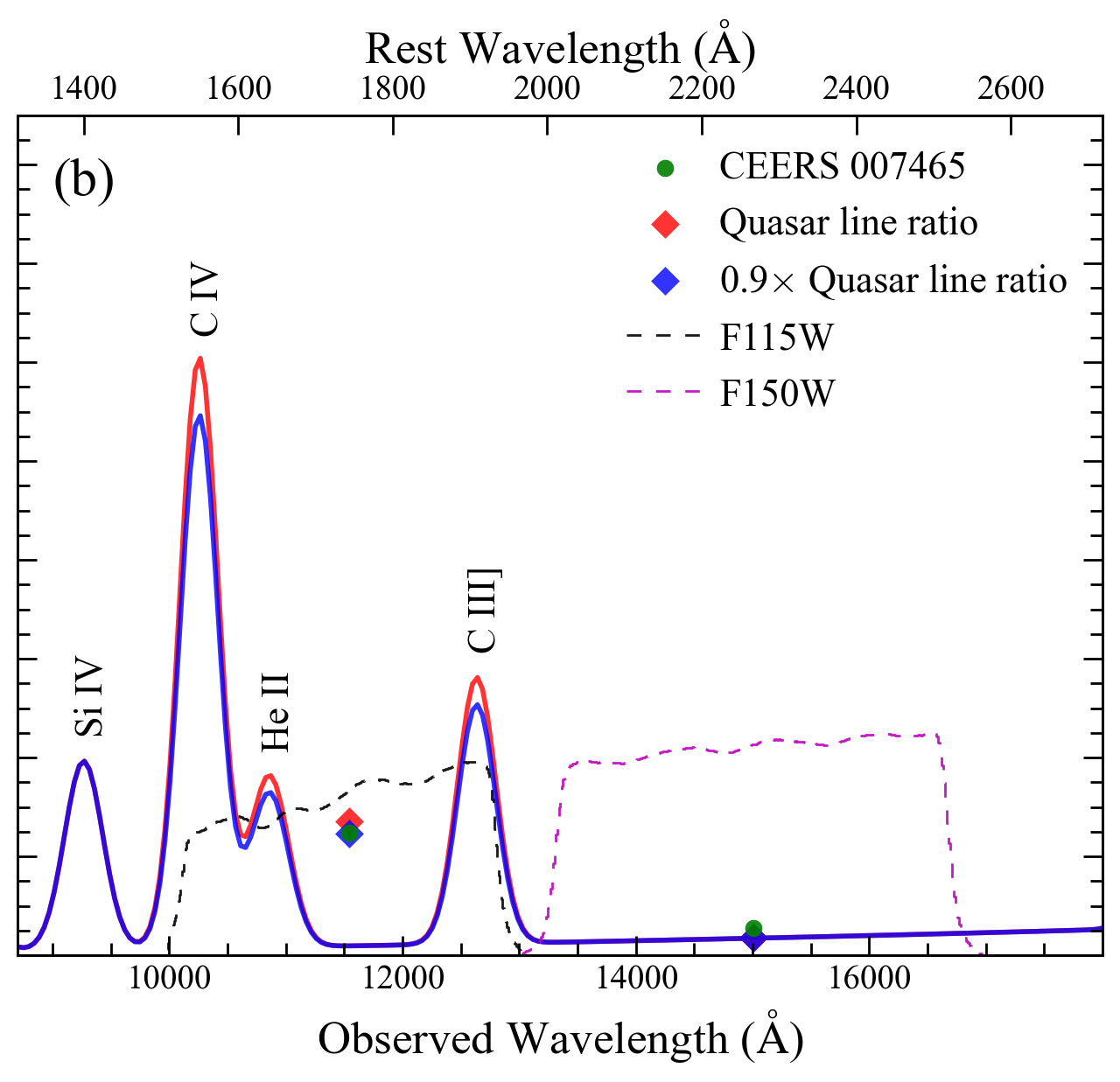}
\caption{Illustration of detailed UV SED analysis for (a) CEERS~006725 and (b) CEERS~007465, following the method in Section~\ref{sec:agnsed}. Each panel shows the observed‐frame SED, plotted as green circles. The response curves for the F115W and F150W filters are shown as black and magenta dashed curves. Key emission lines are labeled (Si\,IV $\lambda 1400$, C\,IV $\lambda 1549$, He\,II $\lambda 1640$, C\,III] $\lambda 1909$). We compare the observed SED with an SED model (red diamonds) that considers the fiducial line ratios from the composite quasar spectrum of \cite{VandenBerk2001AJ}. The data for CEERS~006725 can be matched by an SED model that has 2 times the fiducial line ratios, while CEERS~007465 requires 0.9 times the fiducial line ratios, as indicated by the blue diamonds. }
\label{fig:ceersUV}
\end{figure*}

\subsection{UV SED of CEERS~006725 and CEERS~007465}
\label{sec:UVsed}

CEERS~006725 and CEERS~007465 have strong rest-frame UV emission lines despite their relatively red continua (Figure~\ref{fig:ceersUV}). To quantify the line strengths, we anchored the broad \ha\ flux to the values reported by \citet{Harikane2023ApJ} and scaled the UV emission lines using typical quasar line ratios from \citet{VandenBerk2001AJ}. This approach reproduces the observed UV photometry well. CEERS~006725 requires twice the fiducial quasar line fluxes, whereas CEERS~007465 only needs slightly lower line fluxes (0.9 times) compared to the quasar composite. Nonetheless, both sources have very low continuum luminosities, resulting in extraordinarily high equivalent widths for lines such as \ion{C}{4} (2861\,\AA\ for CEERS~006725 and 4560\,\AA\ for CEERS~007465), far in excess of the equivalent width of $\sim 25$\,\AA\ in typical unobscured quasars \citep{VandenBerk2001AJ}.

In the dust extinction scenario, the red UV spectral indices of CEERS~006725 and CEERS~007465 imply high extinction values of $A_V = 1.93$ and 2.62 mag, respectively. \citet{Harikane2023ApJ} estimate even higher dust extinctions of $A_V = 2.48$ and 4.28 mag for these sources based on the steep Balmer decrements from \cite{Nakajima2023ApJS}. A significant challenge emerges: if the broad-line region (BLR) is embedded within a dusty region, the dust should also extinct the broad emission lines and lead to lower observed line fluxes. Reconciling the extremely high \ion{C}{4} equivalent widths under this scenario would require invoking a configuration in which the BLR lies outside the dusty region, a possibility that contradicts reverberation mapping studies that indicate that the BLR is confined within the dust sublimation radius \citep{Cackett2021iSci}.

Alternatively, the intrinsically red SED scenario posits that the red continuum is an inherent property of the AGN and not a consequence of dust reddening. Although the steep Balmer decrement might be interpreted as evidence for dust extinction, the Balmer decrement in the BLR is highly sensitive to its physical conditions \citep{Li2022ApJ}. The theoretical models of \cite{Korista2004ApJ} show that the H\(\alpha\)/H\(\beta\) ratio generally increases as the BLR cloud density decreases. A cloud density that drops slightly below the critical density for the \ion{C}{4} transition (\(n_e \approx 10^{10}\,\mathrm{cm}^{-3}\)) would yield both an elevated H\(\alpha\)/H\(\beta\) ratio and a higher \(\mathrm{C\,IV}/\mathrm{H\alpha}\) ratio, as observed in CEERS~006725. Of course, this scenario faces the quandary of how an intrinsically red accretion disk can produce a sufficiently hard ionizing spectrum to power the strong UV emission lines.

In summary, we favor the intrinsically red SED scenario. The central regions of active galaxies in general (e.g., \citealt{Xie2021, Zhuang2021}) and those at high redshifts in particular often experience vigorous star formation, which yields a generally blue host continuum (see Figure~\ref{fig:galfits_exp7b}) and supplies ample ionizing photons. The frequent failure by JWST/MIRI and ALMA observations of red high‑\(z\) AGNs to detect the levels of dust emission predicted by energy balance considerations (e.g., \citealt{Setton2025apjl, Wang2025ApJ}) may be regarded as further support for the idea that the \ledit{red continuum is an intrinsic property of the AGN instead of the result of dust extinction \citep{Greene2025arXiv}.}

\begin{figure}
\centering
\includegraphics[width=0.47\textwidth]{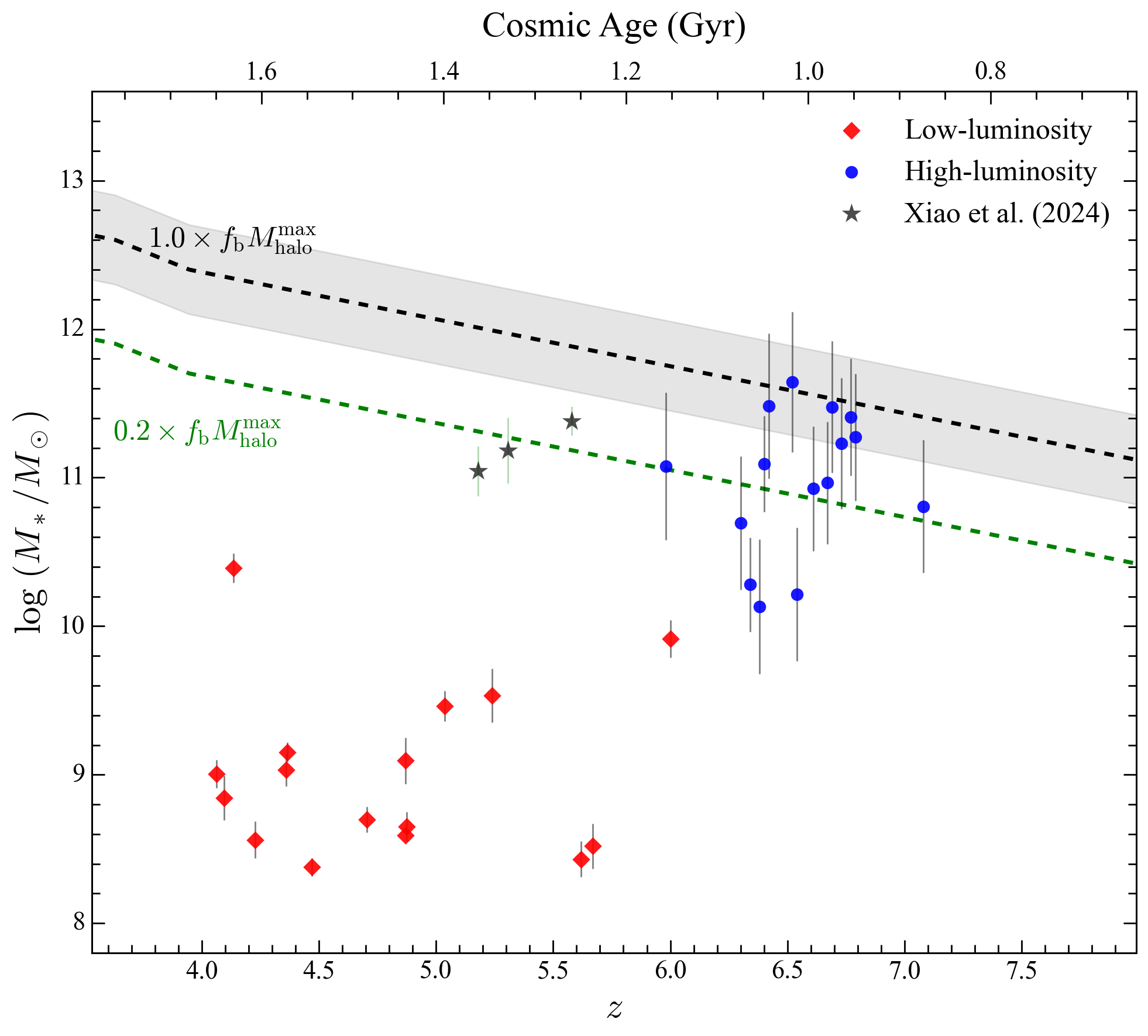}
\caption{Stellar masses of high-redshift quasars compared with halo-based mass estimates. Gray stars indicate three massive galaxies from \citet{Xiao2024Natur}. The bottom axis shows redshift, and the top axis gives the corresponding cosmic age. The black dashed line represents the theoretical stellar mass upper limit, defined as the cosmic baryon fraction ($f_\mathrm{b} = 0.158$; \citealp{Planck2020AA}) multiplied by the maximum halo mass ($M_{\mathrm{halo}}^{\mathrm{max}}$), while the green dashed line shows 20\% of this upper limit. The grey shaded region indicates the estimated scatter of $1\,\sigma \approx 0.3$~dex in $M_{\mathrm{halo}}^{\mathrm{max}}$ induced by cosmic variance.}
\label{fig:halo}
\end{figure}

\subsection{Trigger of Quasar Activity}
\label{sec:trigger}

The mass-weighted effective radii of our quasar host galaxies are systematically smaller by $\sim 0.3$~dex than typical galaxies observed at similar redshift in rest-frame optical images (Section~\ref{sec:morpho}). This discrepancy significantly exceeds the expected wavelength-dependent size variation of $0.15$~dex \citep{Allen2025AA}. Luminous quasars also exhibit high \sersic\ indices ($n \approx 5$) that further implicate morphologies akin to those of bulge-dominated, early-type galaxies. Despite having already built substantial mass in stars ($\bar{M}_\ast \approx 10^{11}\,M_\odot$) and in the central BH ($\bar{M}_{\rm BH} = 10^{9.15}\,M_\odot$), these systems harbor active star formation and ongoing BH accretion, consistent with quasars at high \citep{Salvestrini2025AA} and low \citep{Shangguan2018ApJ, Xie2021, Molina2023ApJ} redshifts.

To investigate how these massive systems formed at high redshifts, we compare the host galaxy stellar mass with estimates of the halo-based maximum mass (Figure~\ref{fig:halo}). Considering the total surveyed cosmic volume derived from an area of $0.088\,\mathrm{deg}^2$ across the redshift range $4 \leq z \leq 7$, we estimate the maximum halo mass ($M_{\mathrm{halo}}^{\mathrm{max}}$) using the halo mass function from \citet{Tinker2008ApJ}. Most quasars lie near the theoretical upper bound (black dashed line) that corresponds to an efficiency of 100\% in converting baryons into stars. The median stellar mass fraction of our luminous quasars is $\sim 20\%$ of the baryonic limit (green dashed line), closely resembling high-redshift massive galaxies \citep{Xiao2024Natur}. \ledit{Notably, J0224$-$4711 from the ASPIRE survey, with $\log (M_\ast/M_\odot)=11.64\pm0.43$, lies $\sim 0.06$ dex above the theoretical 100\% baryon-conversion limit. Given the expected cosmic variance of $\sim 0.3$ dex for our relatively small survey area \citep[cf.][]{Xiao2024Natur}, together with the measurement uncertainty, this object remains consistent with the nominal limit at the $\sim 1\,\sigma$ level and is therefore plausible.} Furthermore, the detection of filamentary structures surrounding the ASPIRE quasars indicates that they inhabit some of the densest known regions at these epochs \citep{Wang2023ApJ}, and thus the actual halo masses of these luminous quasars might exceed our current estimates. Nevertheless, our results clearly indicate extremely high star formation efficiencies among luminous quasars, far in excess of the typical stellar-to-halo mass ratios of $\sim 2\%$ of local massive galaxies \citep{VanUitert2016MNRAS}.

Collectively, luminous quasars exhibit more highly concentrated morphologies ($n \approx 5$) and significantly enhanced star formation efficiencies compared to less luminous quasars and the general galaxy population. The observed compactness suggests rapid central stellar mass assembly, possibly via intense, gas-rich starburst episodes driven by strong central gas inflows concurrently fueling supermassive BH growth \citep{Dekel2009ApJ,Zolotov2015MNRAS}. Such a scenario is reminiscent of the ``blue nugget'' phase, wherein galaxies rapidly accumulate dense stellar cores through starburst events, achieving star formation efficiencies that can exceed $30\%-50\%$ of the available baryonic matter, analogous to efficiencies inferred in local starburst regions and sites of globular cluster formation \citep{Kruijssen2015MNRAS,Elmegreen2018ApJ}.

Given the overdense environments surrounding the luminous quasars \citep{Wang2023ApJ}, it is plausible that major mergers play a complementary role in triggering quasar activity, instead of acting as a competing scenario. The merger-driven scenario traditionally predicts heightened activity at intermediate redshifts ($z \approx 2-3$; \citealt{Hopkins2006ApJS}), but evidence for substantial merger activity at earlier epochs (e.g., $z \gtrsim 6$) increasingly has been reported \citep{Neeleman2019ApJ,Meyer2025ApJ}. Luminous quasars at $z \approx 6-7$ could experience enhanced gas inflows induced by both merger-driven gravitational torques and cosmological cold streams. Under this dual-trigger hypothesis, the intense SFRs and compactness observed in the luminous quasars would naturally arise from a combination of merger-driven dynamical disturbances and continuous inflow-driven gas accretion. Our results reinforce a picture in which both mergers and secular inflows synergistically facilitate early quasar growth and galaxy evolution.

\begin{figure}
\centering
\includegraphics[width=0.48\textwidth]{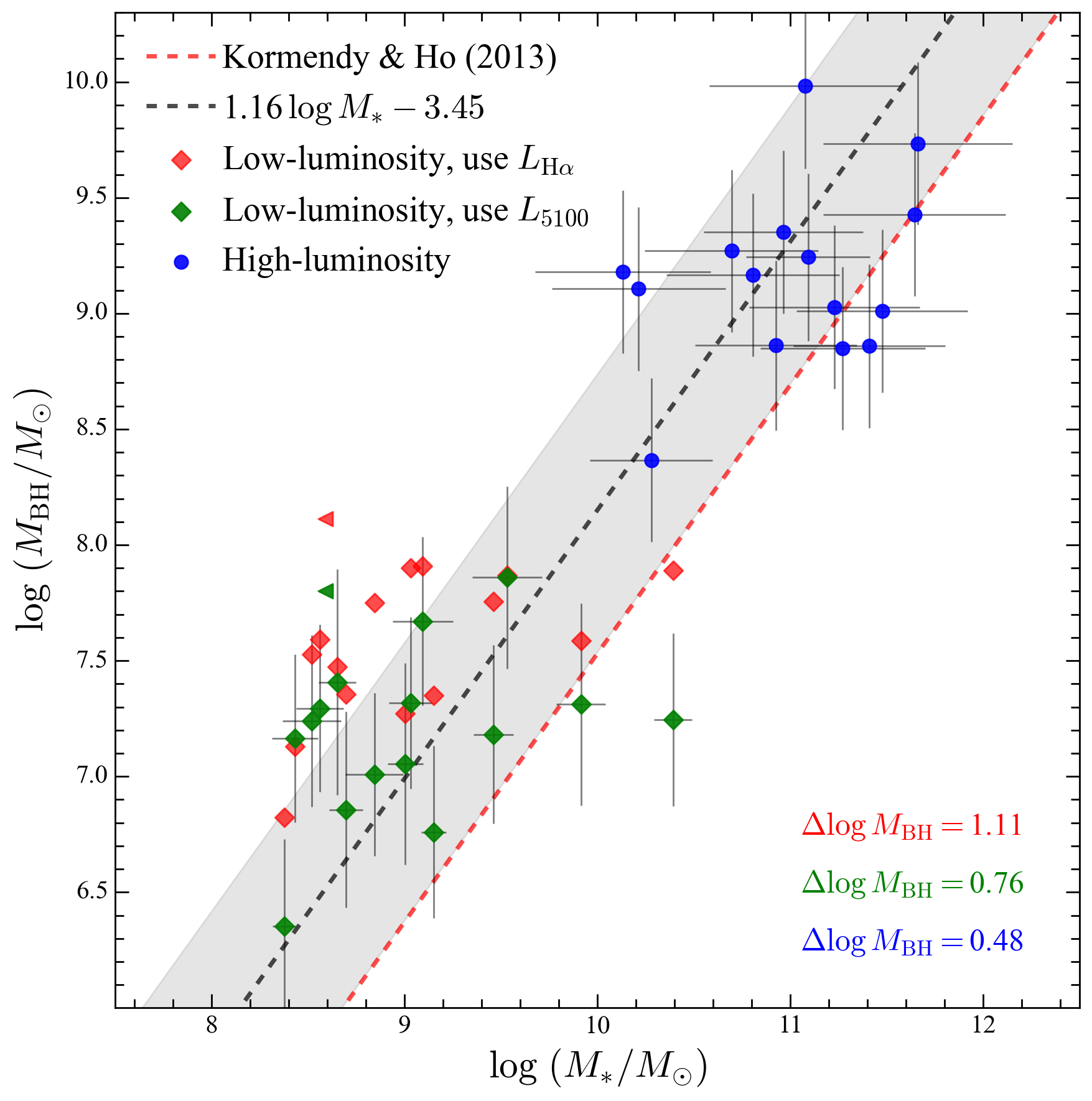}
\caption{The BH mass-stellar mass relation for our high‐redshift quasar sample, compared with the local scaling relation (red dashed line; \citealp{Kormendy2013ARAA}). Blue circles represent high-luminosity quasars, red diamonds low-luminosity quasars with $M_{\mathrm{BH}}$ inferred from $L_{\mathrm{H}\alpha}$, and green diamonds low-luminosity quasars with $M_{\mathrm{BH}}$ derived from $L_{5100}$. Error bars show $1\,\sigma$ uncertainties in both $M_{\mathrm{BH}}$ and $M_{\star}$. The black dashed line gives the best-fit linear regression $\log \, M_{\mathrm{BH}} = 1.16\,\log \, M_{\star} - 3.45$, with the shaded region indicating the intrinsic scatter of $0.6$~dex. The offsets of the subsamples with respect to the local relation are indicated in the lower-right corner. }
\label{fig:mmoffset}
\end{figure}

\subsection{$M_\mathrm{BH}-M_\ast$ Relation and Early AGN Feedback}
\label{sec:offset}

The $M_{\mathrm{BH}}-\sigma_\ast$ and $M_{\mathrm{BH}}-M_{\mathrm{bulge}}$ relations have often been interpreted as evidence for coevolution between supermassive BHs and their host galaxies \citep{Ferrarese2000ApJ, Gebhardt2000ApJ, Kormendy2013ARAA}. However, the physical processes driving this coevolution remain elusive. AGN feedback mechanisms, particularly radiation pressure-driven outflows, may expel gas, suppress star formation, and limit the growth of galaxy size \citep{King2015ARAA, Yuan2018ApJ}. Cosmological simulations (e.g., IllustrisTNG, \citealt{Weinberger2017MNRAS}; EAGLE, \citealt{Schaye2015MNRAS}; SIMBA, \citealt{Dave2019MNRAS}) adopt different implementations of AGN feedback to reproduce these scaling relations, attributing them to the interactions between supermassive BHs and their host galaxies. The deviation of high-redshift quasars from the local $M_\mathrm{BH}-M_\ast$ relation may reflect indirectly the strength of AGN feedback.

High-redshift low-luminosity quasars have systematically higher equivalent widths for broad H$\alpha$ relative to their more luminous counterparts, or compared to nearby AGNs \citep{Maiolino2024AA}. An immediate, practical ramification is that H$\alpha$-based BH masses \citep{Greene2005ApJ} might be overestimated if the BLR radius is inferred from broad H$\alpha$ luminosity. Using BLR radii estimated from $L_{5100}$ instead of $L_{\mathrm{H\alpha}}$ systematically lowers $M_\mathrm{BH}$ and reduces the offset relative to the local $M_{\mathrm{BH}}-M_{\ast,\mathrm{bulge}}$ relation \citep{Kormendy2013ARAA} from 1.11~dex to 0.76~dex (Figure~\ref{fig:mmoffset}). This offset remains larger than the 0.48~dex deviation of the luminous quasar subsample. Fitting a linear relation to the 30 quasars in our sample with detectable host galaxies, with the slope fixed to 1.16 \citep{Kormendy2013ARAA}, yields $\log \, M_{\mathrm{BH}} = 1.16\,\log \, M_\star - 3.45$. The zero point is $\sim 0.6$~dex higher relative to the zero point of the \citet{Kormendy2013ARAA} relation.

Ruling out the possibility that the observed offset reflects selection bias is beyond the scope of this work. Instead, we suggest that AGN feedback in the early Universe may contribute to the observed offset. According to \citet[][see review in \citealt{King2015ARAA}]{King2005ApJ}, the gas that fuels BH growth can be impacted by momentum-driven or energy–driven outflows. The overall gravitational force experienced within the central bulge region can be approximated as

\begin{equation}
F_{\rm grav} \sim \frac{M_\mathrm{dyn} M_{g}}{R^2} = \frac{f_{g} M_\mathrm{dyn}^2}{R^2},
\label{equ:FG1}
\end{equation}

\noindent
where $M_{g}$ is the gas mass and $f_{g} \equiv M_{g}/M_\mathrm{dyn}$ is the gas-to-dynamical mass fraction. For an isothermal galaxy profile with velocity dispersion $\sigma$, $M_\mathrm{dyn} = \frac{2\,\sigma^2 R}{G}$ \citep{Binney1987gady}, with $G$ the gravitational constant. Substituting this expression into Equation~\ref{equ:FG1} yields

\begin{equation}
F_{\rm grav} \sim f_{g} \sigma^4.
\label{equ:FG2}
\end{equation}

\noindent
In an energy-driven outflow, the overall feedback force can be approximated by

\begin{equation}
F_{\rm feed} \sim \frac{L_\mathrm{Edd}}{c} \propto M_\mathrm{BH}.
\end{equation}

\noindent
Balancing $F_{\rm grav}$ with $F_{\rm feed}$ gives $M_\mathrm{BH} \sim f_g \sigma^4$, which, in combination with the \citet{Faber1976} relation $M_{\ast,\mathrm{bulge}} \sim \sigma^4$ then implies a critical mass

\begin{equation}
M_c \propto f_g\, M_{\ast,\mathrm{bulge}}.
\label{equ:feedback}
\end{equation}

\noindent
Strong feedback halts BH growth when $M_\mathrm{BH}$ reaches the critical value $M_c$ \citep{Ding2022ApJ}, and the final BH mass follows a relation roughly consistent with that of \citet{Kormendy2013ARAA}.

As elaborated in Section~\ref{sec:trigger}, quasar activity is likely triggered during a starburst phase, during which a compact ($n \gtrsim 4$), massive host galaxy is assembled. Consequently, the bulge or spheroidal component dominates the host galaxy, such that $M_{\ast,\mathrm{bulge}} \approx M_\ast$. This may explain why our objects align more closely with the $M_{\rm BH} - M_{\ast,\mathrm{bulge}}$ relation of \citet{Kormendy2013ARAA} than with the $M_\mathrm{BH}-M_\ast$ relation of \cite{Zhuang2023NatAs} derived for a broader AGN population (Figure~\ref{fig:mmrelation}).

It is noteworthy from Equation~\ref{equ:feedback} that the zero point of the correlation is related to the gas fraction. At $z\approx 0$, the gas fraction is approximately equal to the cosmic baryon fraction on large scales, $f_{b} = 15.8\%$. Within this context, the 0.6~dex offset observed for high–redshift AGNs (Figure~\ref{fig:mmoffset}) suggests that in these systems the gas fraction can be as high as $f_g \approx 60\%$. At early epochs, it is easy to imagine that the small–scale bulge region can attain such high gas fractions because galaxies are inherently gas–rich, as observed (e.g., \citealp{Decarli2018ApJ, Herrera-Camus2025}), and dynamical processes efficiently drive gas inward \citep[e.g.,][]{Hopkins2008ApJS}. This implies that although the BH in high–redshift AGNs may be overly massive on large scales---thereby effectively influencing the surrounding gas---the gas density in the immediate vicinity of the BH remains high enough so that local feedback cannot completely quench accretion, prolonging the quasar phase until the small–scale gas reservoir is consumed.

\section{Summary}
\label{sec:sec6}

We assemble a representative sample of 31 high‐redshift ($z \gtrsim 4$) AGNs comprising JWST NIRCam observations from multiple surveys, including ASPIRE, CEERS, EIGER, and SHELLQs, to probe the early growth of supermassive BHs and their host galaxies. The sample consist of 15 high-luminosity quasars ($L_\mathrm{5100} \gtrsim 10^{45} \rm \, erg\, s^{-1}$) and 16 less luminous counterparts recently identified by JWST. We uniformly reduce and process the images using state-of-the-art corrections, background subtraction, and astrometric calibration to ensure optimal spatial resolution across the different filters. We perform multi‐band AGN-host decomposition using a newly developed forward-modeling code (\textsc{GalfitS}) to disentangle the nuclear AGN emission from the underlying host light. Simultaneously combining imaging and spectral fitting, this approach incorporates physically motivated priors to facilitate extraction of accurate host galaxy and AGN properties. Extensive input-output tests with mock images validate our methodology, quantify uncertainties related to PSF variations and parameter degeneracies, and place rigorous upper limits on non-detections.

Our key findings are summarized as follows:

\begin{enumerate}

\item We successfully detect the host galaxies of 30 out of the 31 quasars, among them 20 new measurements and five that previously only had published upper limits on stellar mass.

\item Quasar host galaxies are intrinsically more compact than star-forming galaxies of comparable stellar mass, lying $\sim 0.3$~dex below the fiducial stellar mass–size relation at $z = 6.5$.

\item The internal structure of the host galaxies of high-redshift quasars comes in two flavors. Whereas low-luminosity quasars structurally resemble matched star-forming galaxies of similar stellar mass and redshift, both characterized by disk-like profiles with \sersic\ indices $n \approx 1$, the more power quasars predominantly reside in highly concentrated ($n \approx 5$), presumably bulge-dominated, early-type host galaxies.

\item The spectral slopes of the isolated active nuclei exhibit a remarkable dichotomy. Parameterizing the nuclear SEDs with a simple power law reveals that while luminous quasars have a relatively narrow range of blue UV spectral slopes ($\beta_{\mathrm{UV}} = -1.39 \pm 0.36$) similar to that of typical lower redshift quasars, the less luminous quasars span a broad range of UV spectral slopes, from $\beta_{\mathrm{UV}} \approx -2$ to extremely red values of $\beta_{\mathrm{UV}} \approx 4$.

\item High-redshift quasars deviate systematically above the local $M_{\mathrm{BH}} - M_\ast$ relation, by different relative amounts depending on luminosity (or BH mass): high-luminosity sources ($L_\mathrm{5100} \gtrsim 10^{45} \rm \, erg\, s^{-1}$; $M_{\rm BH} \gtrsim 10^9\,M_\odot$) exhibit a median $M_{\mathrm{BH}}/M_\ast = 2.3\%$, to be compared with $M_{\mathrm{BH}}/M_\ast = 4.7\%$ for lower luminosity sources ($L_\mathrm{5100} \lesssim 10^{45} \rm \, erg\, s^{-1}$; $M_{\rm BH} \approx 10^7-10^8\,M_\odot$). After accounting for possible biases in the BH mass estimates of the low-luminosity quasars arising from their higher H$\alpha$ equivalent widths, we estimate an ensemble offset of 0.6~dex in $M_{\mathrm{BH}}/M_\ast$ for our combined sample of high-redshift quasars.

\end{enumerate}

The physical implications of the above empirical findings depend critically on the interpretation of the observed bimodality in nuclear spectral SEDs. While we cannot rule out the possibility that the red spectral slopes arise from dust extinction, we suggest that the red continuum is an intrinsic property of the AGN connected with physical conditions in the BLR or accretion disk. We argue that the offset of high-redshift quasars above the local $M_{\mathrm{BH}}-M_\ast$ relation reflects the influence of AGN feedback in gas-rich environments. With gas fractions as high as $\sim 60\%$, inferred from the offset and theoretical considerations, BHs can grow rapidly in the early Universe. Feedback from energy-driven outflows may be insufficient to quench accretion until the central gas reservoir is depleted. This delayed regulation can explain the overmassive BHs relative to their hosts, suggesting that feedback becomes more effective at later stages as the gas supply diminishes, guiding these systems toward the local scaling relation.

\begin{acknowledgements}
\ledit{We are grateful to referee for constructuve comments. We thank George Rieke and Madeline A. Marshall for insightful discussions that greatly aided in the presentation of our results.} This work was supported by the National Science Foundation of China (12233001) and the China Manned Space Program (CMS-CSST-2025-A09).

\end{acknowledgements}


\software{\textsc{SExtractor} \citep{Bertin1996AAS},\textsc{photutils} \citep{photutils}, \textsc{GalfitS}\citep{GalfitS}}

\appendix

\section{Other PSF Construction Methods}
\label{app:psfs}

\begin{figure*}
\centering
\figurenum{A1}
\includegraphics[width=\textwidth]{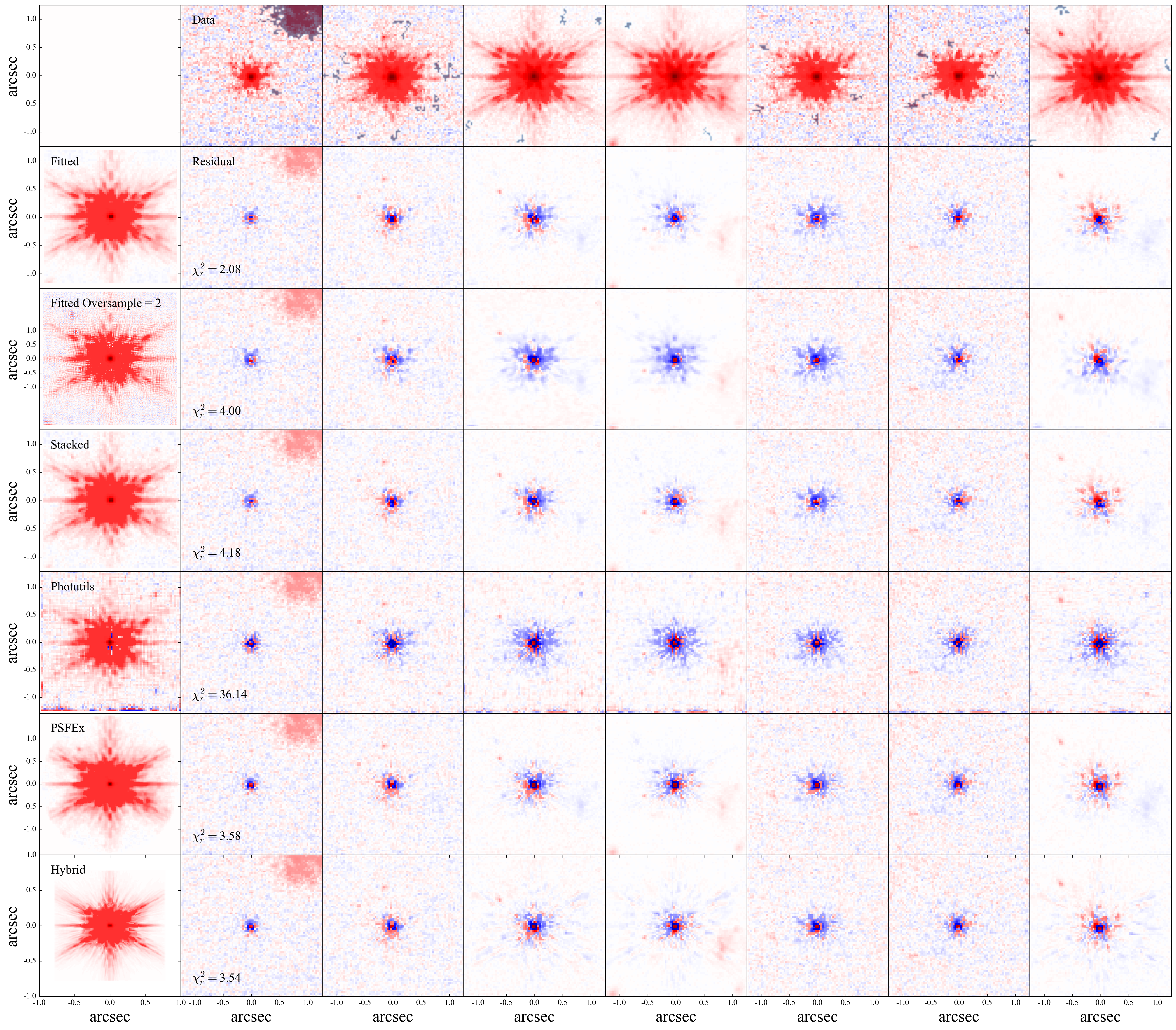}
\caption{Comparison of different PSF models fitted to individual stars in the CEERS pointing 1 field, observed in the F115W band. The first row shows the observed data for several stars. The first column presents the PSF models, and subsequent columns show the residuals from the fitted star images. The methods compared include: the direct-fitting method, the direct-fitting method with an oversampling factor of 2, the stacking method, the \texttt{Photutils} library, the \texttt{PSFEx} tool, and the hybrid PSF \citep{Chen2025}. The reduced chi-squared value ($\chi^2_r$) shown in each row represents the median of all fits using the same model, with lower values indicating better fits..}
\label{fig:psfcompare}
\end{figure*}

\ledit{Figure~\ref{fig:psfcompare} compares the performance of several PSF models in fitting individual stars observed in the F115W band in the CEERS pointing 1 field. The tested models include (1) a direct-fitting empirical PSF, in which we simultaneously fit a set of isolated, unsaturated stars with a common PSF template whose pixel values are free parameters (each star differs only by a shift and a flux normalization); to prevent noise-driven structure, we impose that the recovered PSF image is smooth, an assumption that is particularly important when oversampling is used; (2) a direct-fitting model with an oversampling factor of 2, meaning the PSF is reconstructed on a grid with twice the native sampling in each dimension (i.e., half-pixel spacing), which in principle captures sub-pixel structure but increases the number of free parameters and therefore requires stronger smoothness regularization; (3) a simple stacking PSF obtained by centroiding, normalizing, and combining the stellar cutouts; (4) an empirical ePSF constructed with the \texttt{photutils} \texttt{EPSFBuilder} (using its default sampling; no additional oversampling is applied); (5) an empirical PSF from \texttt{PSFEx} (also using its native sampling); and (6) a hybrid PSF that adopts the high-S/N core from \texttt{PSFEx} and replaces the low-S/N outer wings with a \texttt{WebbPSF} model \citep{Chen2025}. The residuals reveal systematic differences among the methods; while the hybrid PSF performs well at larger radii, the direct-fitting PSF without oversampling consistently yields the lowest reduced chi-squared ($\chi^2_r$) values across the field, and we therefore adopt it for our analysis.}

\section{Notes on Individual Objects}
\label{app:notes}

\figsetstart
\figsetnum{1}
\figsettitle{SED fits and imaging decompositions for the full sample}
\figsetgrpstart

\figsetgrpnum{1.1}
\figsetgrptitle{J2236+0032}
\figsetplot{\detokenize{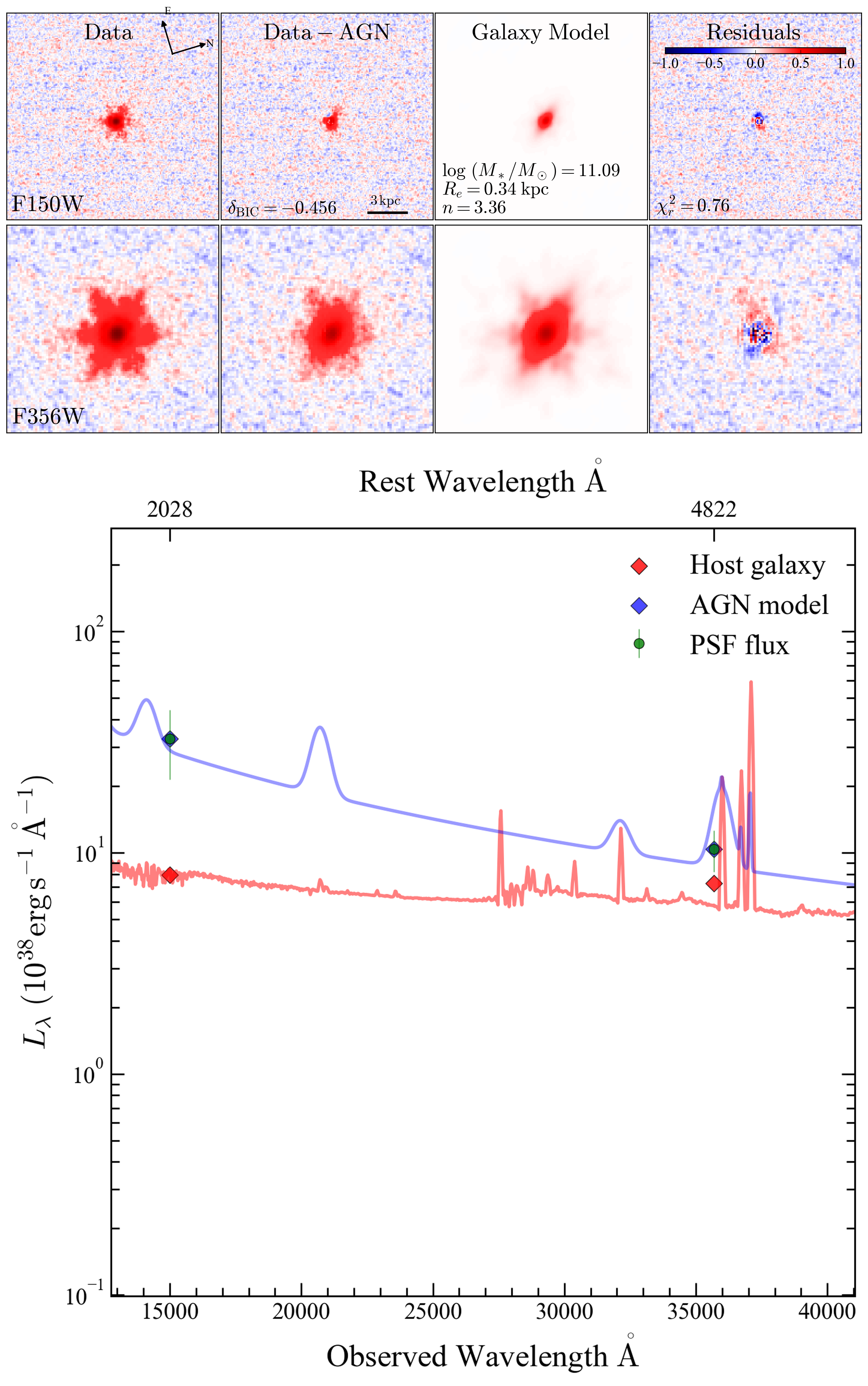}}
\figsetgrpnote{Imaging decomposition (top) and SED/non-parametric flux summary (bottom).}
\figsetgrpend

\figsetgrpnum{1.2}
\figsetgrptitle{J2255+0251}
\figsetplot{\detokenize{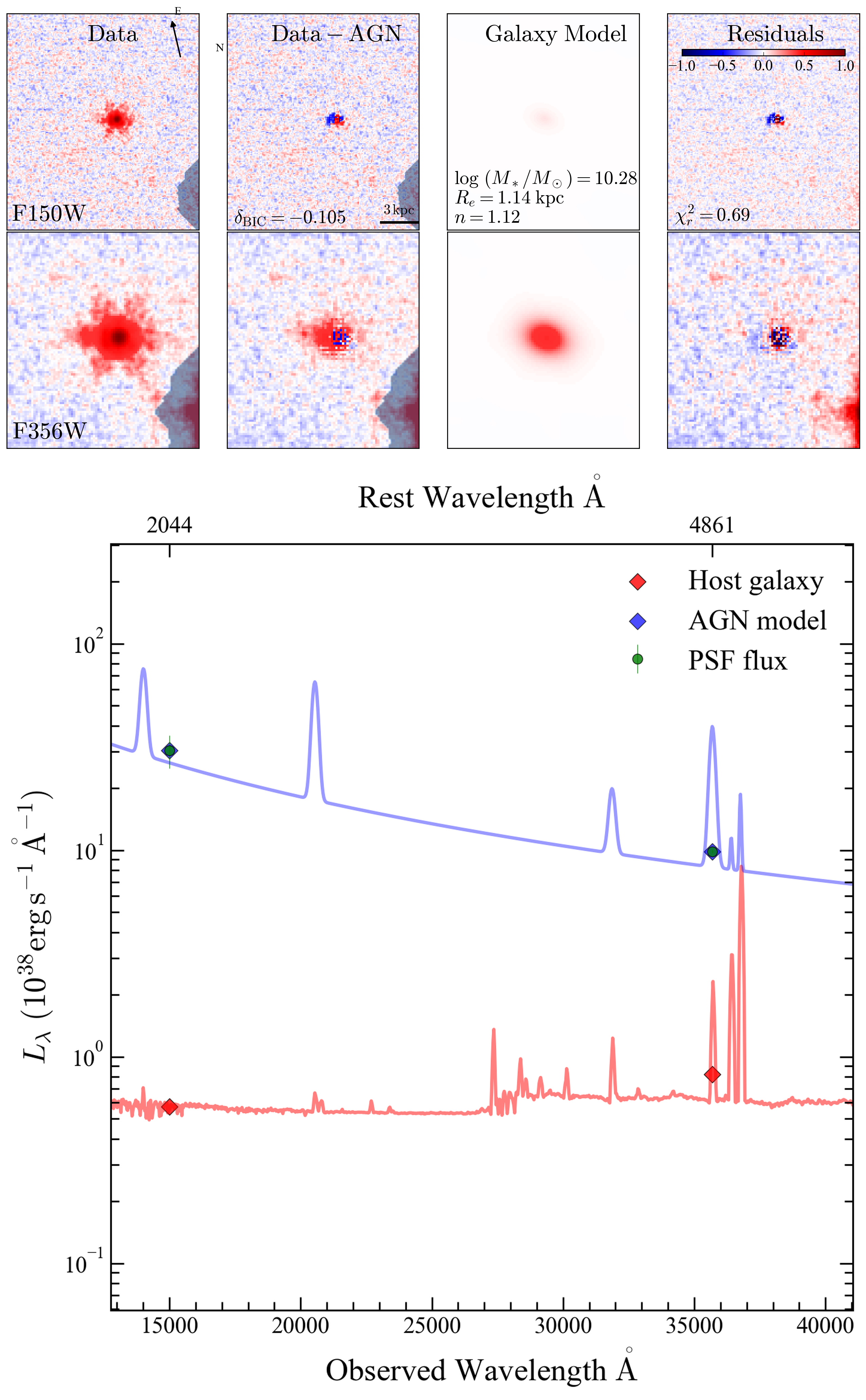}}
\figsetgrpnote{Imaging decomposition (top) and SED/non-parametric flux summary (bottom).}
\figsetgrpend

\figsetgrpnum{1.3}
\figsetgrptitle{J0148+0600}
\figsetplot{\detokenize{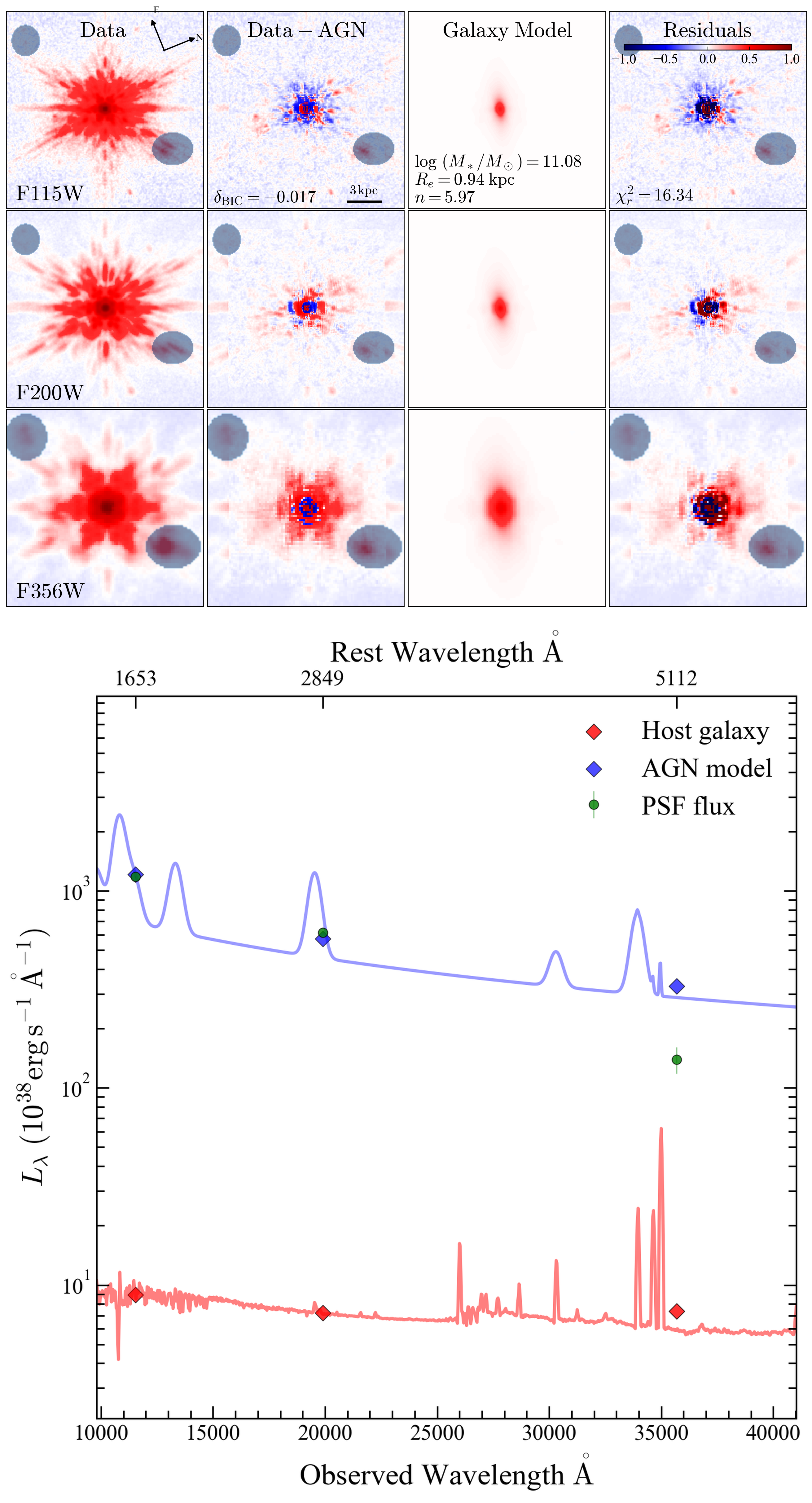}}
\figsetgrpnote{Imaging decomposition (top) and SED/non-parametric flux summary (bottom).}
\figsetgrpend

\figsetgrpnum{1.4}
\figsetgrptitle{J1030+0524}
\figsetplot{\detokenize{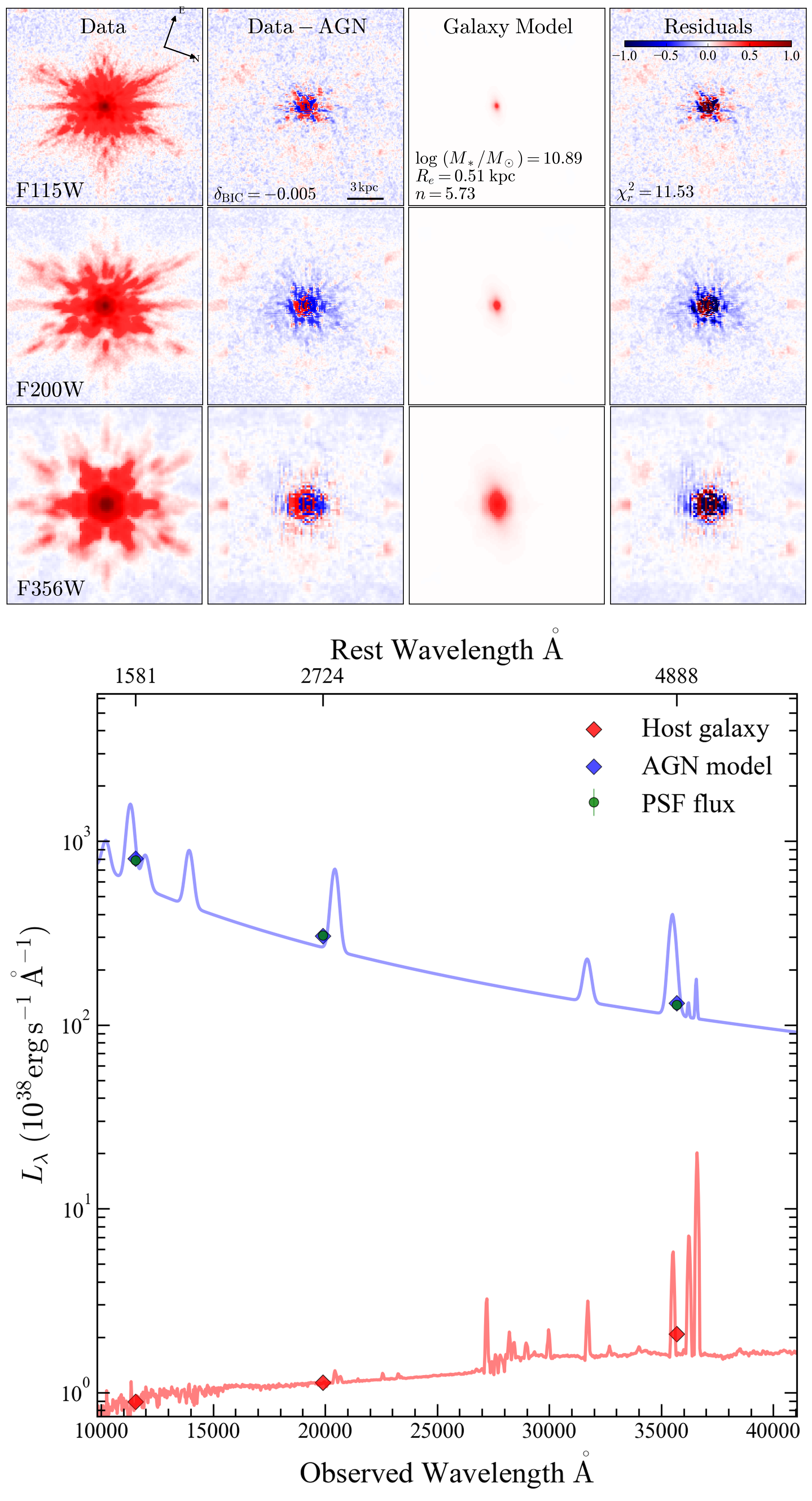}}
\figsetgrpnote{Imaging decomposition (top) and SED/non-parametric flux summary (bottom).}
\figsetgrpend

\figsetgrpnum{1.5}
\figsetgrptitle{J159+02}
\figsetplot{\detokenize{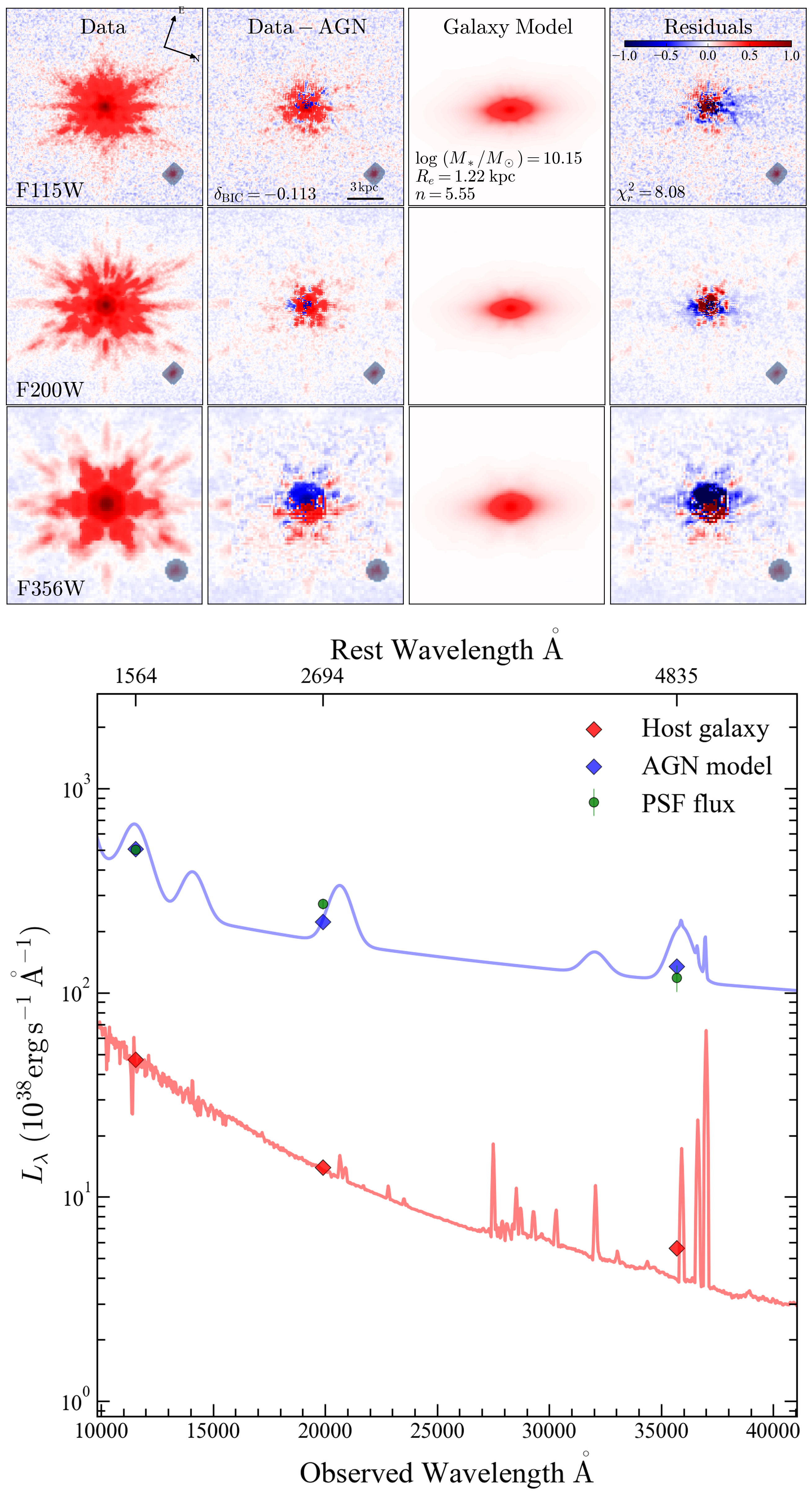}}
\figsetgrpnote{Imaging decomposition (top) and SED/non-parametric flux summary (bottom).}
\figsetgrpend

\figsetgrpnum{1.6}
\figsetgrptitle{J1120+0641}
\figsetplot{\detokenize{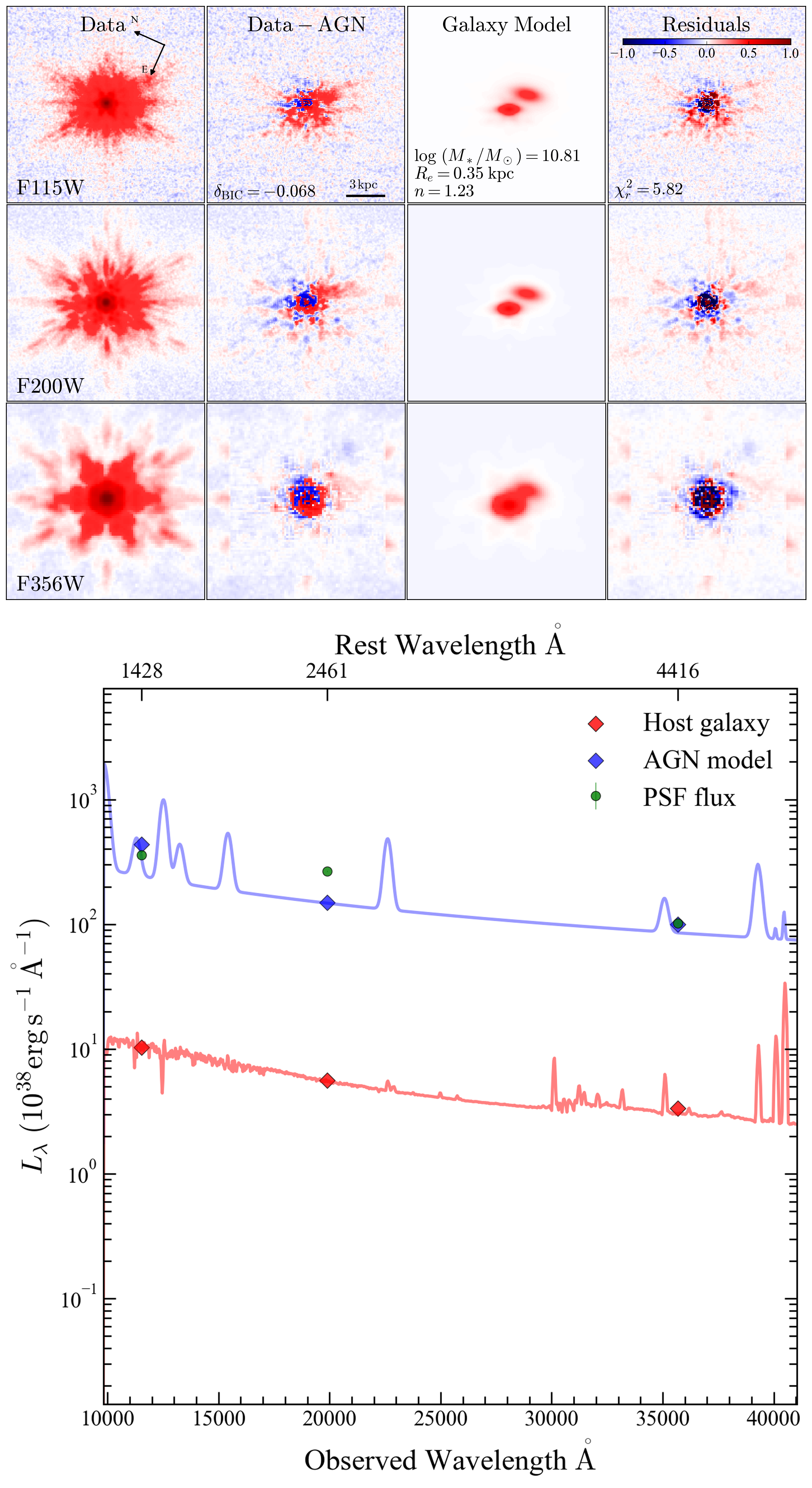}}
\figsetgrpnote{Imaging decomposition (top) and SED/non-parametric flux summary (bottom).}
\figsetgrpend

\figsetgrpnum{1.7}
\figsetgrptitle{J1148+5251}
\figsetplot{\detokenize{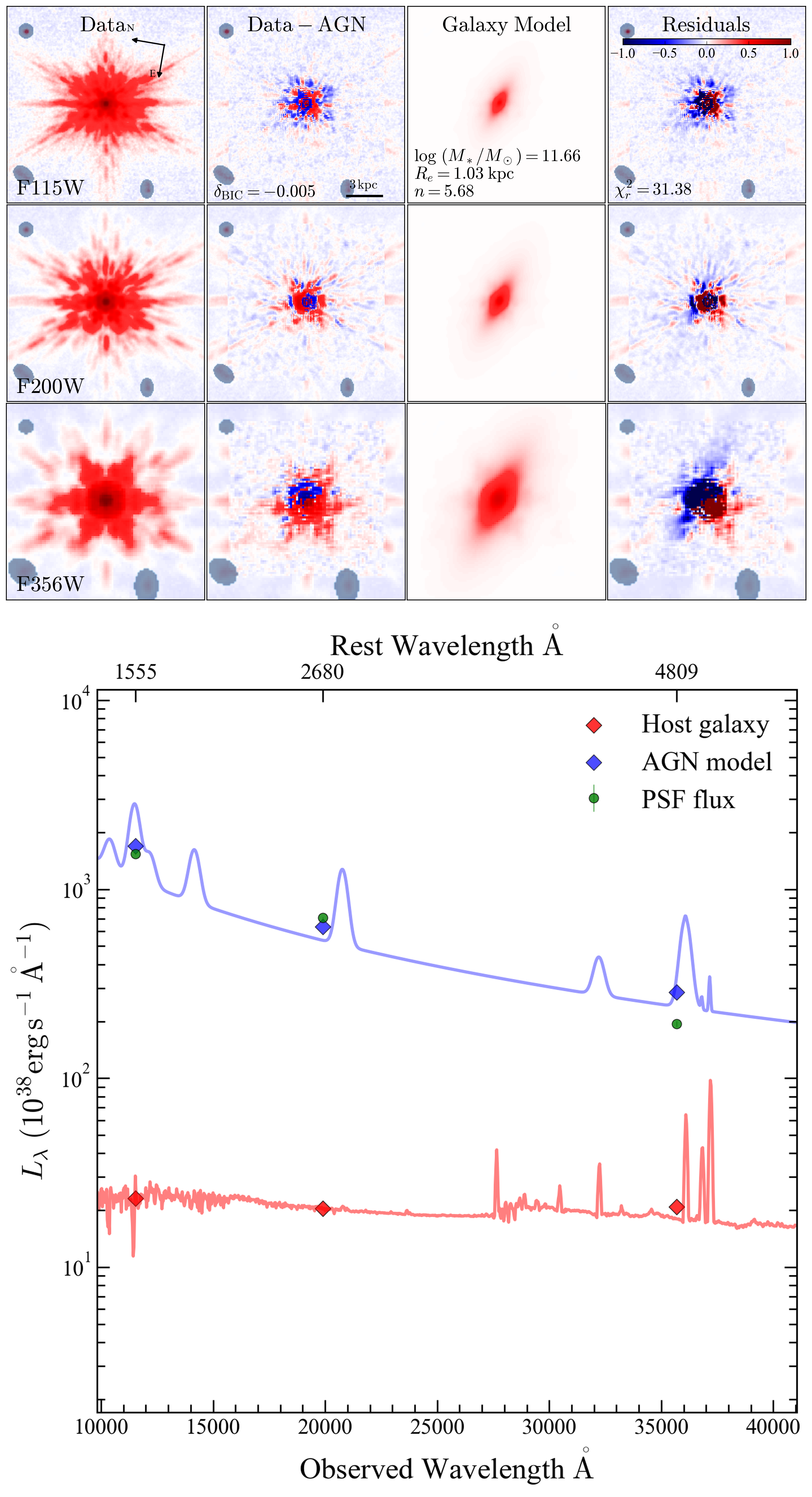}}
\figsetgrpnote{Imaging decomposition (top) and SED/non-parametric flux summary (bottom).}
\figsetgrpend

\figsetgrpnum{1.8}
\figsetgrptitle{J0109-3047}
\figsetplot{\detokenize{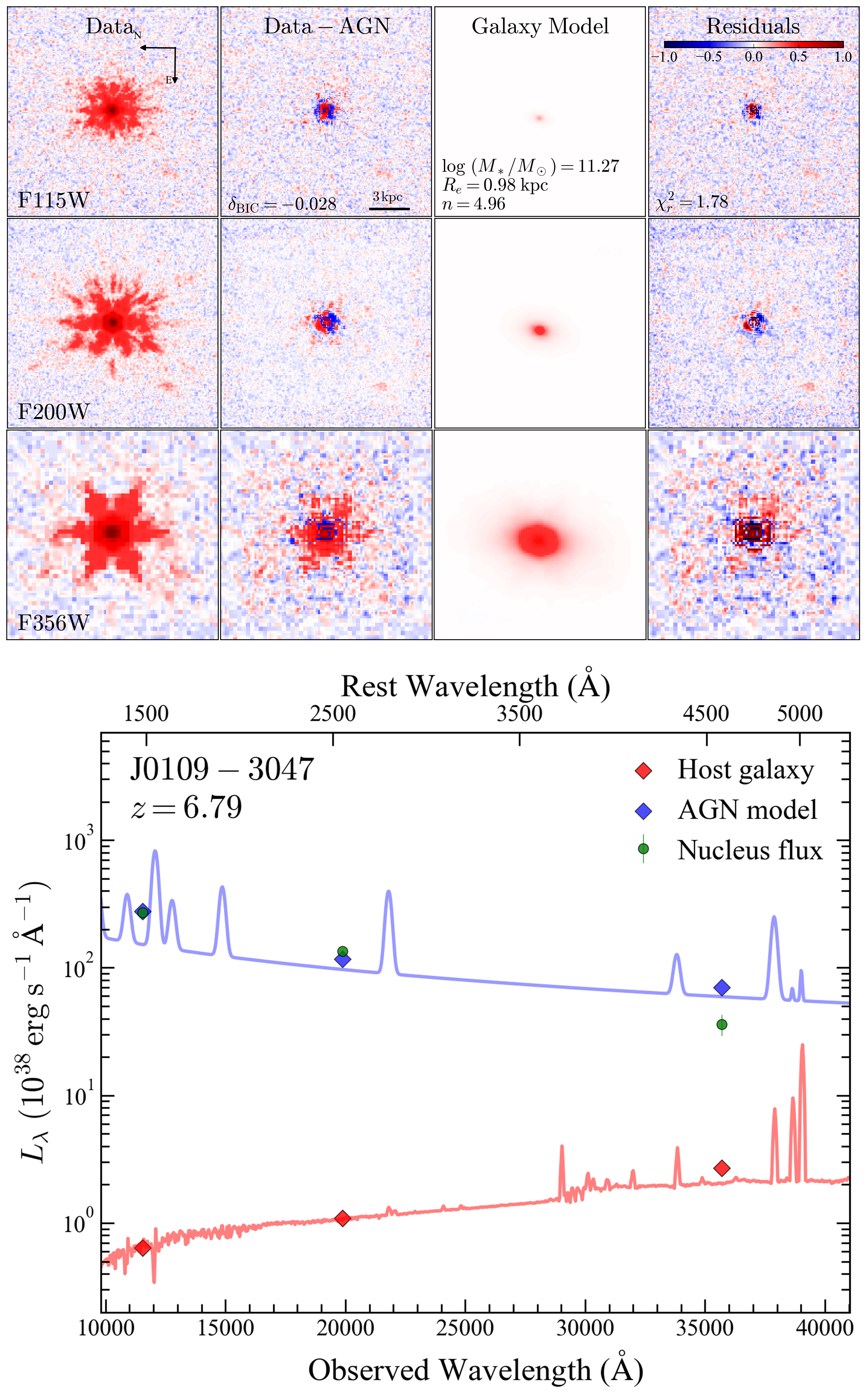}}
\figsetgrpnote{Imaging decomposition (top) and SED/non-parametric flux summary (bottom).}
\figsetgrpend

\figsetgrpnum{1.9}
\figsetgrptitle{J0218+0007}
\figsetplot{\detokenize{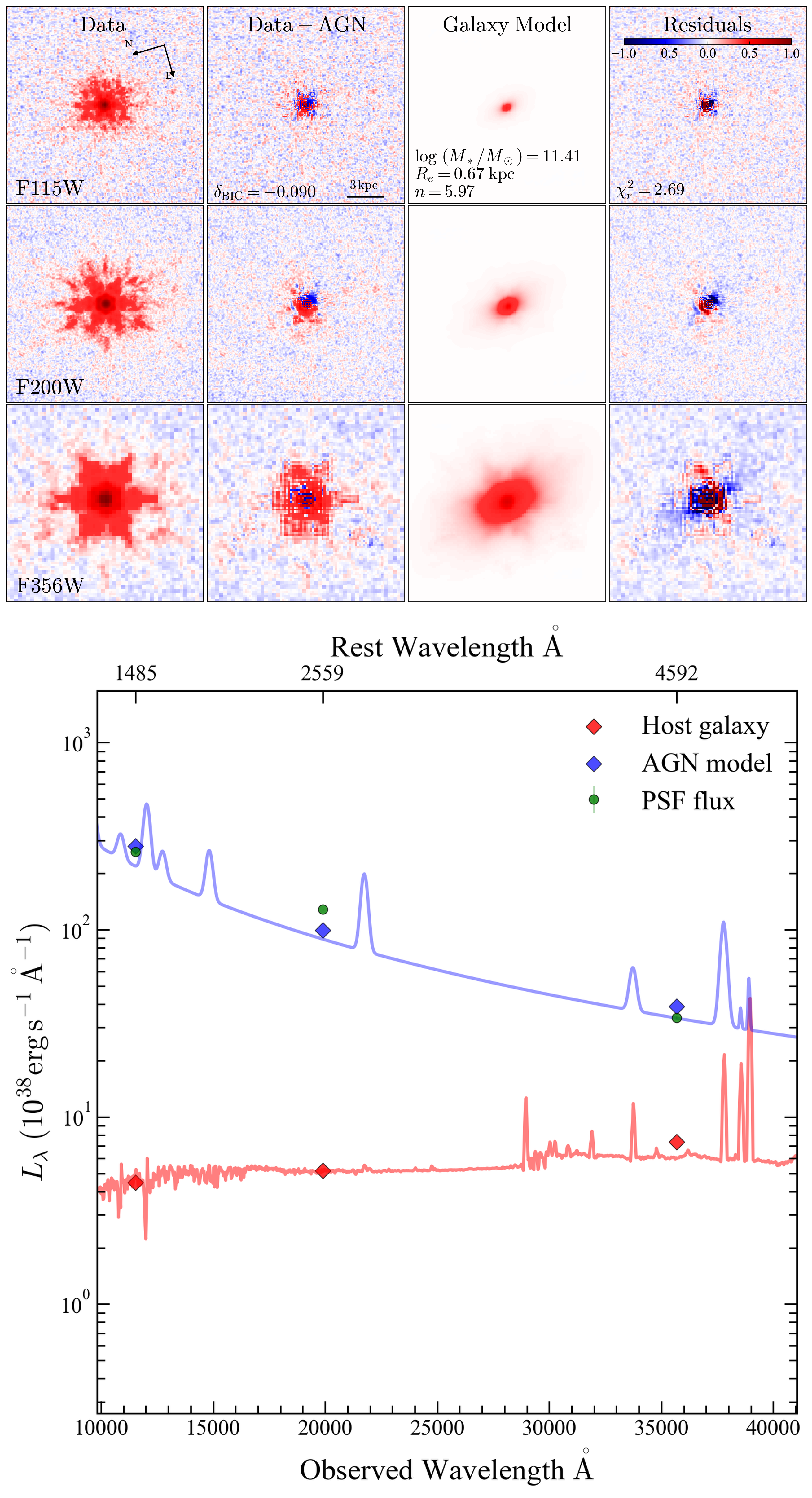}}
\figsetgrpnote{Imaging decomposition (top) and SED/non-parametric flux summary (bottom).}
\figsetgrpend

\figsetgrpnum{1.10}
\figsetgrptitle{J0224-4711}
\figsetplot{\detokenize{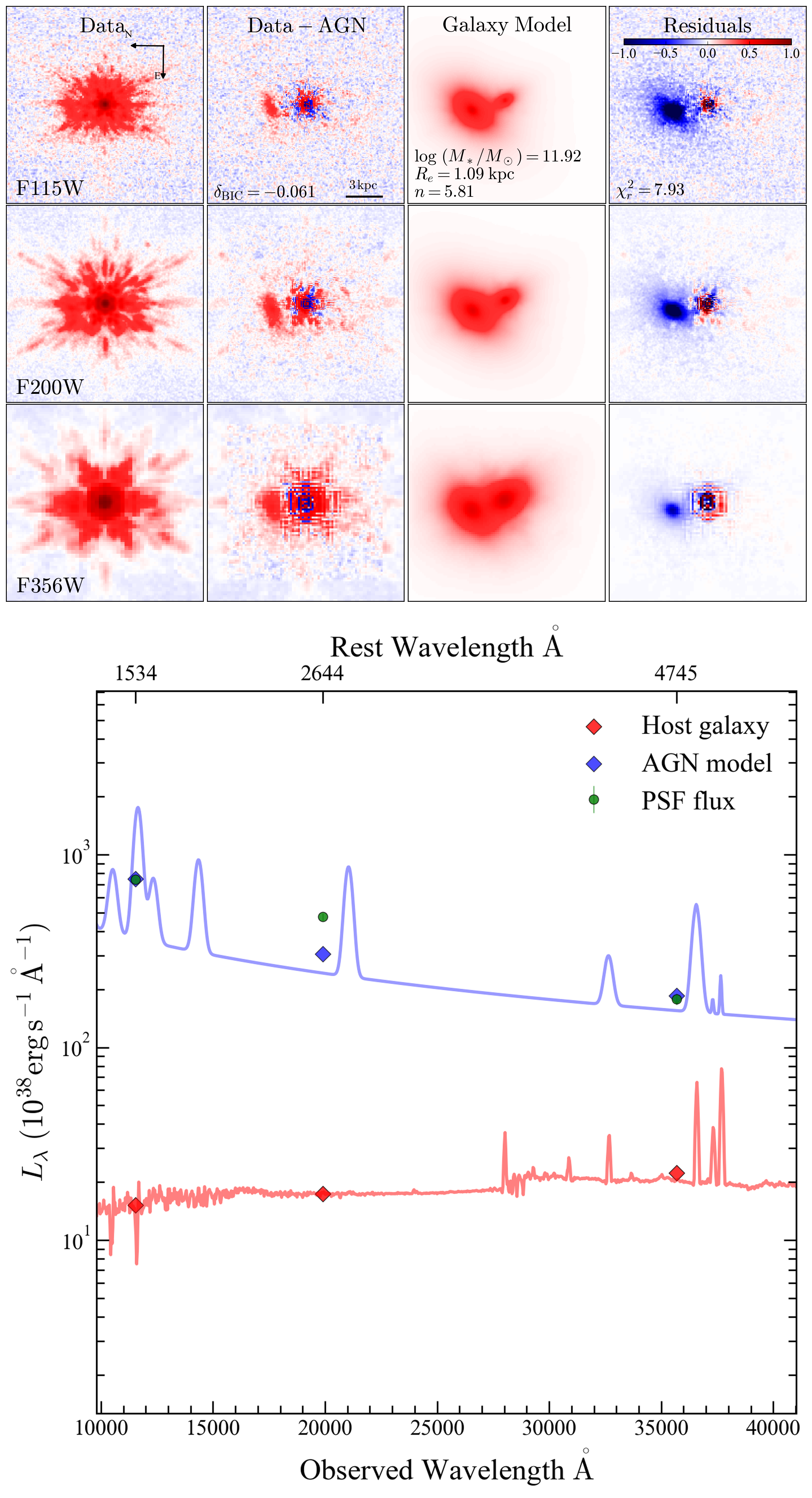}}
\figsetgrpnote{Imaging decomposition (top) and SED/non-parametric flux summary (bottom).}
\figsetgrpend

\figsetgrpnum{1.11}
\figsetgrptitle{J0226+0302}
\figsetplot{\detokenize{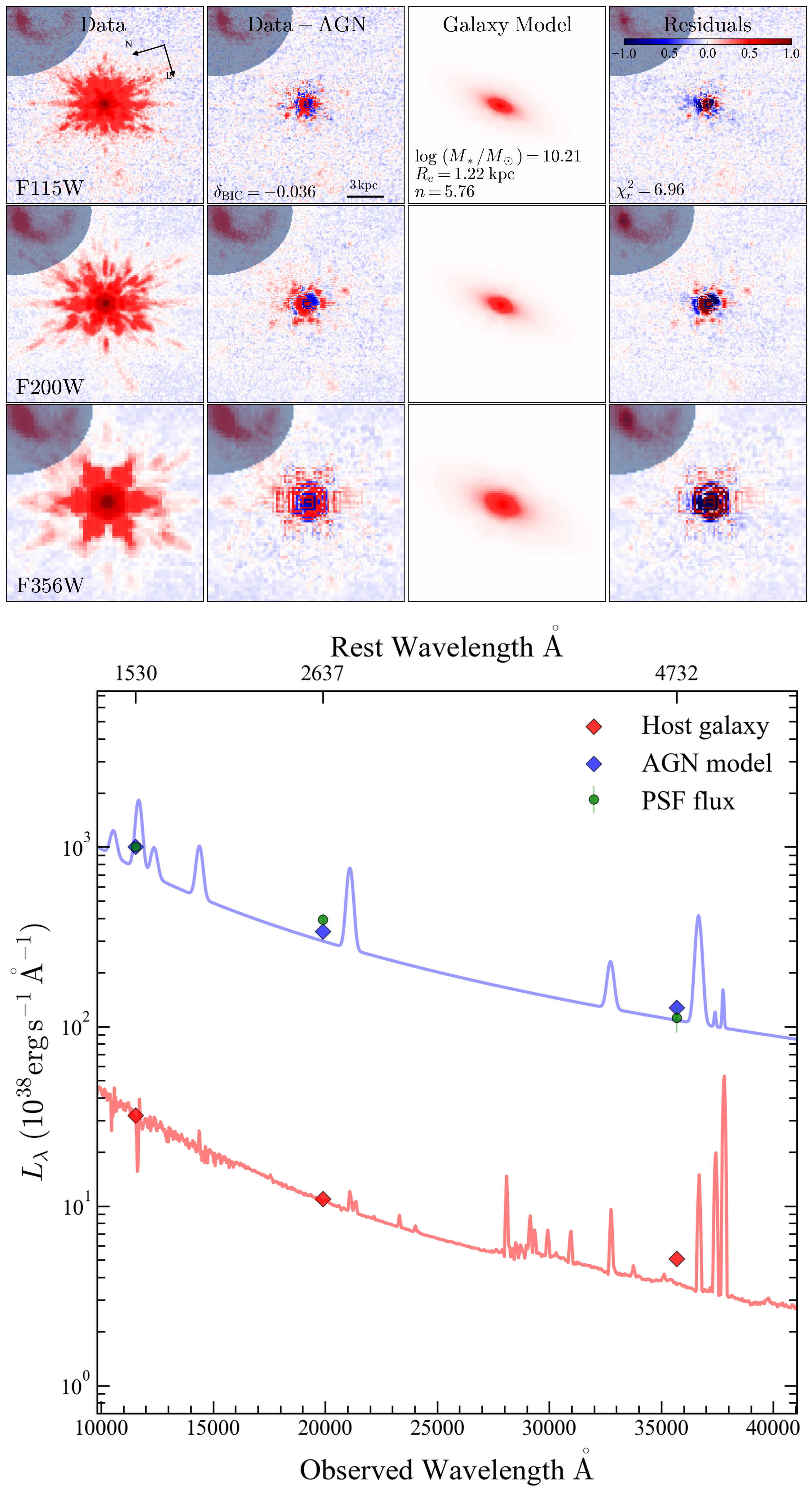}}
\figsetgrpnote{Imaging decomposition (top) and SED/non-parametric flux summary (bottom).}
\figsetgrpend

\figsetgrpnum{1.12}
\figsetgrptitle{J0244-5008}
\figsetplot{\detokenize{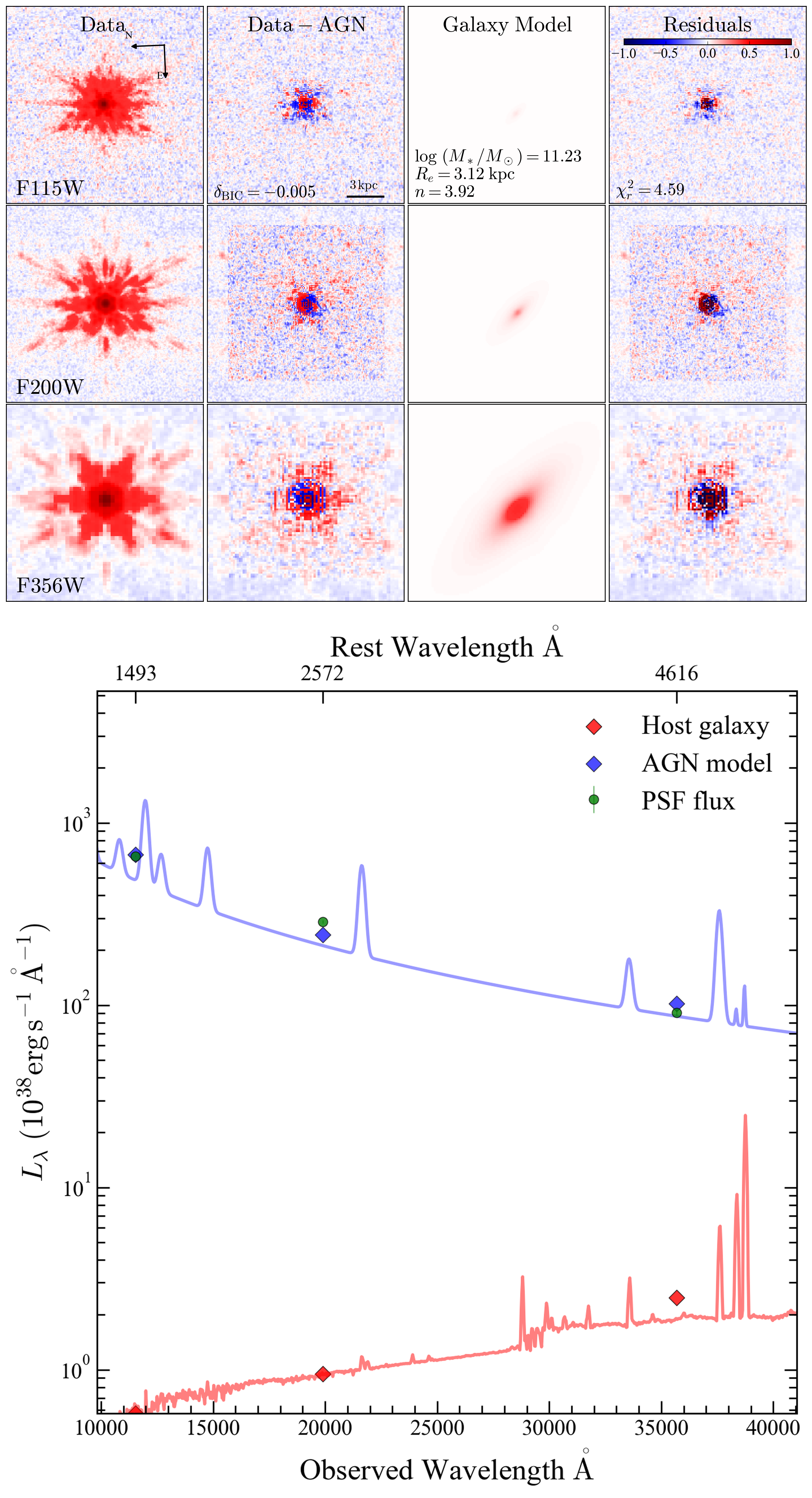}}
\figsetgrpnote{Imaging decomposition (top) and SED/non-parametric flux summary (bottom).}
\figsetgrpend

\figsetgrpnum{1.13}
\figsetgrptitle{J0305-3150}
\figsetplot{\detokenize{figset_J0305-3150.png}}
\figsetgrpnote{Imaging decomposition (top) and SED/non-parametric flux summary (bottom).}
\figsetgrpend

\figsetgrpnum{1.14}
\figsetgrptitle{J2002-3013}
\figsetplot{\detokenize{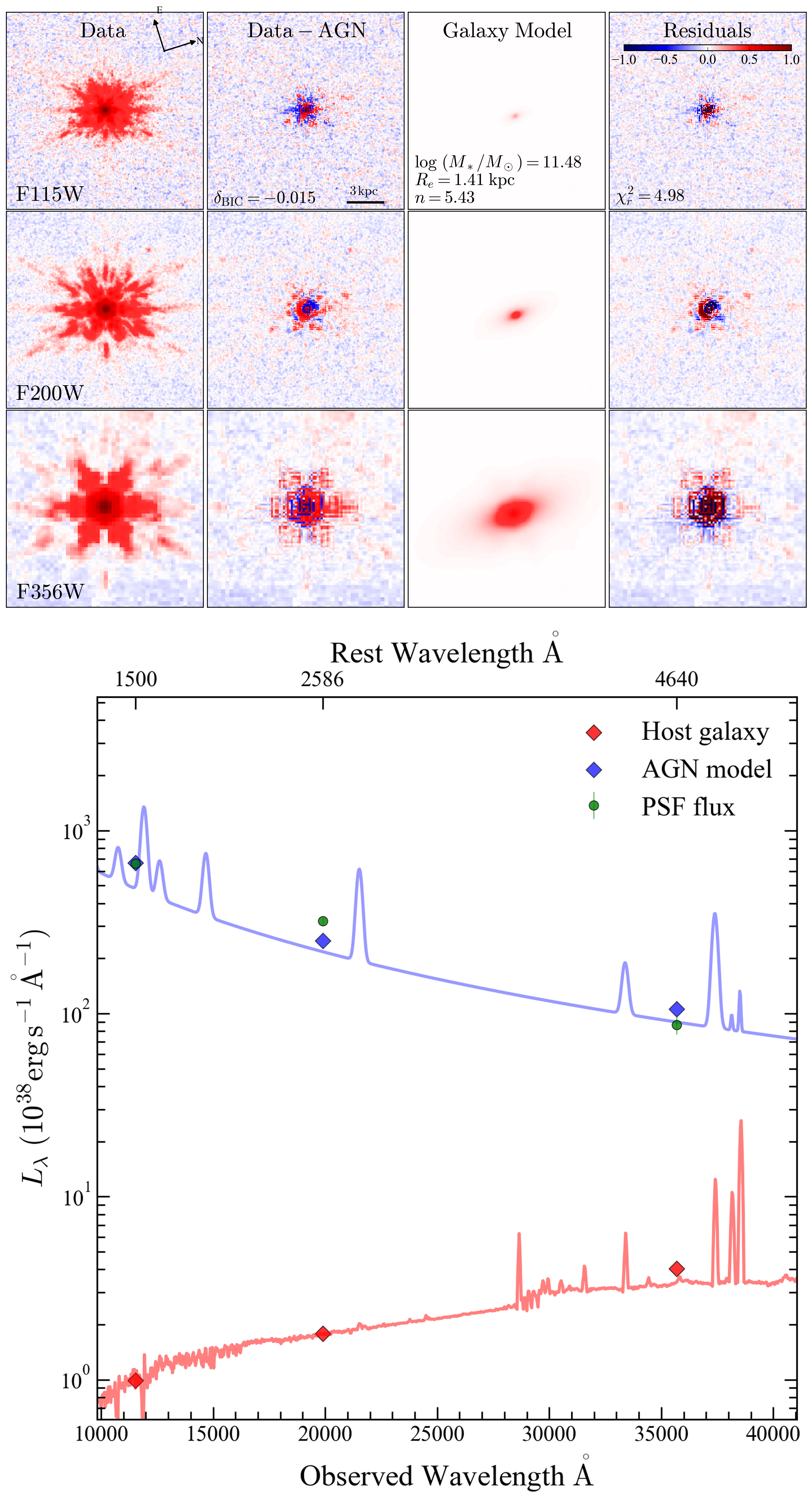}}
\figsetgrpnote{Imaging decomposition (top) and SED/non-parametric flux summary (bottom).}
\figsetgrpend

\figsetgrpnum{1.15}
\figsetgrptitle{J2232+2930}
\figsetplot{\detokenize{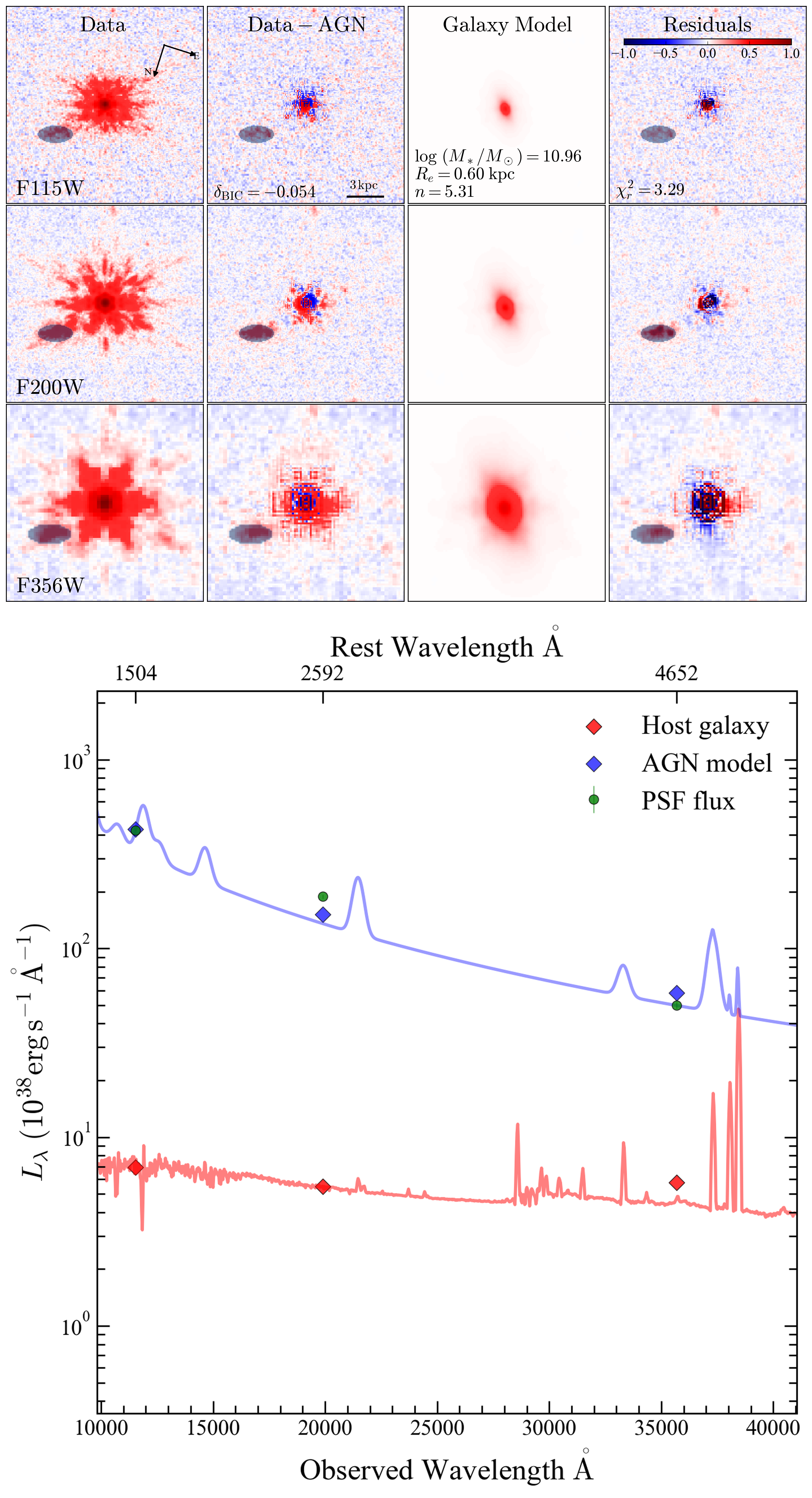}}
\figsetgrpnote{Imaging decomposition (top) and SED/non-parametric flux summary (bottom).}
\figsetgrpend

\figsetgrpnum{1.16}
\figsetgrptitle{CEERS\_007465}
\figsetplot{\detokenize{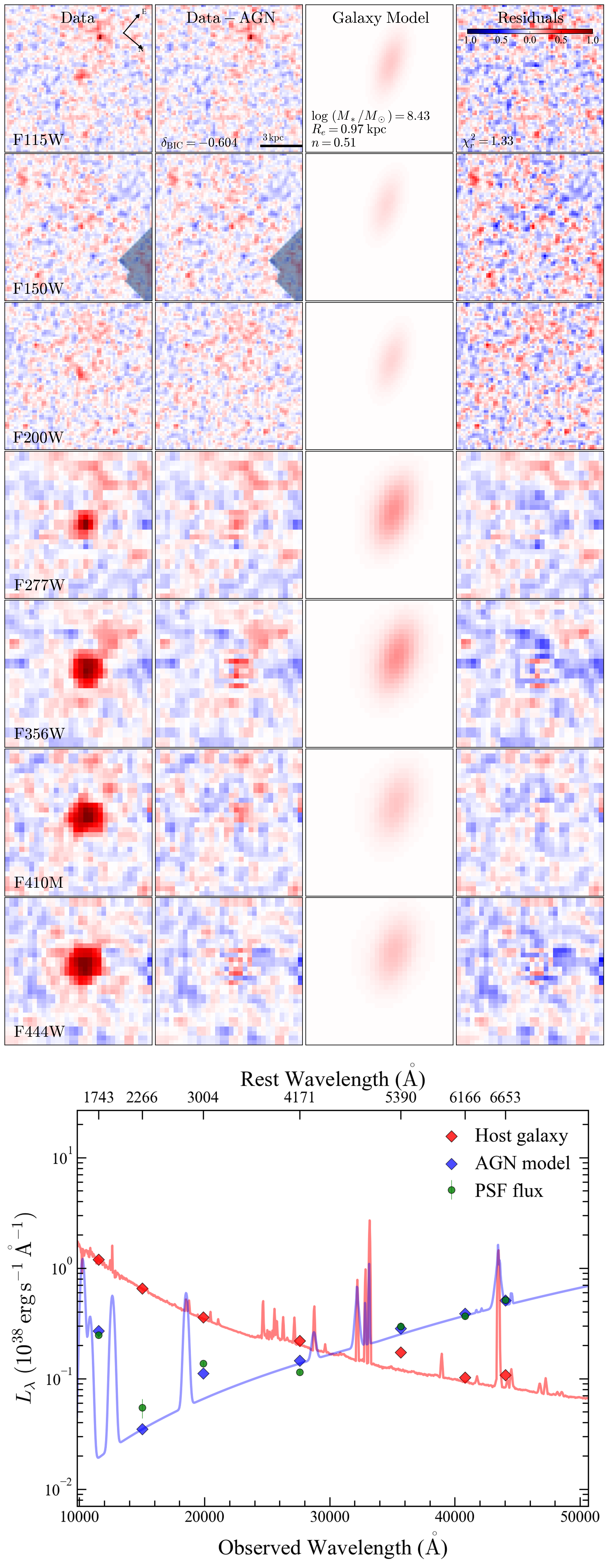}}
\figsetgrpnote{Imaging decomposition (top) and SED/non-parametric flux summary (bottom).}
\figsetgrpend

\figsetgrpnum{1.17}
\figsetgrptitle{CEERS\_006725}
\figsetplot{\detokenize{figset_CEERS_006725.png}}
\figsetgrpnote{Imaging decomposition (top) and SED/non-parametric flux summary (bottom).}
\figsetgrpend

\figsetgrpnum{1.18}
\figsetgrptitle{CEERS\_027825}
\figsetplot{\detokenize{figset_CEERS_027825.png}}
\figsetgrpnote{Imaging decomposition (top) and SED/non-parametric flux summary (bottom).}
\figsetgrpend

\figsetgrpnum{1.19}
\figsetgrptitle{CEERS\_003975}
\figsetplot{\detokenize{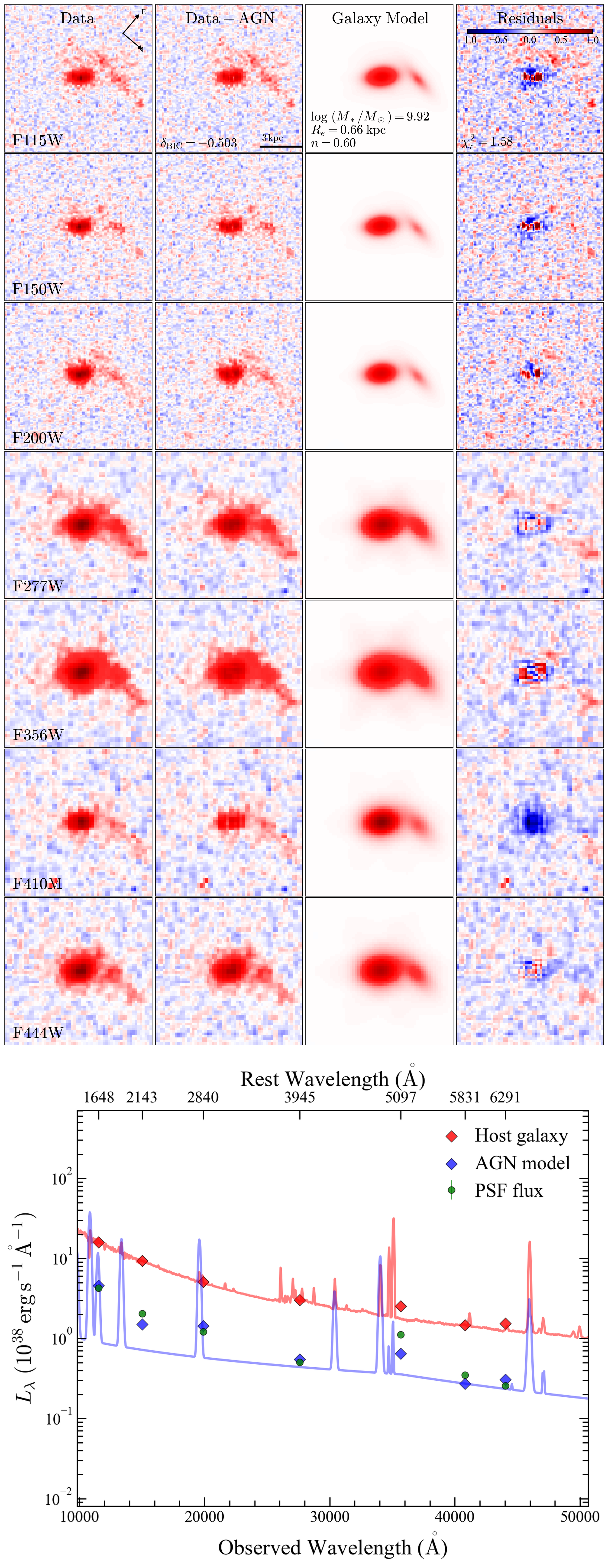}}
\figsetgrpnote{Imaging decomposition (top) and SED/non-parametric flux summary (bottom).}
\figsetgrpend

\figsetgrpnum{1.20}
\figsetgrptitle{J0109-3047-BHAE-1}
\figsetplot{\detokenize{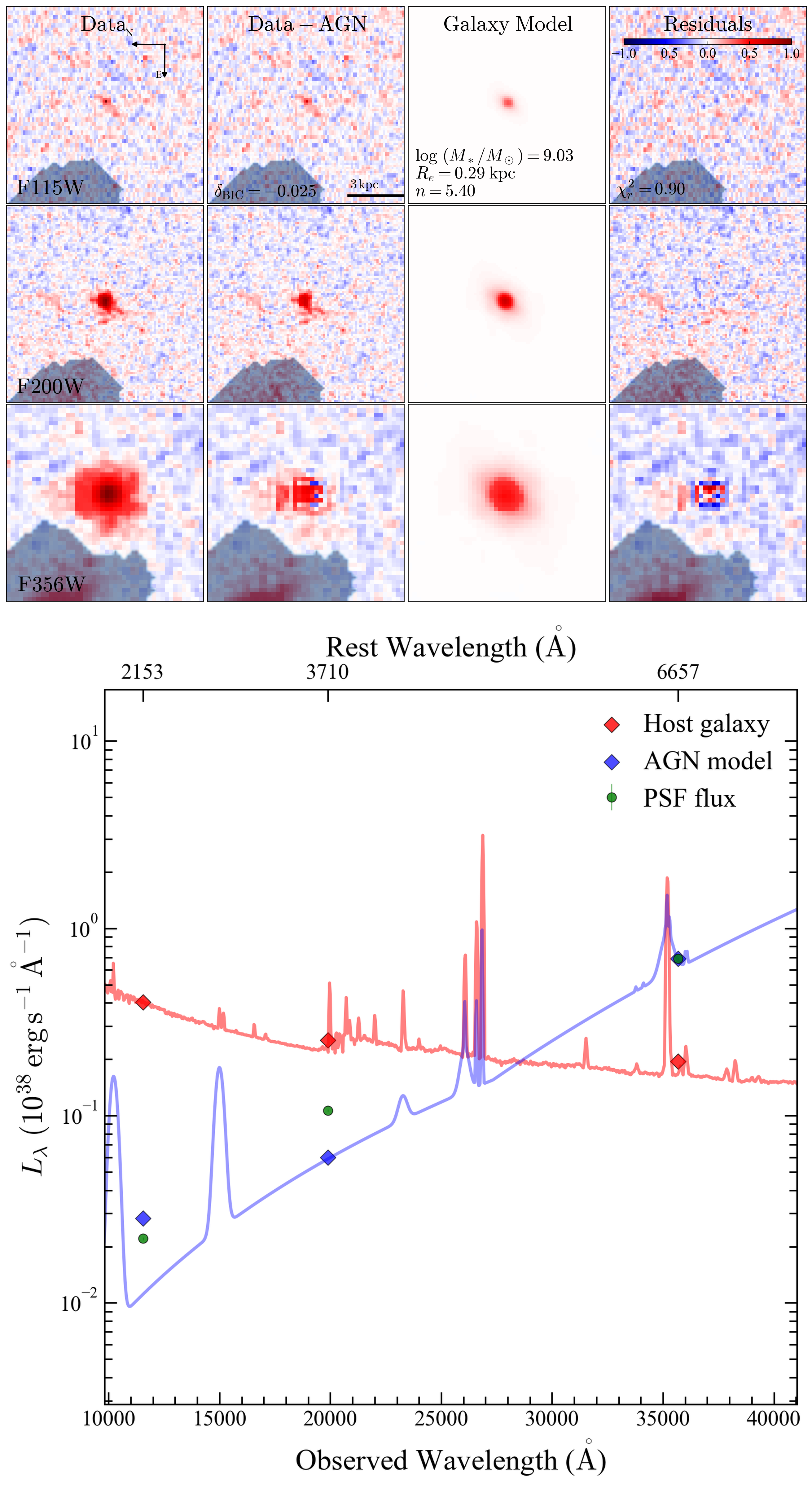}}
\figsetgrpnote{Imaging decomposition (top) and SED/non-parametric flux summary (bottom).}
\figsetgrpend

\figsetgrpnum{1.21}
\figsetgrptitle{J0218+0007-BHAE-1}
\figsetplot{\detokenize{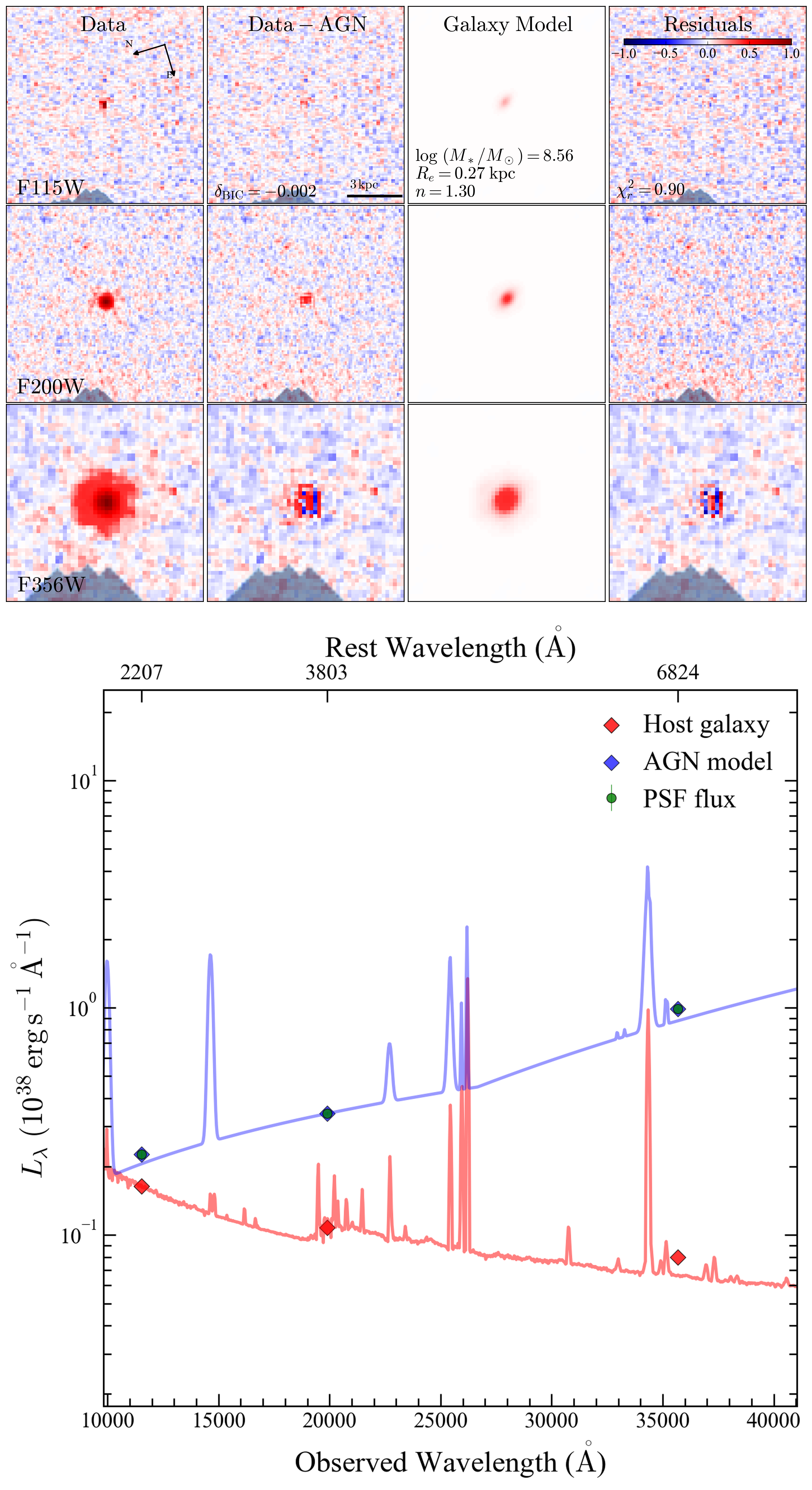}}
\figsetgrpnote{Imaging decomposition (top) and SED/non-parametric flux summary (bottom).}
\figsetgrpend

\figsetgrpnum{1.22}
\figsetgrptitle{J0224-4711-BHAE-1}
\figsetplot{\detokenize{figset_J0224-4711-BHAE-1.png}}
\figsetgrpnote{Imaging decomposition (top) and SED/non-parametric flux summary (bottom).}
\figsetgrpend

\figsetgrpnum{1.23}
\figsetgrptitle{J0229-0808-BHAE-1}
\figsetplot{\detokenize{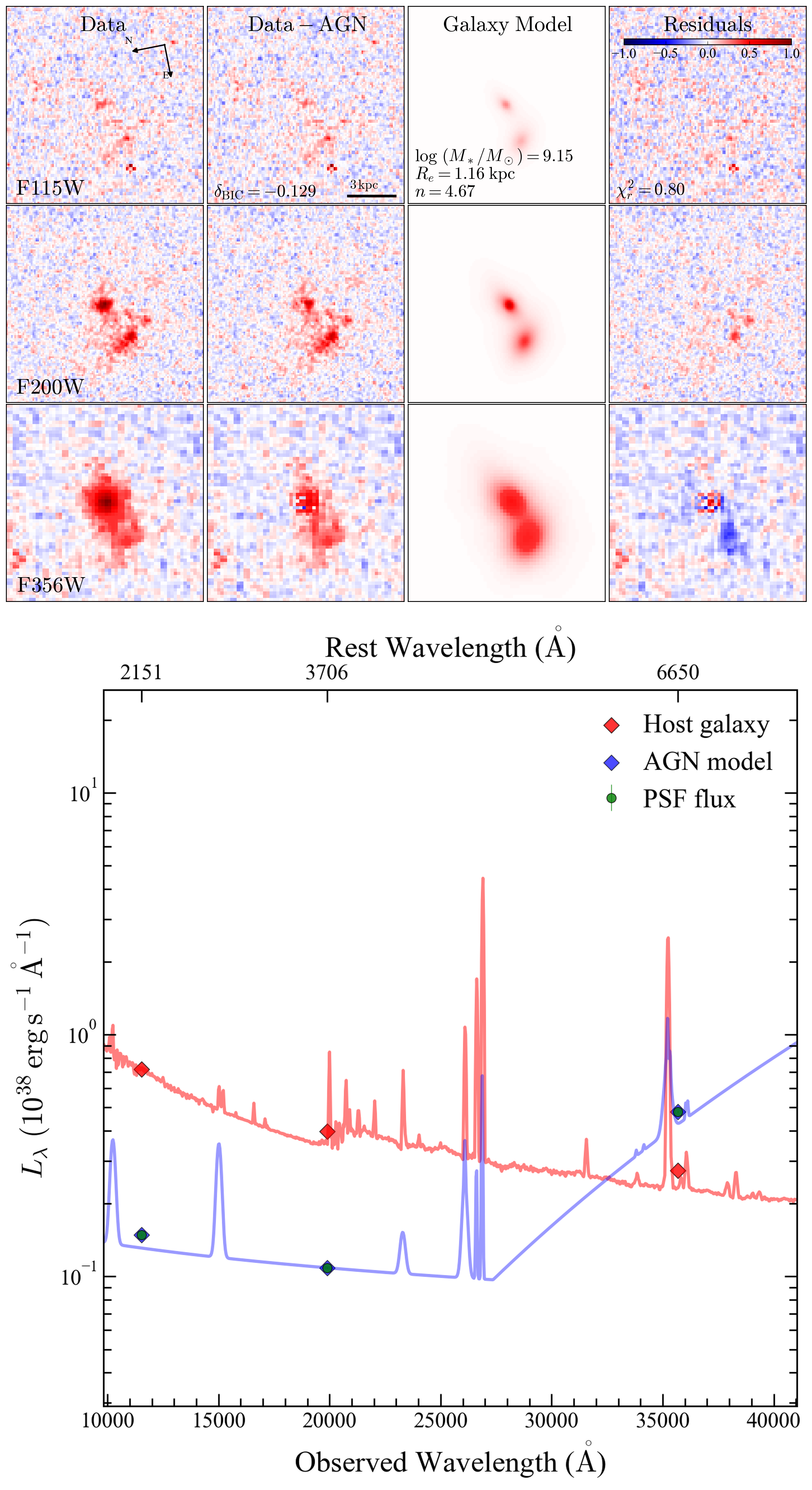}}
\figsetgrpnote{Imaging decomposition (top) and SED/non-parametric flux summary (bottom).}
\figsetgrpend

\figsetgrpnum{1.24}
\figsetgrptitle{J0229-0808-BHAE-2}
\figsetplot{\detokenize{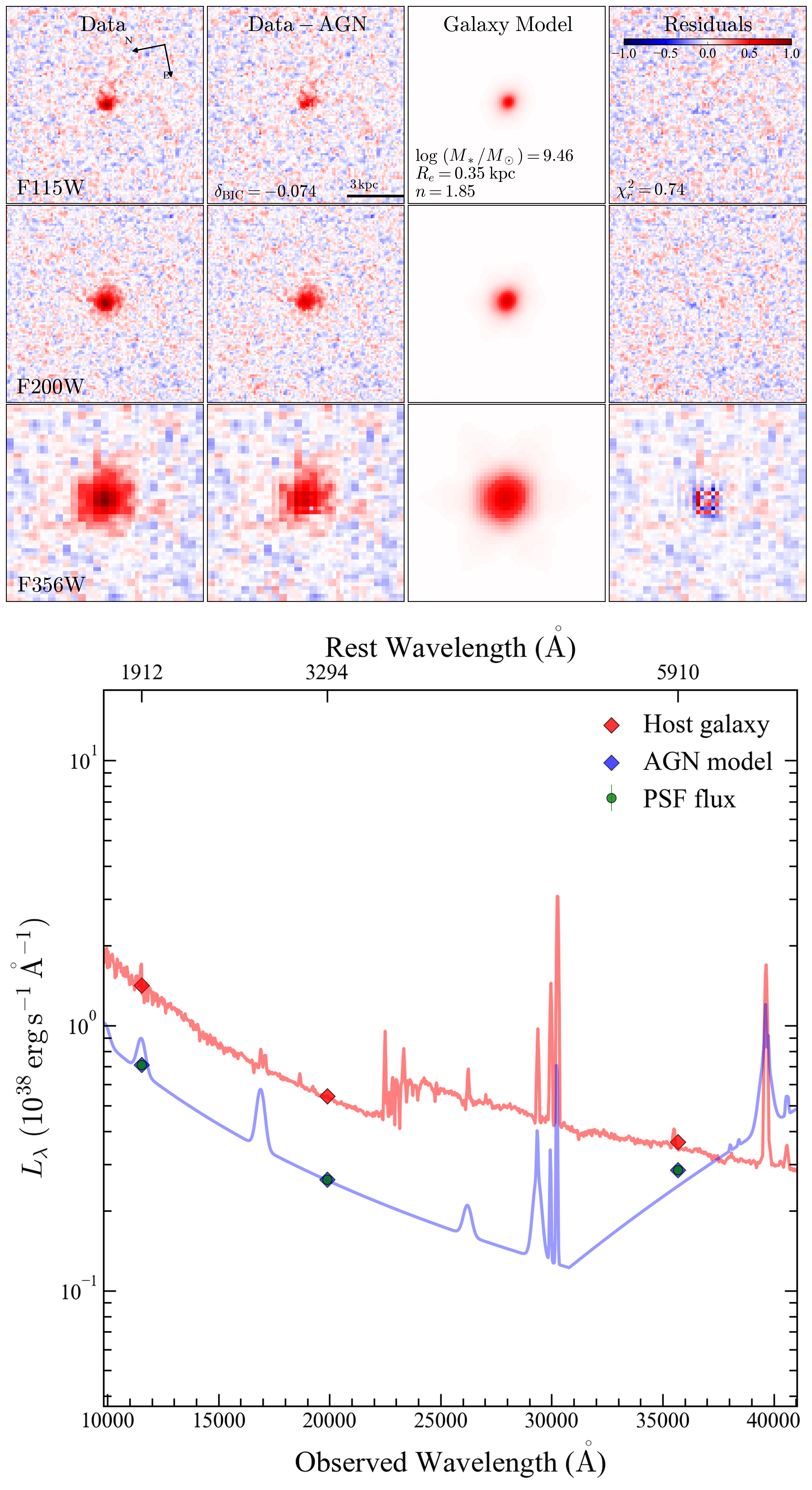}}
\figsetgrpnote{Imaging decomposition (top) and SED/non-parametric flux summary (bottom).}
\figsetgrpend

\figsetgrpnum{1.25}
\figsetgrptitle{J0430-1445-BHAE-1}
\figsetplot{\detokenize{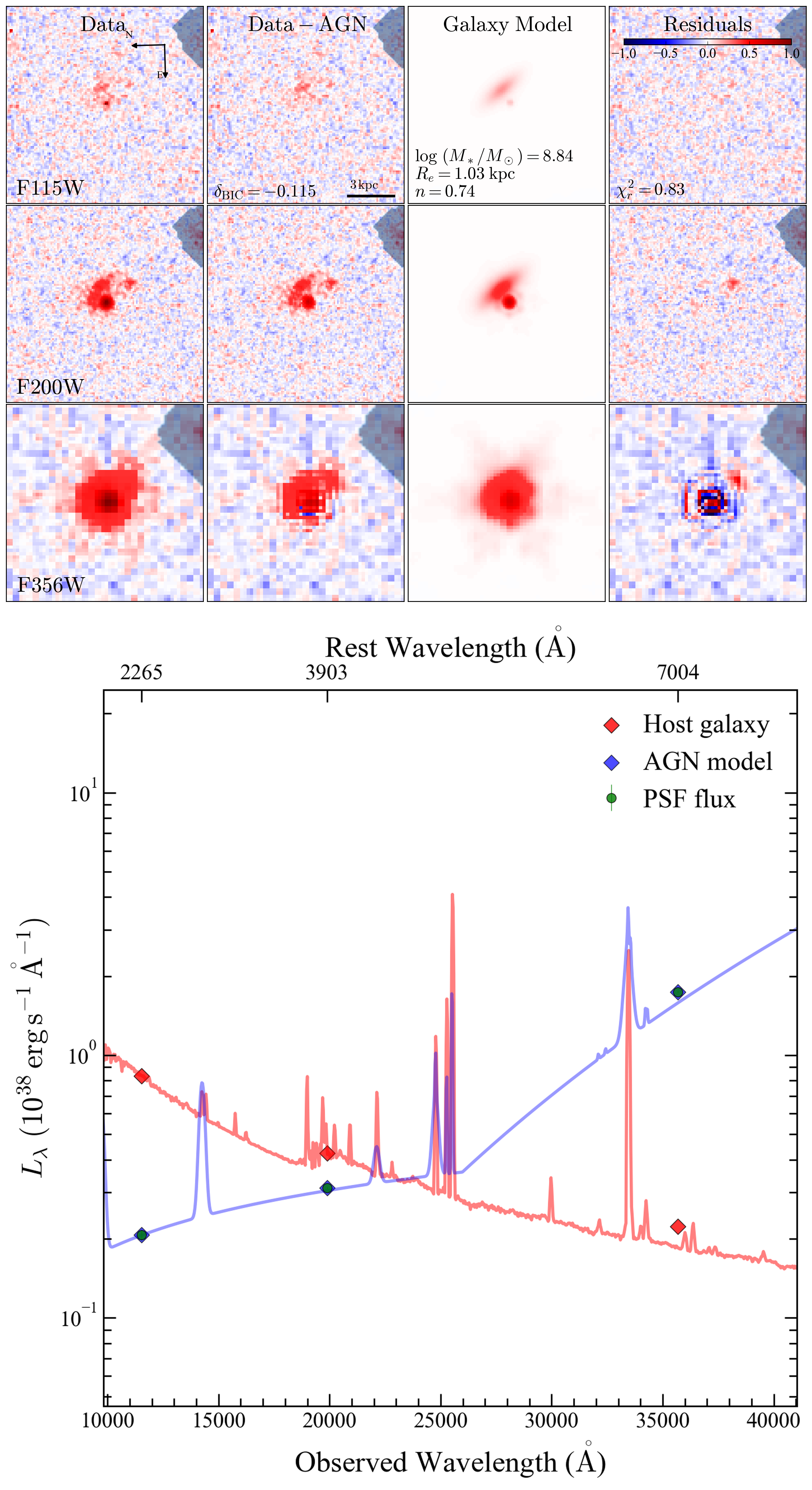}}
\figsetgrpnote{Imaging decomposition (top) and SED/non-parametric flux summary (bottom).}
\figsetgrpend

\figsetgrpnum{1.26}
\figsetgrptitle{J0923+0402-BHAE-1}
\figsetplot{\detokenize{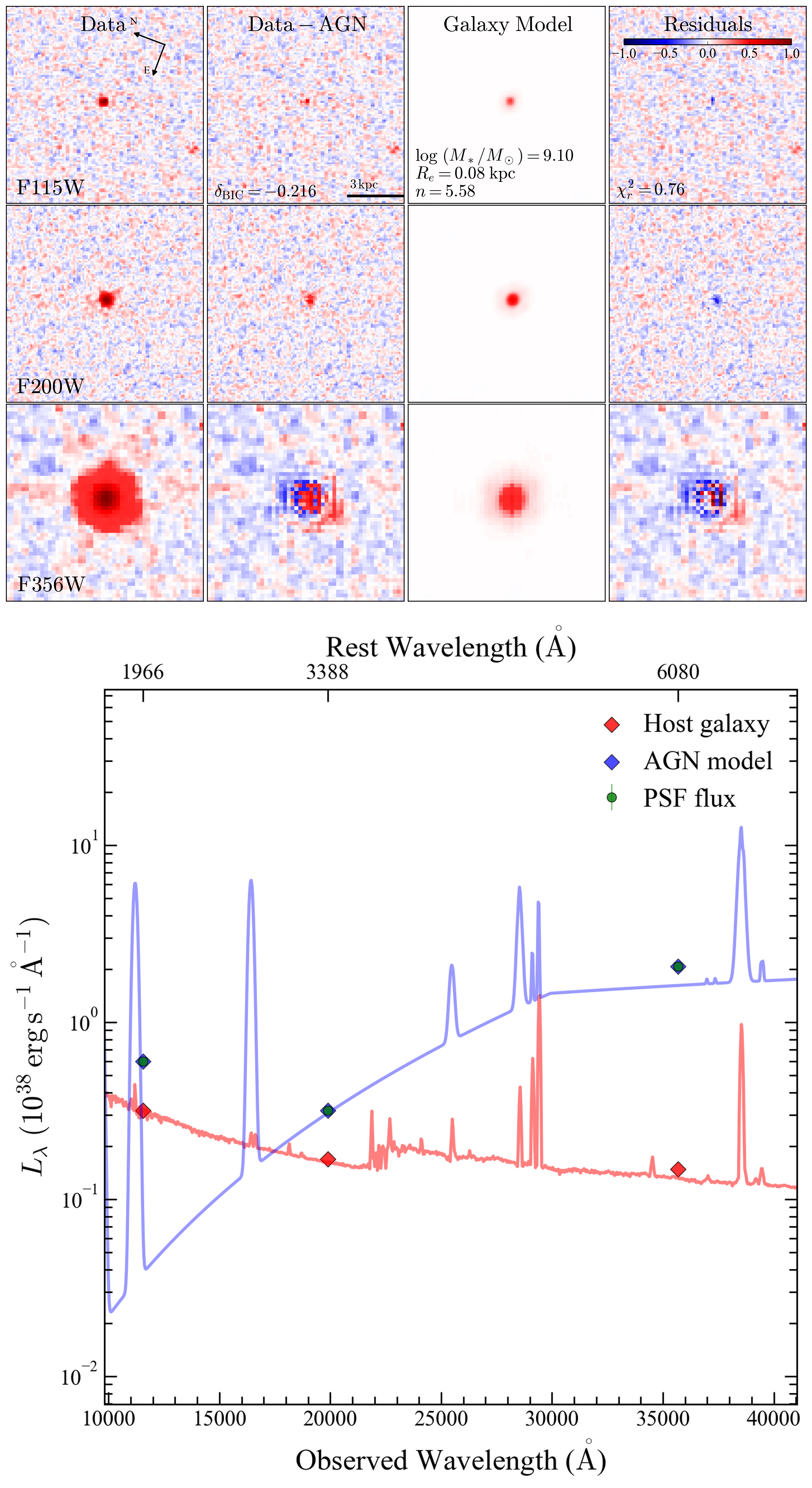}}
\figsetgrpnote{Imaging decomposition (top) and SED/non-parametric flux summary (bottom).}
\figsetgrpend

\figsetgrpnum{1.27}
\figsetgrptitle{J1526-2050-BHAE-2}
\figsetplot{\detokenize{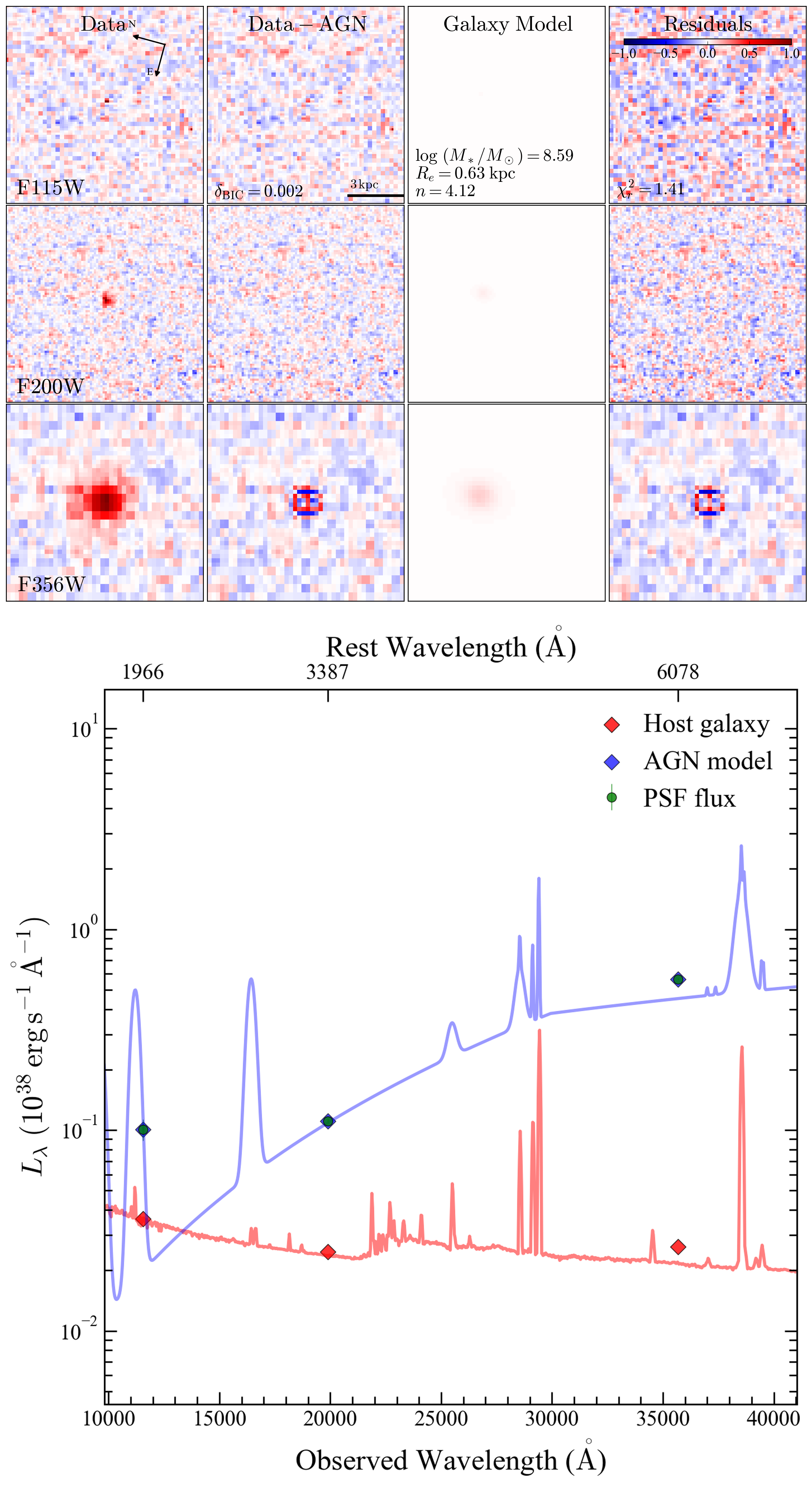}}
\figsetgrpnote{Imaging decomposition (top) and SED/non-parametric flux summary (bottom).}
\figsetgrpend

\figsetgrpnum{1.28}
\figsetgrptitle{J1526-2050-BHAE-3}
\figsetplot{\detokenize{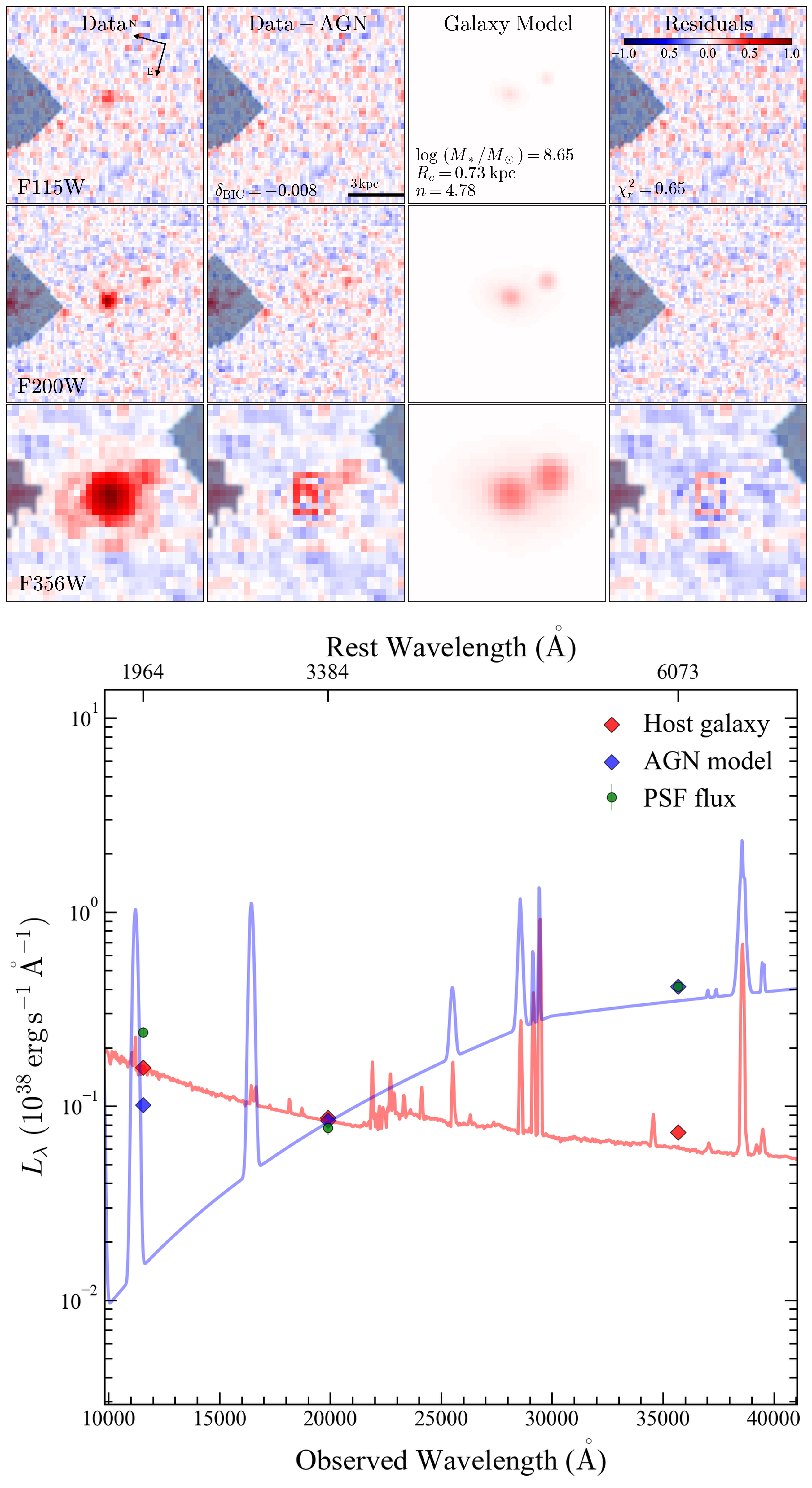}}
\figsetgrpnote{Imaging decomposition (top) and SED/non-parametric flux summary (bottom).}
\figsetgrpend

\figsetgrpnum{1.29}
\figsetgrptitle{J2232+2930-BHAE-1}
\figsetplot{\detokenize{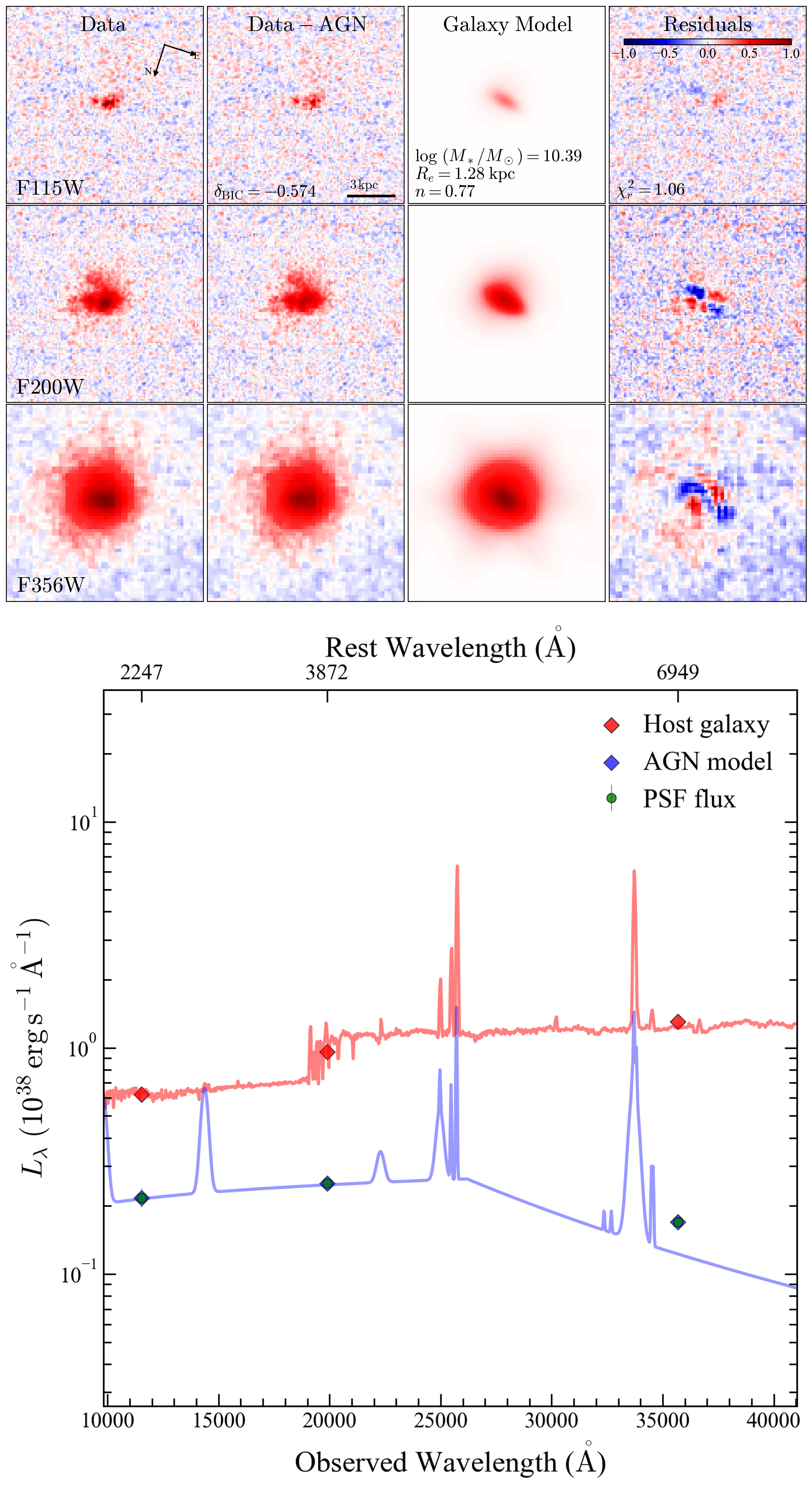}}
\figsetgrpnote{Imaging decomposition (top) and SED/non-parametric flux summary (bottom).}
\figsetgrpend

\figsetgrpnum{1.30}
\figsetgrptitle{J2232+2930-BHAE-2}
\figsetplot{\detokenize{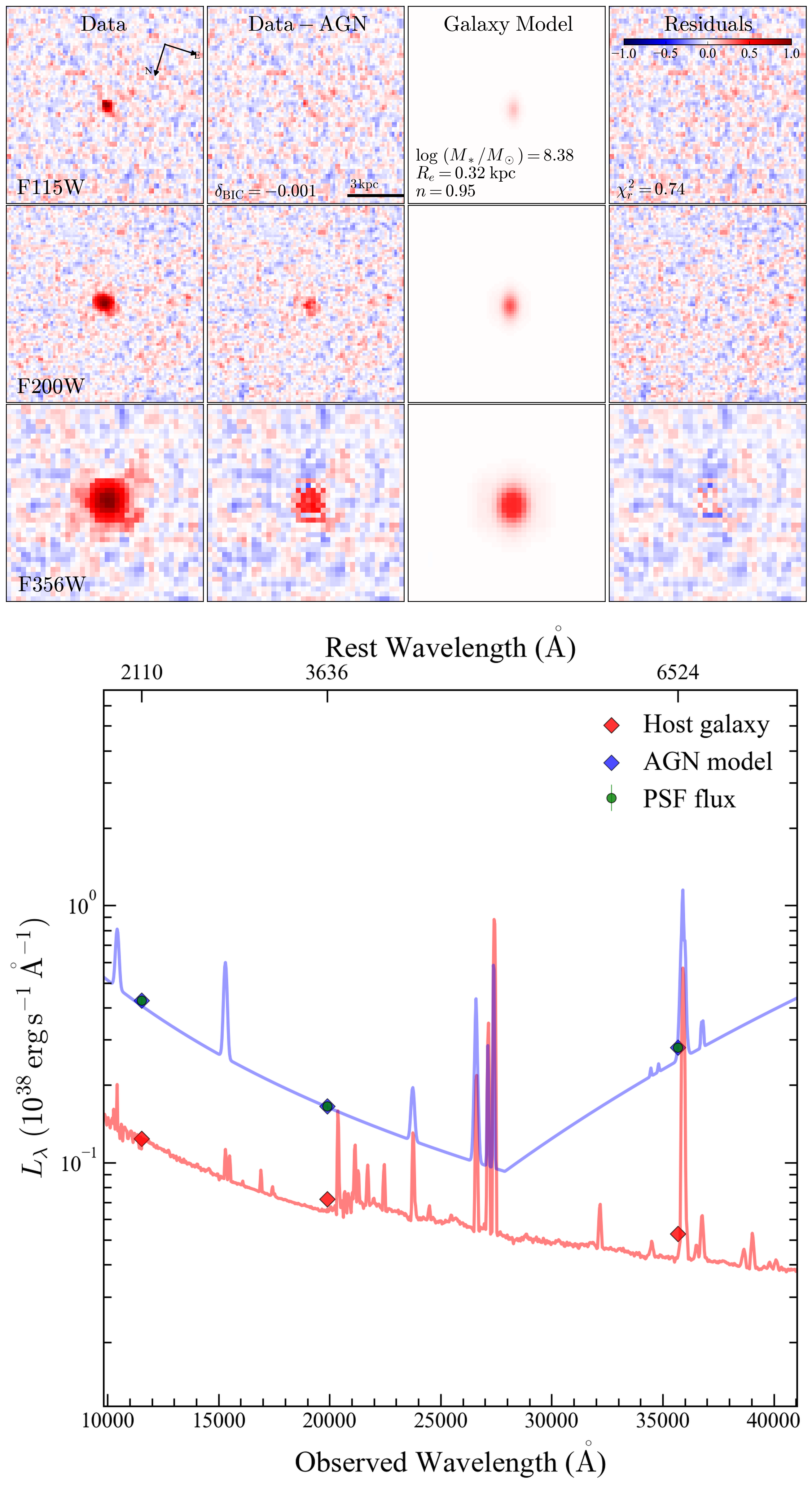}}
\figsetgrpnote{Imaging decomposition (top) and SED/non-parametric flux summary (bottom).}
\figsetgrpend

\figsetgrpnum{1.31}
\figsetgrptitle{J2232+2930-BHAE-3}
\figsetplot{\detokenize{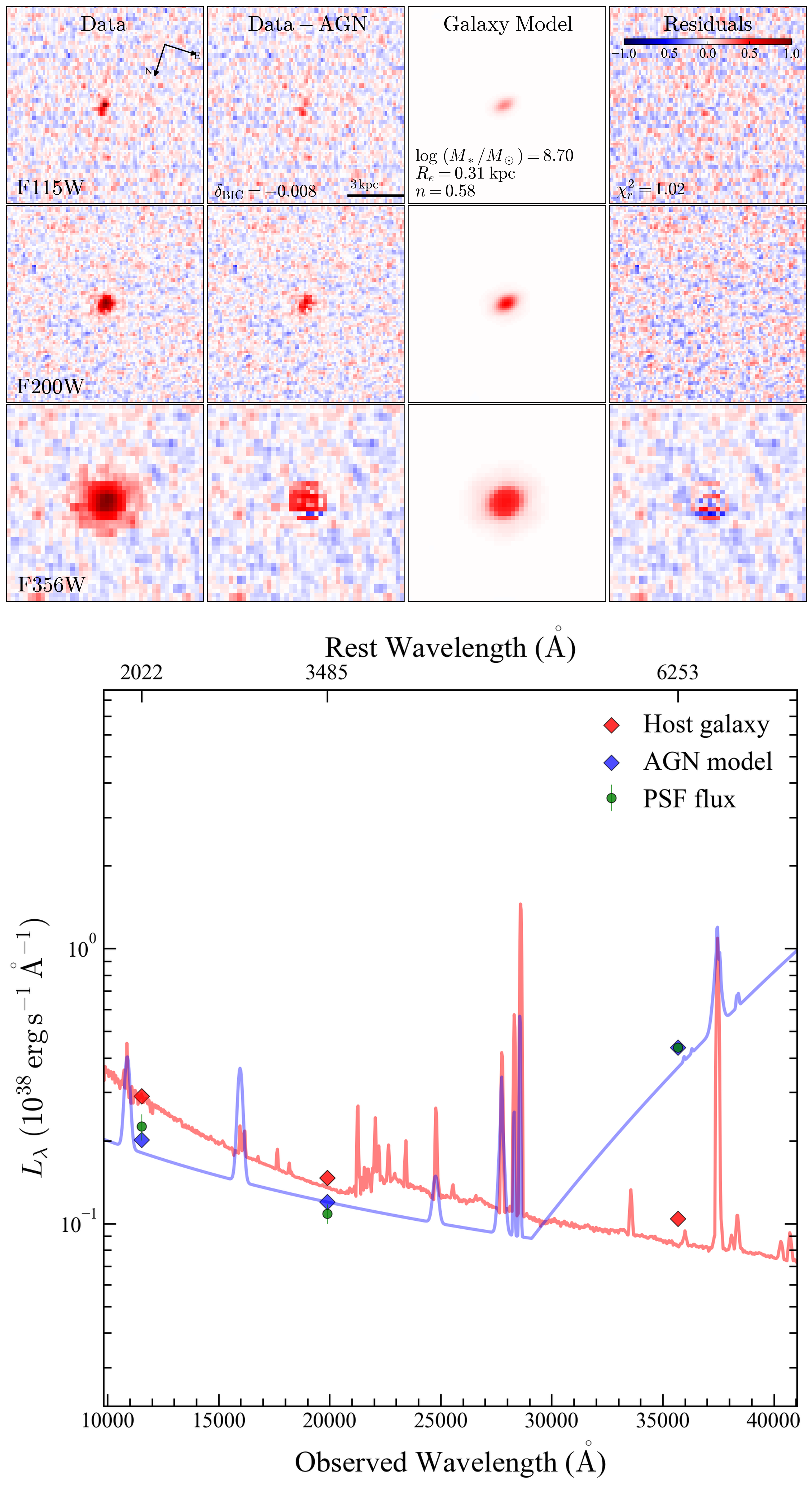}}
\figsetgrpnote{Imaging decomposition (top) and SED/non-parametric flux summary (bottom).}
\figsetgrpend

\figsetend

Below we provide additional details of the fits for several individual objects that required adjustments beyond the models described in Sections~\ref{sec:gssetup} or \ref{sec:agnsed}.

{\it J1120+0641}: This object was studied further by \citet{Marshall2025AA}, who combined NIRSpec integral field spectroscopy with NIRCam photometry of the host continuum emission. Their image decomposition suggests that the host galaxy has a stellar mass of $\log \, (M_\ast/M_\odot) = 9.41$, and that it is accompanied by a companion with $\log \, (M_\ast/M_\odot) = 9.70$. Motivated by their results, we include an additional single \sersic\ galaxy component at the same redshift in our \textsc{GalfitS} analysis. \ledit{ Our results indicate that the companion has a stellar mass of $\log \, (M_\ast/M_\odot) = 9.63$, effective radius $R_e = 0.86$~kpc, axis ratio $q = 0.43$, and a relatively flat surface brightness profile with $n = 0.58$.} The separation between the host and companion is 0\farcs38, or $\sim 2$~kpc at $z = 7.08$. \ledit{While the companion stellar mass is consistent with \citet{Marshall2025AA}, the quasar-host stellar mass inferred from our \textsc{GalfitS} decomposition is higher by 1.4~dex. This discrepancy is driven primarily by the brighter host flux we measure in F356W ($m_\mathrm{F356W} = 23.43 \pm 0.21$) compared to their value ($m_\mathrm{F356W} = 25.11 \pm 0.18$), and may be further amplified by differences in the assumed host-galaxy extinction and star-formation history.}

{\it J0224$-$4711}: A nearby galaxy is blended with the nucleus. With no redshift information available for this galaxy, we add an additional single \sersic\ galaxy component with redshift as a free parameter in our model. With only three filters available, we can obtain only a very loose constraint on the redshift, $z = 3.5$. The fitted galaxy properties are $\log \, (M_\ast/M_\odot) = 10.20$, $R_e = 0\farcs20$, $q = 0.77$, and $n = 2.87$. The separation between the apparent nearby foreground galaxy and J0224$-$4711 is 0\farcs54.

{\it CEERS007465}: The original AGN SED model based on a broken power law (Section~\ref{sec:agnsed}) assumed a fixed ratio of emission-line to continuum luminosity, which resulted in an underestimation of the F115W band flux by $\sim 35\%$. To achieve a good fit, we allowed the strength of He\,II~$\lambda$1640\,\AA\ to vary. A satisfactory fit in the F115W band is reached when the He\,II luminosity is adjusted from $4.41 \times 10^{41}\,\mathrm{erg\,s^{-1}}$ to $1.78 \times 10^{41}\,\mathrm{erg\,s^{-1}}$.

{\it CEERS006725}: Similar to CEERS007465, the AGN SED model in Section~\ref{sec:agnsed} underestimates the F115W band flux by about 10\%. To improve the fit, we again allow the strength of He\,II emission to vary. In this case, the model achieves a good fit in the F115W band (Figure~\ref{fig:galfits_exp7b}) when the He\,II luminosity increases from $6.52 \times 10^{41}\,\mathrm{erg\,s^{-1}}$ to $8.43 \times 10^{41}\,\mathrm{erg\,s^{-1}}$.

{\it CEERS027825}: A galaxy of unknown redshift is located nearby. We include an additional single \sersic\ galaxy component with redshift as a free parameter in our model. The seven available filters suggest that the galaxy is a foreground object, although the redshift is only loosely constrained at $z = 0.3$. Its fitted properties are $\log \, (M_\ast/M_\odot) = 7.64$, $R_e = 0\farcs05$, $q = 0.85$, and $n = 0.54$. The separation between this nearby foreground galaxy and CEERS027825 is 0\farcs26.

{\it CEERS003975}: A galaxy of unknown redshift is located nearby. We include an additional single \sersic\ galaxy component with redshift as a free parameter in our model. The seven filters suggest that this galaxy may be at a similar redshift to CEERS003975 ($z = 6.00$). The fitted properties are $\log \, (M_\ast/M_\odot) = 9.44$, $R_e = 0\farcs16$, $q = 0.31$, and $n = 1.09$. The separation between this galaxy and CEERS003975 is 0\farcs58, or $3.41$~kpc.

{\it J0229$-$0808-BHAE-1}: A nearby galaxy is seen close to our target of interest. As only three filters are available and the nearby galaxy exhibits a roughly similar SED based on visual inspection, we include an additional single \sersic\ component with $z = 4.36$ in our model. The fitted properties are $\log \, (M_\ast/M_\odot) = 9.06$, $R_e = 0\farcs13$, $q = 0.38$, and $n = 1.11$. The separation between this galaxy and J0229$-$0808-BHAE-1 is 0\farcs36, or 2.46~kpc.

{\it J0430$-$1445-BHAE-1}: A nearby galaxy is located adjacent to our target of interest. As only three filters are available and the nearby galaxy exhibits a roughly similar SED based on visual inspection, we include an additional single \sersic\ component with $z = 4.09$ in our model. The fitted properties are $\log \, (M_\ast/M_\odot) = 8.94$, $R_e = 0\farcs02$, $q = 0.30$, and $n = 2.37$. The separation between this galaxy and J0430$-$1445-BHAE-1 is about 0\farcs28, or $1.95$~kpc.

{\it J1526$-$2050-BHAE-3}: As only three filters are available and the nearby galaxy exhibits a roughly similar SED based on visual inspection, we include an additional single \sersic\ component with $z = 4.87$ in our model. The fitted properties are $\log \, (M_\ast/M_\odot) = 8.81$, $R_e = 0\farcs02$, $q = 0.58$, and $n = 3.75$. The separation between this galaxy and J1526$-$2050-BHAE-3 is 0\farcs24, or 1.58~kpc.

{\it J2232+2930-BHAE-1}: As only three filters are available and the nearby galaxy exhibits a roughly similar SED based on visual inspection, we include an additional single \sersic\ component with $z = 4.13$ in our model. The fitted properties are $\log \, (M_\ast/M_\odot) = 9.97$, $R_e = 0\farcs11$, $q = 0.36$, and $n = 0.50$. The separation between this galaxy and J2232+2930-BHAE-1 is 0\farcs11, or 0.75~kpc. This galaxy might represent a substructure within the host of J2232+2930-BHAE-1, effectively increasing the host's original mass from $\log \, (M_\ast/M_\odot) = 10.80$ to 10.86.

\section{Input-output Tests}
\label{app:iotest}

We perform input–output tests to address several key objectives: (1) assess the detectability of host galaxies using our method (Section~\ref{sec:hostdec}); (2) evaluate the reliability of the derived key host–galaxy parameters (Section~\ref{sec:reliability}); (3) quantify the uncertainties arising from our choice of empirical PSF (Section~\ref{sec:uncert}); and (4) validate the use of a non-parametric AGN SED compared to a parametric AGN method (Section~\ref{sec:agnsed}). Here we provide details on the generation of our mock AGNs.

\begin{figure*}
\centering
\figurenum{C1}
\includegraphics[width=0.95\textwidth]{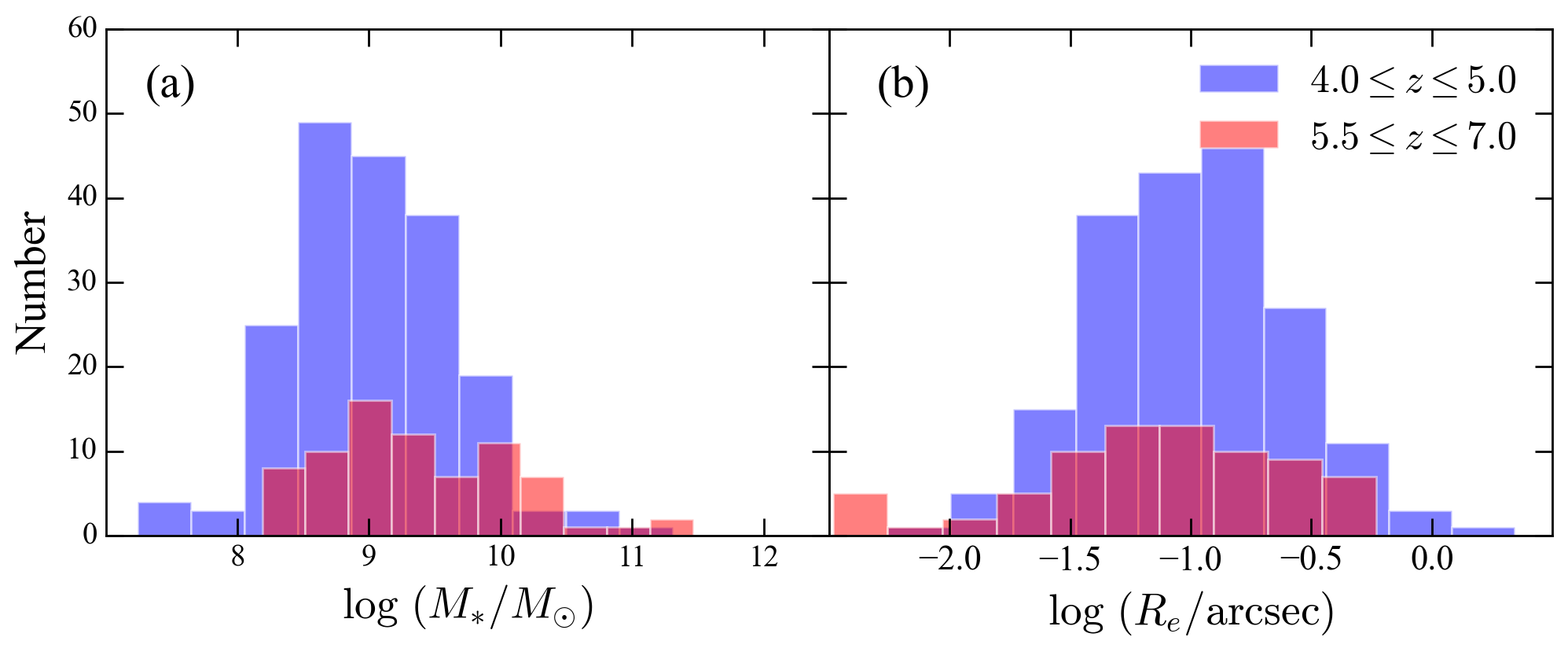}
\caption{Distribution of (a) stellar masses and (b) effective radius, for the galaxies from \citet{Sun2024ApJ} used for the input-output tests, separated into the redshift ranges $5.5 \leq z \leq 7.0$ (red) and $4.0 \leq z \leq 5.0$ (blue).}
\label{fig:galaxy}
\end{figure*}

We generate mock AGNs by manually adding a nucleus to the multiband images of real galaxies. We use galaxy images from the CEERS field, where imaging in all seven filters is available, to avoid potential systematics associated with a parametric multiband galaxy model, especially given the limited prior constraints at high redshift. We select two sets of galaxies samples from the parent sample of \citet{Sun2024ApJ}: an subsample of 82 galaxies at $5.5 \leq z \leq 7.0$, which primarily matches our high-luminosity quasars,
and a subsample of 190 galaxies at $4.0 \leq z \leq 5.0$, which mainly corresponds to the lower luminosity quasars. We use all available galaxies in the given redshift range to maximize the accuracy of our detection rate determination. The division into two groups is motivated by the different nuclear SEDs of the two subsamples of quasars \citep{Yang2023ApJ,Lin2024ApJ} (see also Section~\ref{sec:ncolor}). We first perform a single-\sersic\ \textsc{GalfitS} fit on all 272 galaxies (denoted as \texttt{run1} in Table~\ref{tab:gssetup}); \ledit{the derived stellar mass and effective radius distribution is shown in Figure~\ref{fig:galaxy}}.

For each galaxy, we manually add a point source at the best-fit galaxy center obtained from \texttt{run1}. For the 82 galaxies in the luminous quasar subsample, the point source SED is derived from the quasar SED template of \citet{Temple2021MNRAS}, which accounts for both redshift evolution and host galaxy subtraction. We scale this template to four luminosity levels, $\log \, L_{5100} = 43, 44, 45$, and $46~\mathrm{erg\,s^{-1}}$, covering the full luminosity range of our sample. Mock images are generated in seven NIRCam bands (F115W, F150W, F200W, F277W, F356W, F410M, F444W). Instead of using the fitted PSF model constructed from the CEERS field, for each AGN luminosity we select five foreground stars whose fluxes are similar to those expected for an AGN at $z=6.5$ in the F356W filter, and create five distinct mock AGN images based on these stars. Finally, we update the error maps with appropriate Poisson noise. In total, this procedure produces $82 \times 4 \times 5 = 1640$ mock AGN images with seven-band coverage.

\begin{figure}
\centering
\figurenum{C2}
\includegraphics[width=0.48\textwidth]{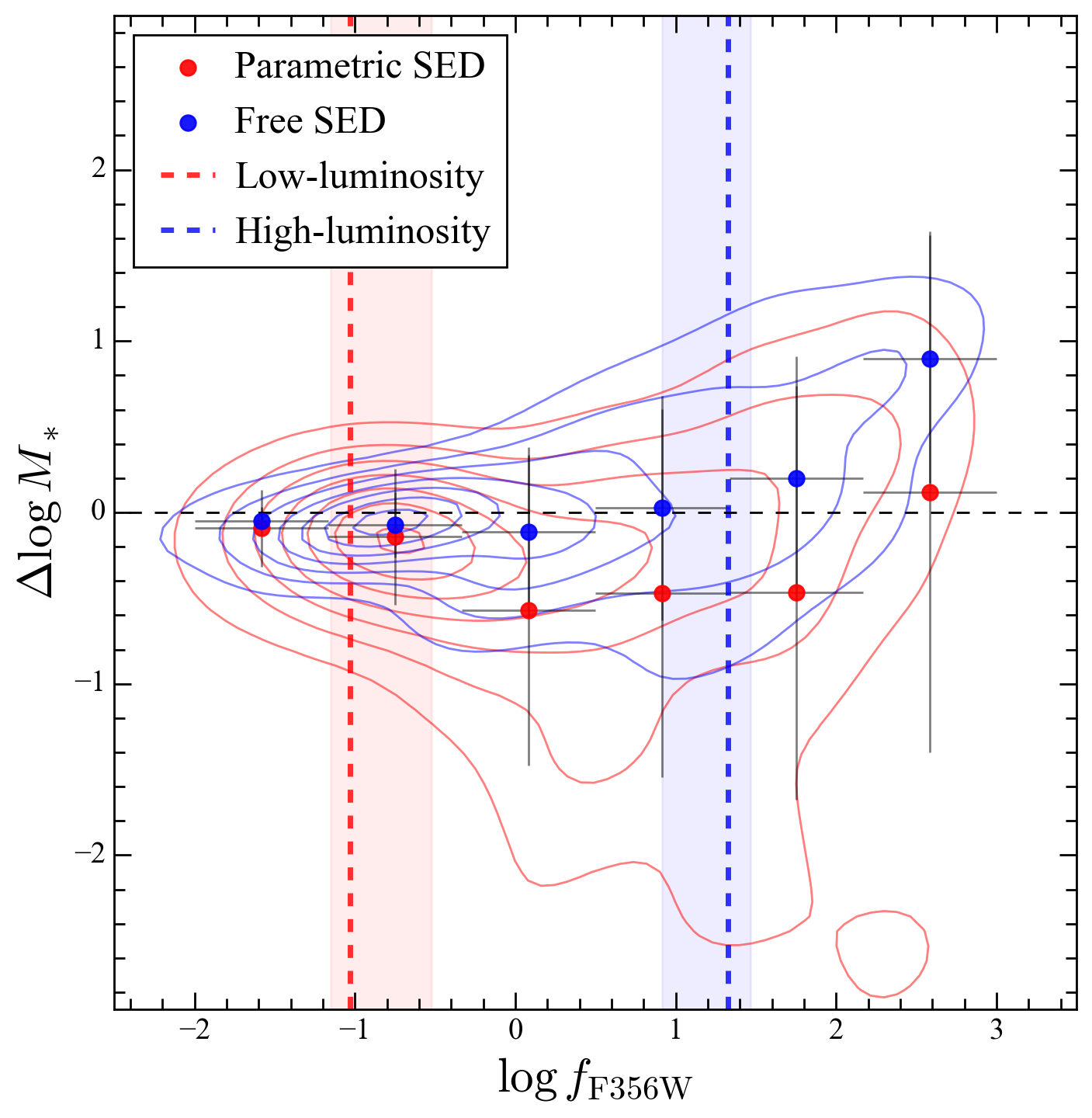}
\caption{Comparison of the systematic bias in host galaxy stellar mass measurements as a function of $f_{\rm F356W}$, the AGN-to-host flux ratio in the F356W band. The blue points and contours represent the fiducial 7-band fit to mock AGNs (Section~\ref{sec:galfits}), which considers the AGN luminosity in each band as free, independent parameters. The red points and contours correspond to fits that assume a parametric AGN SED (Section~\ref{sec:agnsed}). Error bars indicate the $1\,\sigma$ scatter. The median $f_{\mathrm{F356W}}$ and the 16th and 84th percentiles are marked by the dashed line and shaded region for the low-luminosity (red) and high-luminosity (blue) quasars.  The horizontal dashed line indicates $\Delta \log \, M_* = 0$.}
\label{fig:compare2}
\end{figure}

\begin{figure}
\centering
\figurenum{C3}
\includegraphics[width=0.48\textwidth]{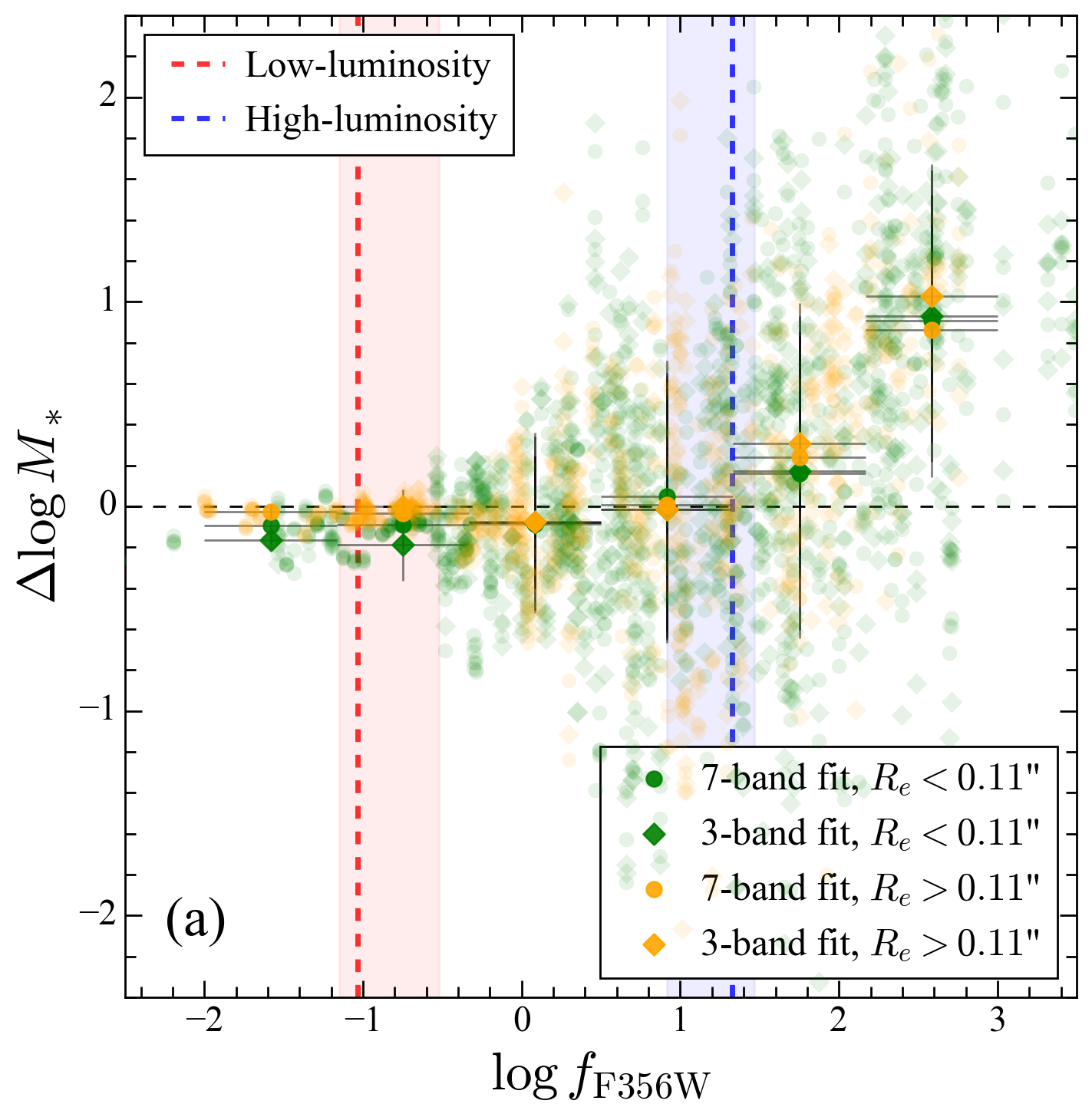}
\includegraphics[width=0.48\textwidth]{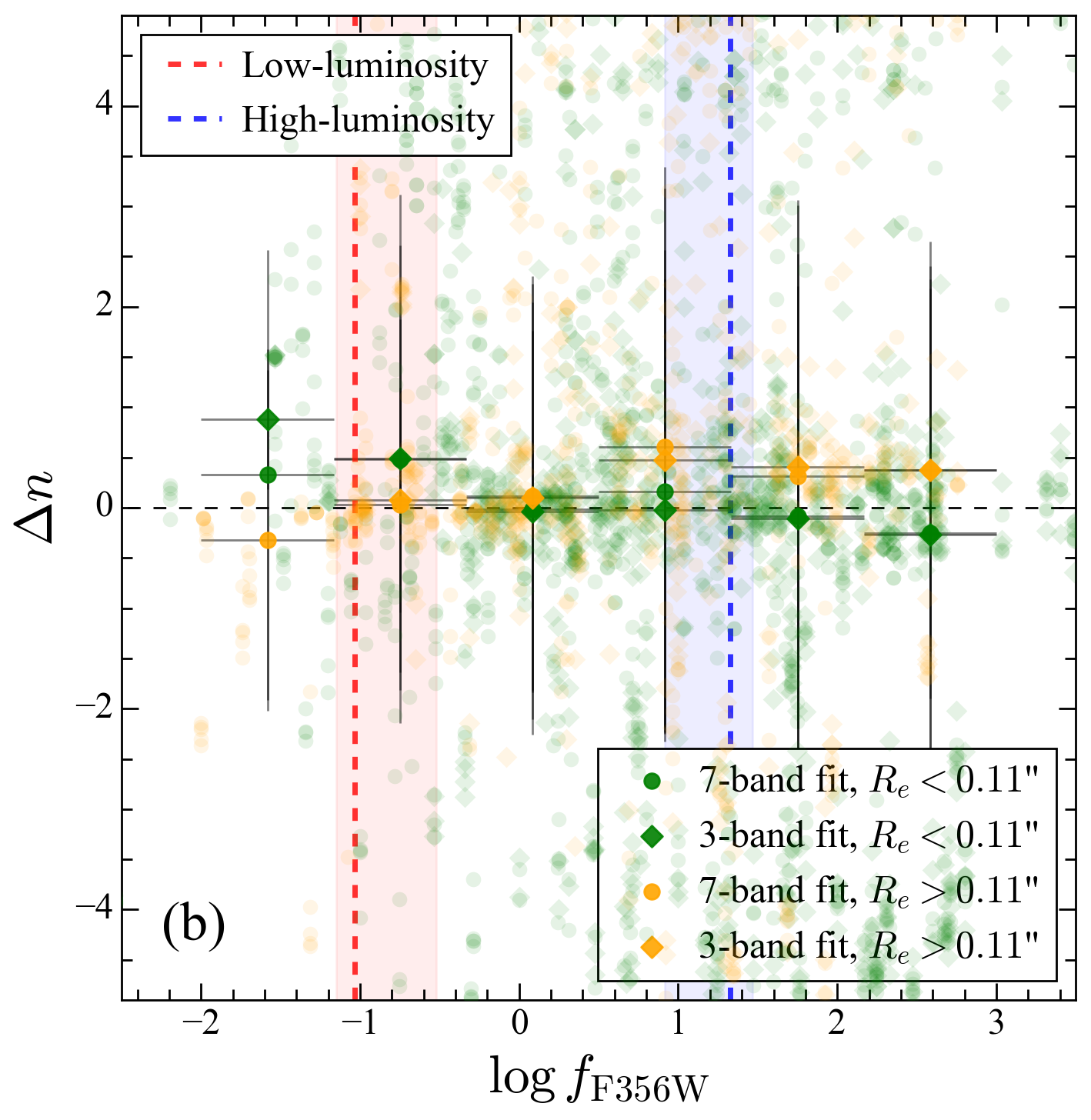}
\caption{Systematic bias in the recovery of host-galaxy properties from the mock AGN tests as a function of $f_{\mathrm{F356W}}$, the AGN-to-host flux ratio in the F356W band. Panel (a) shows input-output difference of stellar mass, and panel (b) shows \sersic\ index. Results are shown separately for compact input hosts ($R_e \le 0\farcs11$; green) and extended input hosts ($R_e > 0\farcs11$; orange). Circles denote the fiducial 7-band fits, while diamonds denote the 3-band fits. Faint symbols show individual mock realizations; large symbols show the median values in bins of ($f_{\mathrm{F356W}}$), with error bars indicating the ($1\,\sigma$) scatter. The median ($f_{\mathrm{F356W}}$) and the 16th and 84th percentiles are marked by the dashed line and shaded region for the low-luminosity (red) and high-luminosity (blue) quasars. The horizontal dashed line indicates zero bias.}
\label{fig:compare3}
\end{figure}

We perform three sets of fits on the mock AGNs. In \texttt{run2}, we perform a point source–only fit. In \texttt{run3}, we perform a point source plus galaxy fit following our optimal strategy described in Section~\ref{sec:gssetup}. In \texttt{run4}, we perform a seven-band fit using the AGN SED model from Section~\ref{sec:agnsed} instead of the non-parametric AGN SED model used in \texttt{run3}. Analysis of the results from \texttt{run1}, \texttt{run2}, and \texttt{run3} shows that:

\begin{enumerate}

\item The \textsc{GalfitS} method, when applied to multiband images, achieves a higher detection rate than single-band AGN–host decomposition (Figure~\ref{fig:detect}).

\item The key host galaxy parameters are reliably recovered using our optimal fitting strategy (Figure~\ref{fig:compare}).

\item The uncertainty arising from the PSF can be quantified following $\sigma_{\log\, {M_\ast} \, \rm PSF} = 0.12\,\log\, (L_{5100}/\rm erg\, s^{-1}) - 5.11$ (Figure~\ref{fig:psferror}).

\item Adopting a non-parametric AGN SED for the nucleus is consistent with the parametric approach for $\log \, f_{\mathrm{F356W}} \lesssim 0$, and results in a smaller systematic bias when the AGN is brighter (Figure~\ref{fig:compare2}).

\item \ledit{Splitting the mock AGNs by host size at $R_e=0.11^{\prime\prime}$ (the median of our sample) yields consistent median biases and scatters in both $\Delta\log M_\ast$ and $\Delta n$ at fixed $f_{\rm F356W}$, indicating that the recovery accuracy is primarily driven by flux contrast rather than by $R_e$. (Figure~\ref{fig:compare3})}

\end{enumerate}

For the 190 galaxies designed for the low-luminosity quasars, our primary goal is to determine the stellar mass upper limit for the non-detection case of J1526$-$2050-BHAE-2. To this end, we directly use the SED of J1526$-$2050-BHAE-2 to generate mock AGNs in the F115W, F200W, and F356W bands. We employ the fitted PSF model as the point source, resulting in a total of 190 mock AGN images. We then perform both a point source–only run and a point source plus galaxy run (denoted as \texttt{uplim} in Table~\ref{tab:gssetup}) to calculate the detection rate as a function of the stellar mass $M_\ast$. The resulting detection rate, shown in Figure~\ref{fig:uplim}, is used to estimate the stellar mass upper limit for J1526$-$2050-BHAE-2.

\section{Host Galaxy Magnitudes}
\label{app:hostmag}

\begin{figure}
\centering
\figurenum{D1}
\includegraphics[width=0.7\textwidth]{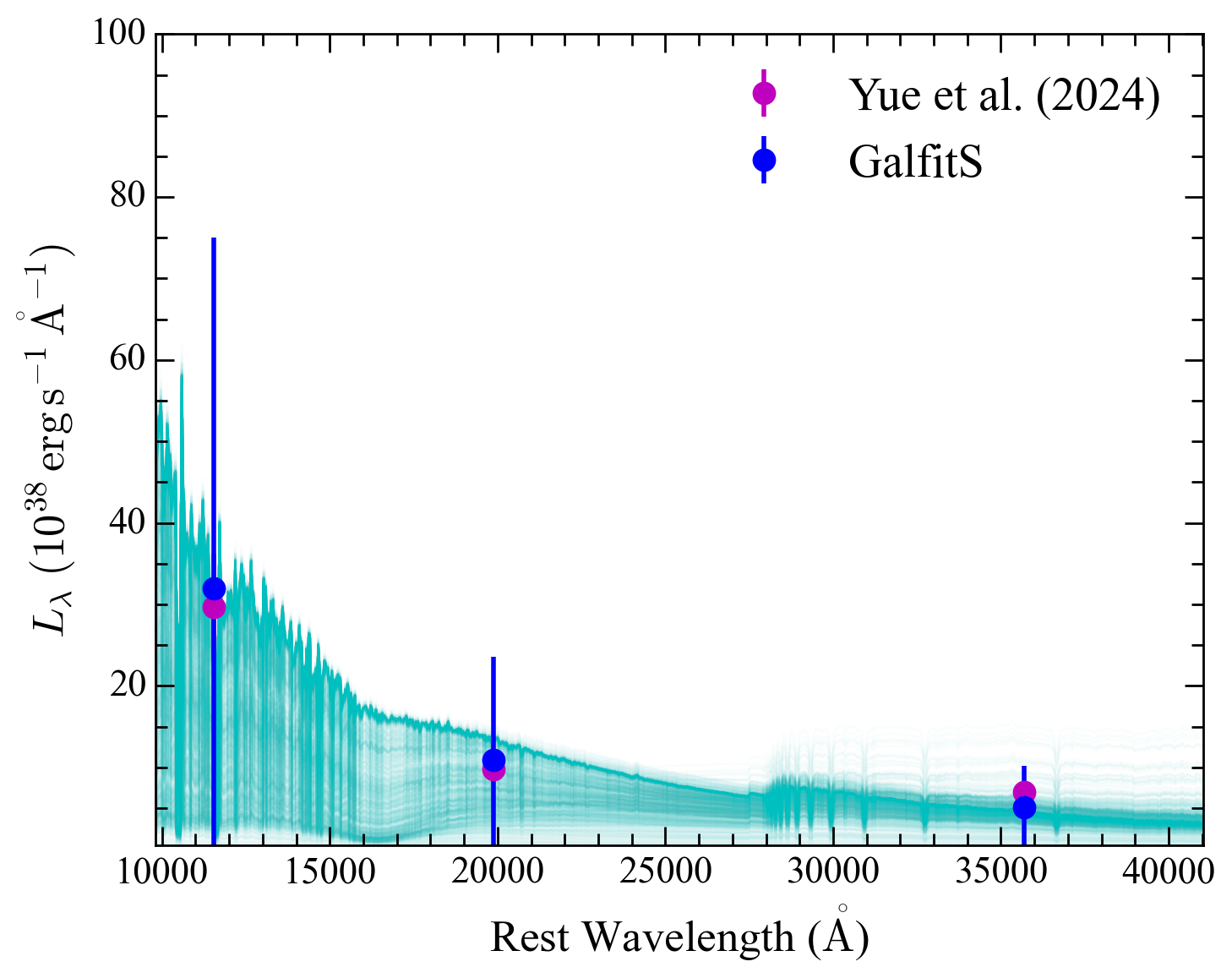}
\caption{Example posterior sampling of the host galaxy SED for J0148+0600. The cyan curves show 1000 host SED realizations drawn from the posterior distribution of the \textsc{GalfitS} SED parameters. Blue points show the observed luminosity ($L \equiv 4\pi f D_{c}^2$) of the host at the three NIRCam filters, with error bars indicating the $1\,\sigma$ uncertainties derived from the posterior sampling (16th--84th percentiles). Magenta points correspond to the host measurements from \citet{Yue2024ApJ}.}
\label{fig:hostflux}
\end{figure}

\ledit{ \textsc{GalfitS} fits the multiband images by forward-modeling the host galaxy with an explicit SED parameterization (Section~\ref{sec:galfits}), rather than fitting independent fluxes in each filter. We therefore derive the host-galaxy apparent magnitudes by computing synthetic photometry from the best-fit host SED through the relevant filter transmission curves. To propagate the full parameter covariance into flux uncertainties, we Monte Carlo sample the posterior distribution of the SED parameters: for each object, we draw 1000 parameter sets from the posterior from Section~\ref{sec:posterior} , generate the corresponding host SED, and recompute the host magnitudes in each band. We adopt the median of the resulting magnitude distribution as the reported value, and take the 16th and 84th percentiles as the $1\,\sigma$ uncertainties. The derived host-galaxy magnitudes for the full sample are listed in Table~\ref{tab:hostmag}. Figure~\ref{fig:hostflux} illustrates this procedure for J0148+0600.}

\startlongtable
\begin{deluxetable*}{cccccccccc}
\setlength{\tabcolsep}{2.5pt}
\tablecaption{Host Galaxy Magnitude from \textsc{GalfitS} Analysis} \label{tab:hostmag}
\tabletypesize{\footnotesize}
\tablehead{
      \colhead{Name}             &
      \colhead{$\Delta$R.A.}       &
      \colhead{$\Delta$Decl.}      &
      \colhead{$m_\mathrm{F115W}$} &
      \colhead{$m_\mathrm{F150W}$} &
      \colhead{$m_\mathrm{F200W}$} &
      \colhead{$m_\mathrm{F277W}$} &
      \colhead{$m_\mathrm{F356W}$} &
      \colhead{$m_\mathrm{F410M}$} &
      \colhead{$m_\mathrm{F444W}$} \\
&
(\arcsec) &
(\arcsec) &
(mag)&
(mag)&
(mag)&
(mag)&
(mag)&
(mag)&
(mag) \\
\colhead{(1)} &
\colhead{(2)} &
\colhead{(3)} &
\colhead{(4)} &
\colhead{(5)} &
\colhead{(6)} &
\colhead{(7)} &
\colhead{(8)} &
\colhead{(9)} &
\colhead{(10)} 
}
\startdata
J2236$+$0032         & $-0.01\pm0.01$  & $-0.03\pm0.01$  & ---               & $24.33\pm0.26$ & ---               & ---                &$22.54\pm0.19$ & ---               & ---               \\
J2255$+$0251         & $-0.05\pm0.01$  & $0.15\pm0.01$   & ---               & $27.17\pm0.40$ & ---               & ---                &$24.90\pm0.06$ & ---               & ---               \\
J0148$+$0600         & $-0.05\pm0.01$  & $0.09\pm0.01$   & $24.73\pm0.20$    & ---            & $23.78\pm0.10$    & ---                &$22.48\pm0.19$ & ---               & ---               \\
J1030$+$0524         & $0.04\pm0.01$   & $0.15\pm0.01$   & $27.26\pm0.19$    & ---            & $25.82\pm0.12$    & ---                &$23.89\pm0.07$ & ---               & ---               \\
J159$+$02            & $0.02\pm0.03$   & $-0.06\pm0.03$  & $22.97\pm1.31$    & ---            & $23.10\pm1.07$    & ---                &$22.82\pm0.85$ & ---               & ---               \\
J1120$+$0641         & $-0.05\pm0.01$  & $0.09\pm0.01$   & $25.17\pm0.33$    & ---            & $24.68\pm0.21$    & ---                &$23.43\pm0.21$ & ---               & ---               \\
J1148$+$5251         & $0.01\pm0.01$   & $-0.11\pm0.01$  & $23.74\pm0.32$    & ---            & $22.69\pm0.22$    & ---                &$21.40\pm0.14$ & ---               & ---               \\
J0109$-$3047         & $-0.12\pm0.02$  & $0.00\pm0.02$   & $27.66\pm0.30$    & ---            & $25.91\pm0.21$    & ---                &$23.65\pm0.13$ & ---               & ---               \\
J0218$+$0007         & $-0.04\pm0.01$  & $0.00\pm0.01$   & $25.56\pm0.20$    & ---            & $24.22\pm0.12$    & ---                &$22.57\pm0.04$ & ---               & ---               \\
J0224$-$4711         & $0.09\pm0.01$   & $-0.02\pm0.01$  & $24.20\pm0.20$    & ---            & $22.88\pm0.14$    & ---                &$21.34\pm0.07$ & ---               & ---               \\
J0226$+$0302         & $0.01\pm0.05$   & $-0.09\pm0.06$  & $23.40\pm1.46$    & ---            & $23.38\pm1.26$    & ---                &$22.95\pm1.10$ & ---               & ---               \\
J0244$-$5008         & $-0.22\pm0.03$  & $0.16\pm0.02$   & $27.78\pm0.57$    & ---            & $26.06\pm0.47$    & ---                &$23.74\pm0.52$ & ---               & ---               \\
J0305$-$3150         & $-0.06\pm0.01$  & $0.02\pm0.01$   & $26.53\pm0.16$    & ---            & $25.32\pm0.09$    & ---                &$23.47\pm0.07$ & ---               & ---               \\
J2002$-$3013         & $0.14\pm0.02$   & $-0.10\pm0.02$  & $27.18\pm0.45$    & ---            & $25.37\pm0.30$    & ---                &$23.21\pm0.20$ & ---               & ---               \\
J2232$+$2930         & $-0.00\pm0.01$  & $-0.07\pm0.01$  & $25.08\pm0.12$    & ---            & $24.15\pm0.07$    & ---                &$22.82\pm0.07$ & ---               & ---               \\
CEERS 007465         & $-0.20\pm0.04$  & $0.01\pm0.01$   & $26.87\pm0.48$    & $26.95\pm0.50$ & $26.99\pm0.41$    & $26.81\pm0.39$     &$26.52\pm0.37$ & $26.80\pm0.36$    & $26.57\pm0.36$    \\
CEERS 006725         & $-0.01\pm0.02$  & $-0.01\pm0.02$  & $27.20\pm0.76$    & $27.31\pm0.80$ & $27.27\pm0.66$    & $27.02\pm0.67$     &$26.72\pm0.66$ & $26.99\pm0.66$    & $26.76\pm0.65$    \\
CEERS 027825         & $0.91\pm0.07$   & $-0.42\pm0.04$  & $26.06\pm0.17$    & $25.95\pm0.17$ & $25.85\pm0.18$    & $25.33\pm0.18$     &$25.30\pm0.18$ & $24.91\pm0.18$    & $25.01\pm0.17$    \\
CEERS 003975         & $0.78\pm0.06$   & $0.99\pm0.04$   & $24.10\pm0.08$    & $24.11\pm0.09$ & $24.16\pm0.06$    & $24.00\pm0.04$     &$23.64\pm0.03$ & $23.95\pm0.02$    & $23.73\pm0.02$    \\
J0109$-$3047-BHAE-1  & $0.01\pm0.01$   & $-0.02\pm0.01$  & $27.86\pm0.20$    & ---            & $27.18\pm0.16$    & ---                &$26.20\pm0.18$ & ---               & ---               \\
J0218$+$0007-BHAE-1  & $0.04\pm0.02$   & $0.01\pm0.03$   & $28.81\pm1.42$    & ---            & $28.09\pm0.98$    & ---                &$27.14\pm0.75$ & ---               & ---               \\
J0224$-$4711-BHAE-1  & $0.04\pm0.01$   & $0.03\pm0.01$   & $27.39\pm0.17$    & ---            & $26.73\pm0.10$    & ---                &$25.88\pm0.12$ & ---               & ---               \\
J0229$-$0808-BHAE-1  & $0.01\pm0.01$   & $0.06\pm0.01$   & $27.23\pm0.22$    & ---            & $26.69\pm0.17$    & ---                &$25.83\pm0.15$ & ---               & ---               \\
J0229$-$0808-BHAE-2  & $-0.27\pm0.60$  & $0.95\pm0.46$   & $26.61\pm0.10$    & ---            & $26.47\pm0.09$    & ---                &$25.63\pm0.10$ & ---               & ---               \\
J0430$-$1445-BHAE-1  & $0.15\pm0.01$   & $-0.05\pm0.01$  & $27.02\pm0.11$    & ---            & $26.58\pm0.04$    & ---                &$26.00\pm0.05$ & ---               & ---               \\
J0923$+$0402-BHAE-1  & $0.03\pm0.01$   & $0.04\pm0.01$   & $28.20\pm0.23$    & ---            & $27.71\pm0.16$    & ---                &$26.58\pm0.18$ & ---               & ---               \\
J1526$-$2050-BHAE-2\tablenotemark{$\star$}   & ---             & ---             & ---               & ---            & ---               & ---                & ---           & ---               & ---               \\
J1526$-$2050-BHAE-3  & $0.03\pm0.02$   & $0.00\pm0.01$   & $28.96\pm0.62$    & ---            & $28.43\pm0.50$    & ---                &$27.35\pm0.40$ & ---               & ---               \\
J2232$+$2930-BHAE-1  & $0.10\pm0.01$   & $0.08\pm0.01$   & $27.34\pm0.29$    & ---            & $25.70\pm0.07$    & ---                &$24.09\pm0.07$ & ---               & ---               \\
J2232$+$2930-BHAE-2  & $0.43\pm0.19$   & $0.78\pm0.35$   & $29.16\pm0.18$    & ---            & $28.57\pm0.09$    & ---                &$27.64\pm0.05$ & ---               & ---               \\
J2232$+$2930-BHAE-3  & $0.01\pm0.01$   & $0.01\pm0.01$   & $28.28\pm0.29$    & ---            & $27.84\pm0.22$    & ---                &$26.94\pm0.19$ & ---               & ---
\enddata
\tablecomments{The host galaxy magnitude derived from the \textsc{GalfitS} analysis of the high-redshift quasars.
Col. (1): Name of the quasar.
Cols. (2)--(3): Offset between the host galaxy center relative to the nucleus.
Cols. (4)--(10): Host galaxy magnitude derived according to Appendix~\ref{app:hostmag}.
\tablenotetext{\star}{The host galaxy of J1526$-$2050-BHAE-2 is not detected (Section~\ref{sec:hostdec}).}
}
\end{deluxetable*}

\end{document}